\documentclass[journal,letterpaper]{IEEEtran}

\usepackage[utf8]{inputenc}
\usepackage[T1]{fontenc}
\usepackage{graphicx}
\DeclareGraphicsExtensions{.pdf}
\usepackage{xcolor}

\usepackage{titling}
\pretitle{\begin{center}\fontsize{24bp}{24bp}\selectfont}
    \posttitle{\par\end{center}}
\preauthor{\begin{center}\fontsize{10bp}{14bp}\selectfont}
    \postauthor{\par\end{center}\vspace{24bp}}
\predate{}
\date{}
\postdate{}

\usepackage[cmex10]{amsmath}
\usepackage{amsthm,amssymb,amsfonts,mathrsfs,bm,mathtools}
\mathtoolsset{centercolon=true}

\DeclareMathAlphabet{\mathpzc}{OT1}{pzc}{m}{it}

\usepackage{newfloat}
\usepackage[caption=false,labelsep=quad,indention=1pt]{subfig}
\usepackage{setspace}
\usepackage[inline]{enumitem}
\usepackage{algorithm}
\usepackage{algpseudocode}

\usepackage{etoolbox}
\makeatletter
\expandafter\patchcmd\csname\string\algorithmic\endcsname{\itemsep\z@}{\itemsep=.6ex plus1pt}{}{}
\makeatother
\AtBeginEnvironment{algorithmic}{\setstretch{1}}
\usepackage{float}
\usepackage[hyphens]{url}
\usepackage[noadjust,sort,compress]{cite}
\usepackage{xspace}
\usepackage{xcolor}
\usepackage[bwr]{callouts}

\usepackage{pgfplots}
\pgfplotsset{compat=newest}
\pgfplotsset{select coords between index/.style 2 args={
    x filter/.code={
        \ifnum\coordindex<#1\def\pgfmathresult{}\fi
        \ifnum\coordindex>#2\def\pgfmathresult{}\fi
    }
}}

\newcommand{\Real}{\mathbb{R}}

\newcommand{\Complex}{\mathbb{C}}
\newcommand\given{{\mathbin{}\mid\mathbin{}}}
\newcommand\vect[1]{\mathbf{#1}}

\providecommand\given{} 
\newcommand\SetSymbol[1][]{
  \nonscript\,#1\vert \allowbreak \nonscript\,\mathopen{}}
\DeclarePairedDelimiterX\Set[1]{\lbrace}{\rbrace}%
{ \renewcommand\given{\SetSymbol[\delimsize]} #1 }
\DeclarePairedDelimiterX\innerp[2]{\langle}{\rangle}{#1
  \mathop{}\delimsize\vert\mathop{} #2}
\DeclarePairedDelimiterX\norm[1]\lVert\rVert{\ifblank{#1}{\:\cdot\:}{#1}}

\DeclareMathOperator{\trace}{trace}

\DeclareMathOperator{\diag}{diag}

\DeclareMathOperator{\prox}{Prox}

\DeclareMathOperator{\Id}{Id}

\DeclareMathOperator*{\Argmin}{arg\,min}

\DeclareMathOperator{\Vect}{vec}
\newcommand*{\conj}[1]{\overline{#1}}
\newcommand*{\hermconj}{\mathsf{H}}

\let\oldsqrt\sqrt
\def\sqrt{\mathpalette\DHLhksqrt}
\def\DHLhksqrt#1#2{%
\setbox0=\hbox{$#1\oldsqrt{#2\,}$}\dimen0=\ht0
\advance\dimen0-0.2\ht0
\setbox2=\hbox{\vrule height\ht0 depth -\dimen0}%
{\box0\lower0.4pt\box2}}

\newtheoremstyle{kostasstyle}
{2pt}
{2pt}
{\normalfont}
{0pt}
{\bfseries}
{.}
{1ex}
{}

\theoremstyle{kostasstyle}

\newtheorem{assumption}{Assumption}

\makeatletter

\newcommand*{\ie}{%
  \@ifnextchar{,}%
  {\textit{i.e.}}%
  {\textit{i.e.,}\@\xspace}%
}
\newcommand*{\eg}{%
  \@ifnextchar{,}%
  {\textit{e.g.}}%
  {\textit{e.g.,}\@\xspace}%
}
\newcommand*{\etc}{%
  \@ifnextchar{.}%
  {\textit{etc}}%
  {\textit{etc.}\@\xspace}%
}
\newcommand*{\etal}{%
  \@ifnextchar{.}%
  {\textit{et al}}%
  {\textit{et al.}\@\xspace}%
}
\newcommand*{\cf}{%
  \@ifnextchar{.}%
  {\textit{cf}}%
  {\textit{cf.}\@\xspace}%
}
\newcommand*{\aka}{%
  \@ifnextchar{,}%
  {\textit{a.k.a.}}%
  {\textit{a.k.a.}\@\xspace}%
}
\makeatother

\makeatletter
\def\endthebibliography{%
  \def\@noitemerr{
  \@latex@warning{Empty `thebibliography' environment}}%
  \endlist
}
\newcommand*{\balancecolsandclearpage}{%
  \close@column@grid
  \clearpage
  \twocolumngrid
}
\makeatother

\renewcommand\footnotemark{}
\MakeRobust{\eqref}
\MakeRobust{\cf}

\usepackage{authblk}

 \title{Bi-Linear Modeling of Data Manifolds for
  Dynamic-MRI Recovery} \author[1]{Gaurav N. Shetty}
\author[1]{Kostas Slavakis}
\author[2]{Abhishek Bose}
\author[3]{Ukash Nakarmi}
\author[4]{Gesualdo Scutari}
\author[5]{Leslie Ying}
\affil[1]{Department of Electrical Engineering, University at Buffalo. Emails:~\{gauravna,kslavaki\}@buffalo.edu.}
\affil[2]{Department of Computer Science, INRIA, Université Côte d’ Azur. Email:~abhishek.bose@inria.fr.}
\affil[3]{Department of Electrical Engineering and Radiology, Stanford University. Email:~nakarmi@stanford.edu.}
\affil[4]{Department of Industrial Engineering, Purdue University. Email: gscutari@purdue.edu.}
\affil[5]{Department of Electrical Engineering and Biomedical Engineering, University at Buffalo. Email:~leiying@buffalo.edu.}

\begin{document}
\maketitle
\begin{abstract} 
  This paper puts forth a novel bi-linear modeling framework for data recovery
  via manifold-learning and sparse-approximation arguments and considers its
  application to dynamic magnetic-resonance imaging (dMRI). Each temporal-domain
  MR image is viewed as a point that lies onto or close to a smooth manifold,
  and landmark points are identified to describe the point cloud concisely. To
  facilitate computations, a dimensionality reduction module generates
  low-dimensional/compressed renditions of the landmark points. Recovery of
  high-fidelity MRI data is realized by solving a non-convex minimization task
  for the linear decompression operator and affine combinations of
  landmark points which locally approximate the latent manifold geometry. An
  algorithm with guaranteed convergence to stationary solutions of the
  non-convex minimization task is also provided. The aforementioned framework
  exploits the underlying spatio-temporal patterns and geometry of the acquired
  data without any prior training on external data or information. Extensive
  numerical results on simulated as well as real cardiac-cine
  MRI data illustrate noteworthy improvements of the advocated machine-learning
  framework over state-of-the-art reconstruction techniques.
\end{abstract}


\section{Introduction}\label{sec:intro}

Magnetic-resonance imaging (MRI), a non-invasive, non-ionizing and high-fidelity
visualization technology, has found widespread applications in cardiac-cine,
dynamic contrast-enhanced and neuro-imaging, playing a key role in medical
research and diagnosis~\cite{Liang.Lauterbur.book}. MRI's inherent limitations
and various physiological constraints often incur slow data acquisition, while
long scanning times make MRI an expensive process, may cause patient discomfort,
and hinder thus its usefulness.

Raw MRI data are observed in the \textit{k-space}\/ domain; the image (visual)
data are computed by the inverse Fourier transform of the k-space
ones~\cite{Liang.Lauterbur.book}. A prominent way to speed up data acquisition
is to sample the k-space or frequency domain densely enough to ensure
reconstruction of a high-fidelity and artifact-free image via Fourier-transform
arguments~\cite{Petersen.Middleton.62}. In the case of dynamic (d)MRI, where an
extra-temporal dimension is added to the spatial domain, sampling is also
performed along the time axis. Due to MRI's slow scanning times, it becomes
difficult for the data acquisition process to keep up with the motion of the
organs or the fluid flow in the field of view
(FOV)~\cite{Liang.Lauterbur.book}. It is a usual case to not be able to reach
the necessary sampling density, not only in the temporal direction but also in
the k-space domain, to guarantee artifact-free reconstructed images. This
``under-sampling'' inflicts signal aliasing and
distortion~\cite{Liang.Lauterbur.dMRI.94}.

Naturally, a lot of the MRI-research effort has been focusing on developing
reconstruction algorithms that improve the spatio-temporal resolution of MR
images given the highly under-sampled k-space data. To this end,
artificial-intelligence (AI) approaches have been very recently placed at the
focal point of MRI research; examples are convolutional neural networks
(CNNs){~\cite{wang2016accelerating, chen2017low,
    zhu2018image, jin2017deep, aggarwal2018model,schlemper2018deep}}, deep
variational networks(VNs)~\cite{hammernik2018learning}, and generative
adversarial neural networks (GANs)~\cite{mardani2017recurrent,
  mardani2017deep}. AI methods learn non-linear mappings via extensive offline
training on large-scale datasets, different from or in addition to the acquired
data, and use those learned non-linear mappings to map the observed
low-resolution, or, undersampled data to their high-fidelity counterparts. In
contrast to the \textit{data-driven}\/ AI methods, the present work, as well as
the following prior-art schemes, assume \textit{no}\/ offline training on
large-scale datasets and resort \textit{solely}\/ to the observed data.

A popular approach to exploit the underlying spatio-temporal patterns within
dMRI data is compressed sensing (CS)~\cite{lustig2006kt, Jung.07,
  otazo2010combination, liang2012k}. Low-rank structures~\cite{lingala2011ktslr,
  zhao2012pssparse} and total-variation-based
schemes~\cite{block2007undersampled, knoll2011second, feng2014golden} have also
been explored at length and found to produce promising results for slow varying
dynamic data. For instance, \cite{zhao2012pssparse} proposes the estimation,
first, of a temporal basis of image time series via singular-value
decomposition, prior to formulating a sparsity inducing convex-recovery
task. Nevertheless, these schemes seem to be less effective when it comes to
dMRI with extensive inter-frame motion or with a low number of temporal frames,
as mentioned in~\cite{poddar2016dynamic, ravishankar2017.lassi}. {Acceleration capabilities of CS combined with the motion-robustness of radial
  imaging techniques have also found their place in MRI reconstruction and have
  shown promising results in cardiac and respiratory motion correction. The
  motion resolved strategy~\cite{feng2016xd} introduces extra motion dimensions
  by sorting the continuously acquired k-space data into distinct motion states
  followed by a sparsity and total variation based CS technique for
  reconstruction.} There has also been a growing interest in MRI-recovery by
dictionary-learning (DL) schemes~\cite{awate2012spatiotemporal,
  wang2014compressed, caballero2014dictionary, Nakarmi2016accelerating,
  Wang2017parallel}. In~\cite{ravishankar2017.lassi}, for example, the dMRI data
are decomposed in two components: A low-rank one, that captures the temporal
(video) background, and a sparse one, described via spatio-temporal patch-based
DL that models the (dynamic) foreground.

Manifold-learning techniques have also been employed to recover dMRI data from
highly under-sampled observations~\cite{nakarmi2017m, poddar2016dynamic,
  poddar2018free, poddar2018recovery, Usman.Manifold.15, nakarmi2018mls,
  ahmed2019free}. In \cite{poddar2016dynamic}, a graph-Laplacian matrix is
formed via the Euclidean distances between points of a data cloud and is
subsequently used as a regularizer in a convex-recovery
task. {Improving on~\cite{poddar2016dynamic}, a bandlimited modeling
  of the data points, coupled with an improved estimation of the Laplacian
  matrix, was proposed in~\cite{poddar2018free} to make the reconstruction
  resilient to noise and patient motion, computationally inexpensive and less
  demanding memory-wise. The work in \cite{poddar2018recovery} proposes a kernel
  variation to the graph Laplacian estimation in~\cite{poddar2016dynamic}.}  A
popular path followed by manifold-learning schemes is to perform dimensionality
reduction of the collected high-dimensional data prior to applying a
reconstruction algorithm, \eg, \cite{Usman.Manifold.15, nakarmi2018mls}. In the
non-MRI context, \cite{shen2017nonlinear} capitalizes, also, on a
Euclidean-distance-based Laplacian matrix to perform dimensionality reduction
prior to reconstructing data by local principal component analysis. In the
previous schemes, \textit{all}\/ of the observed points participate in the
dimensionality-reduction task, raising thus computational burdens, especially in
cases where the number of data is excessively large. Methods that perform
dimensionality reduction on properly chosen small-cardinality subsets of the
observed data cloud have been introduced for clustering and classification, but
\textit{not}\/ for regression tasks~\cite{silva2006selecting, chen2006improved,
  Landmark.MinMax}.

This paper follows the manifold-learning path and serves a two-fold objective:
\begin{enumerate*}[label={\bfseries\roman*)}]
\item Present a machine-learning framework, the bi-linear modeling of data manifolds (BiLMDM), that contributes novelties to exploiting local and latent data structures via a sparsity-aware and bi-linear optimization task; and
\item apply BiLMDM to the dMRI-data recovery problem.
\end{enumerate*}
In a nutshell, BiLMDM can be described as follows. Each vector of data, observed
from an undersampled dMRI temporal frame, is modeled as a point
onto or close to an unknown manifold, embedded in a Euclidean space. The only
assumption imposed on the manifold is smoothness~\cite{Tu.book.08}. {Landmark points are chosen to concisely describe the
  observed data-vector cloud, and are mapped to a lower-dimensional space, as
  in~\cite{nakarmi2018mls, RSE.13}, to effect compression and enable
  low-computational footprints in the proposed algorithmic solutions. Motivated
  by the smooth-manifold hypothesis, the proposed work approximates each data
  vector as an affine combinations of neighboring \textit{landmark}\/ points
  (\cf~Fig.~\ref{fig:kspace.LandmarkPoints}).} A \textit{locally
  bi-linear}\/ factorization model, novel for data representations, is then
formed to model/fit the point cloud: One factor gathers the coefficients of the
previous affine combinations of the landmark points, while the other one serves
as the linear \textit{decompression}\/ operator that unfolds the points back to
the image dimensions. Improving on our preliminary
results~\cite{slavakis2017bi}, a highly modular bi-linear optimization task is
tailored to the dMRI-data recovery problem, penalized by terms which account for
sparsity along the temporal axis and other modeling assumptions. A
successive-convex-approximation algorithm is proposed to guarantee convergence
to a stationary solution of the previous bi-linear optimization task.

{ The proposed work is validated
  against the following state-of-the-art schemes: Partially separable sparsity
  aware model (PS-Sparse)~\cite{zhao2012pssparse}, joint manifold learning and
  sparsity aware (MLS) framework~\cite{nakarmi2018mls}, smoothness
  regularization on manifolds (SToRM)~\cite{poddar2016dynamic}, low rank and
  adaptive sparse signal model (LASSI)~\cite{ravishankar2017.lassi} and extra
  dimensional golden angle radial sparse parallel
  (XD-GRASP)~\cite{feng2016xd}. In spite of sharing with the previous
  state-of-the-art schemes the principle of identifying low-dimensional
  structures for high-dimensional data and exploiting this structure to
  reconstruct data, BiLMDM's original contributions in data approximations can
  be summarized as follows: 
  \begin{enumerate*}[label={\bfseries\roman*)}]

  \item BiLMDM departs from the mainstream manifold-learning approach of
    identifying a data-graph Laplacian matrix that penalizes an optimization
    (data-recovery) task as in~\cite{poddar2018free, poddar2016dynamic, ahmed2019free}, and uses instead the fundamental geometric principle
    of tangent spaces of smooth manifolds to search for data
    dependencies/patterns through local affine combinations of landmark points
    (\cf~Fig.~\ref{fig:kspace.LandmarkPoints});

  \item it offers a novel bi-linear factorization model, where one linear factor
    captures the local smoothness of the manifolds, via local affine
    combinations, while the other one accounts for the linear decompression
    operator which maps the low-dimensional affine approximations back to the
    high-dimensional raw-data/input space. The affine combinations and the
    decompression operator are jointly identified via a highly modular
    non-convex data recovery task, which may accommodate also additional prior
    information, \eg, periodicity over the temporal axis of the dMRI data. Unlike the
    traditional bi-linear schemes, BiLMDM doesn't follow the classical dictionary learning approach
    of decomposing the image into two components as discussed earlier and model the dynamic foreground of the image as a sparse combination of adaptive dictionary atoms 
    (which act as image basis) of MR image signals. 

  \end{enumerate*}
  A more detailed description of the differences of BiLMDM with the
  state-of-the-art dMRI recovery schemes is deferred to
  Sec.~\ref{sec:Task.Algo}. The efficacy of BiLMDM was observed on both
  synthetically generated and experimentally acquired cardiac cine MR data,
  undersampled via both Cartesian and radial trajectories. BiLMDM has
  consistently outperformed all the competing techniques and produced
  reconstructions which were sharp, free of undesirable artifacts, deformations
  and temporal bleeding.}

The rest of the paper is organized as follows: Section~\ref{sec:data} describes,
in short, the dMRI acquisition scheme in the k-space
domain. Section~\ref{sec:assume} details the BiLMDM's modeling assumptions,
while Sec.~\ref{sec:Task.Algo} describes the algorithm to solve the proposed
non-convex minimization task. The extensive numerical tests of
Section~\ref{sec:results} showcase that BiLMDM outperforms state-of-the-art dMRI
recovery schemes. The manuscript {summarises the numerical results in
  Section~\ref{sec:discussion}} and concludes in
Section~\ref{sec:conclusion}. Finally, the appendix gathers basic mathematical
facts and expressions that are essential for the implementation of the algorithm
described in Sec.~\ref{sec:Task.Algo}.

\section{Bi-Linear Modeling of Data Manifolds}\label{sec:BiLMDM}

\subsection{dMRI data description}\label{sec:data}

\begin{figure}[t]
  \centering
  \subfloat[\label{fig:sampling.strategy.cartesian}] {\includegraphics[width =
    .3\linewidth]{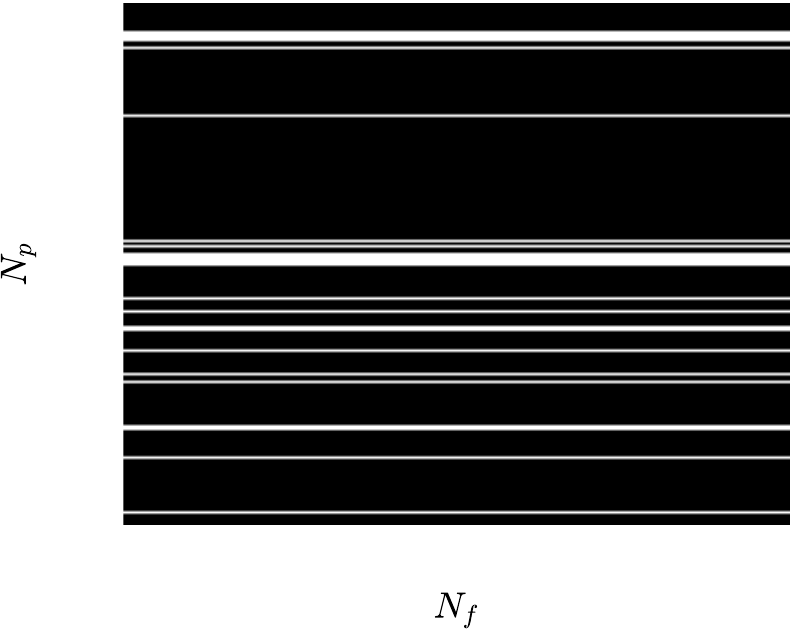}} \quad 
  \subfloat[\label{fig:sampling.strategy.radial}] {\includegraphics[width =
    .3\linewidth]{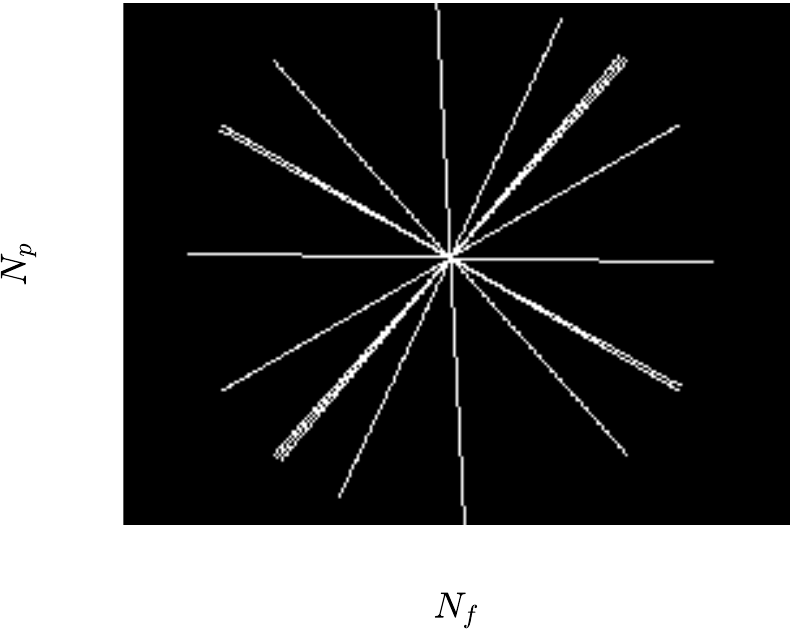}} \quad
  \subfloat[\label{fig:DataCube}]{\includegraphics[width = .3\linewidth]{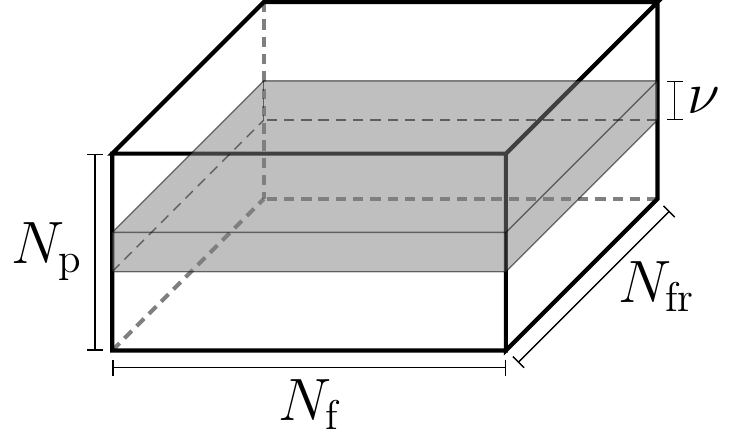}}
  \caption{\protect\subref{fig:sampling.strategy.cartesian} k-space with 1-D Cartesian sampling
    pattern; \protect\subref{fig:sampling.strategy.radial} k-space with radial sampling pattern;
    \protect\subref{fig:DataCube} The $N_{\text{p}} \times N_{\text{f}} \times N_{\text{fr}}$
    (k,t)-space. ``Navigator (pilot) data'' comprise the gray-colored
    $\nu \times N_{\text{f}} \times N_{\text{fr}}$ area of the (k,t)-space ($\nu\ll
    N_{\text{p}}$).} \label{fig:ktspace}
\end{figure}

MRI data $\bm{\mathcal{Y}} \in \Complex^{N_{\text{p}} \times N_{\text{f}}}$
($\Complex$ denotes the set of all complex-valued numbers) are observed in
\textit{k-space}\/ (frequency domain), which spans an area of size
$N_{\text{p}} \times N_{\text{f}}$
(\cf~Figs.~\ref{fig:sampling.strategy.cartesian} and
\ref{fig:sampling.strategy.radial}), with $N_{\text{p}}$ standing for the number
of phase-encoding lines and $N_{\text{f}}$ for the number of frequency-encoding
ones~\cite{Liang.Lauterbur.book}. Data $\bm{\mathcal{Y}}$ can be considered as
the two-dimensional (discrete) Fourier transform $\mathcal{F}(\cdot)$ of the
image-domain data
$\bm{\mathcal{X}} \in \Complex^{N_{\text{p}} \times N_{\text{f}}}$, \ie,
$\bm{\mathcal{Y}} =
\mathcal{F}(\bm{\mathcal{X}})$~\cite{Liang.Lauterbur.book}. Without any loss of
generality, this study assumes that the ``low-frequency'' part of
$\bm{\mathcal{Y}}$ is located around the center of the
$N_{\text{p}} \times N_{\text{f}}$ area. Availability of the data over the whole
k-space is infeasible in practice; k-space is usually severely
under-sampled~\cite{Liang.Lauterbur.dMRI.94}. There exist several strategies to
sample the k-space; examples are the 1-D Cartesian
(Fig.~\ref{fig:sampling.strategy.cartesian}) and the radial
(Fig.~\ref{fig:sampling.strategy.radial}) ones, where the ``white'' lines in
Figs.~\ref{fig:sampling.strategy.cartesian} and
\ref{fig:sampling.strategy.radial} denote the available/sampled data, while data
in the ``black'' areas are not observed. A general trend among sampling
strategies is to put more emphasis on low-frequency components, which carry
contrast information and with high SNR, and select few high-frequency
components, which comprise high-resolution image details. The 1-D Cartesian
sampling pattern emulates the acquisition of k-space pixels via the 1-D Gaussian
distribution, acquiring a large number of samples in the central k-space area
while sampling few ones from the ``high-frequency'' area
(\cf~Fig.~\ref{fig:sampling.strategy.cartesian}). The radial-sampling pattern
consists of radial spokes which yield dense sampling at the center of k-space,
while the sampling density is decreased as the spokes move away from the center
(\cf~Fig.~\ref{fig:sampling.strategy.radial}).

In dMRI, an additional dimension is added to the MRI k-space to accommodate time (the axis vertical
on the paper in Fig.~\ref{fig:DataCube}), resulting in the augmented (k,t)-space. The dMRI
(k,t)-space can be viewed, in other words, as the $N_{\text{fr}}$-fold Cartesian product of the
$(N_{\text{p}} \times N_{\text{f}})$-sized MRI k-space, where $N_{\text{fr}}$ represents the number
of observed MRI frames over time. In dMRI, k-space $\bm{\mathcal{Y}}_j$ and image-domain
$\bm{\mathcal{X}}_j$ data are connected via $\bm{\mathcal{Y}}_j = \mathcal{F}(\bm{\mathcal{X}}_j)$,
$j\in\Set{1, \ldots, N_{\text{fr}}}$. The (k,t) space is usually highly under-sampled. To extract
reliable information from the (k,t)-space data, this work follows~\cite{poddar2016dynamic, zhao2012pssparse, nakarmi2018mls} and considers a small number $\nu$ ($\ll N_{\text{p}}$) of phase-encoding lines, coined "navigator (pilot) data" (the gray-colored area in Fig.~\ref{fig:DataCube}), to learn the
intrinsic low-dimensional structure of the data.

To facilitate processing, the (k,t)-space data are vectorized. More specifically,
$\Vect(\bm{\mathcal{Y}}_j)$ stacks one column of $\bm{\mathcal{Y}}_j$ below the other to yield the
complex-valued $N_{\text{k}}\times 1$ vector $\vect{y}_j :=
\Vect(\bm{\mathcal{Y}}_j)$ where $N_{\text{k}} := N_{\text{p}}N_{\text{f}}$ denotes the number of pixels in every frame. To avoid notation clutter, $\mathcal{F}$ still denotes the
two-dimensional (discrete) Fourier transform even when applied to vectorized versions of image
frames:
$\mathcal{F}[\Vect(\bm{\mathcal{X}}_j)] := \Vect[\mathcal{F}(\bm{\mathcal{X}}_j)]=
\Vect(\bm{\mathcal{Y}}_j)$. All vectorized k-space frames are gathered in the
$N_{\text{k}}\times N_{\text{fr}}$ matrix
$\vect{Y} := [\vect{y}_1, \vect{y}_2, \ldots, \vect{y}_{N_{\text{fr}}}]$ so that the vectorized
image-domain data are
$\vect{X} := \mathcal{F}^{-1}(\vect{Y}) := [\mathcal{F}^{-1}(\vect{y}_1),
\mathcal{F}^{-1}(\vect{y}_2), \ldots, \mathcal{F}^{-1}(\vect{y}_{N_{\text{fr}}})]$, where
$\mathcal{F}^{-1}(\cdot)$ denotes the inverse two-dimensional (discrete) Fourier transform. The
navigator data of the $j$th k-space frame (\cf~Fig.~\ref{fig:DataCube}),
$j \in \{1, 2, \ldots, N_{\text{fr}} \}$, are gathered into a $\nu N_{\text{f}}\times 1$ vector
$\vect{y}_j^{\text{nav}}$. All navigator data comprise the $\nu N_{\text{f}}\times N_{\text{fr}}$
matrix
$\vect{Y}_{\text{nav}} := [\vect{y}_1^{\text{nav}}, \vect{y}_2^{\text{nav}}, \ldots,
\vect{y}_{N_{\text{fr}}}^{\text{nav}}]$.

\begin{figure}[!tp]
  \centering
  \includegraphics[width=.5\linewidth]{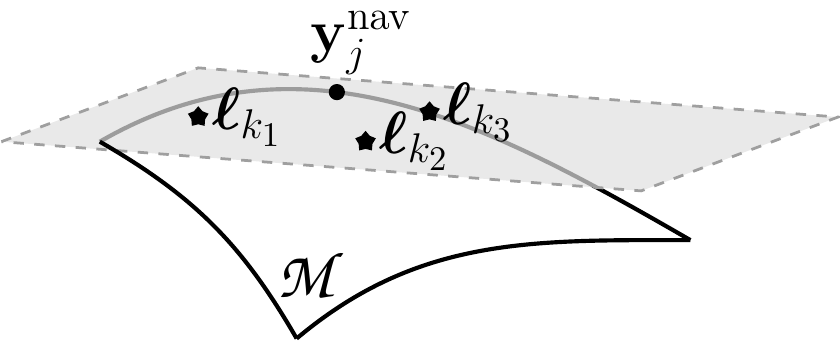}\label{fig:LandAffineComb}
  \caption{Landmark points $\Set{\bm{\ell}_{k_i}}_{i=1}^3$ are affinely combined to describe
    $\vect{y}_j^{\text{nav}}$. All affine combinations of $\Set{\bm{\ell}_{k_i}}_{i=1}^3$ are
    depicted by the gray-colored plane.}\label{fig:kspace.LandmarkPoints}
\end{figure}

\subsection{Modeling assumptions}\label{sec:assume}

The high-dimensional navigator data $\Set{\vect{y}_j^{\text{nav}}}_{j=1}^{N_{\text{fr}}}$ carry
useful information about spatio-temporal dependencies in the (k,t)-space. To promote parsimonious
data representations, especially in cases where $N_{\text{fr}}$ attains large values, it is
desirable to extract a subset
$\Set{\bm{\ell}_k}_{k=1}^{N_{\ell}} \subset \Set{\vect{y}_j^{\text{nav}}}_{j=1}^{N_{\text{fr}}}$
($N_{\ell}\leq N_{\text{fr}}$), called \textit{landmark}\/ points, which provide a ``concise
description,'' in a user-defined sense, of the data cloud
$\Set{\vect{y}_j^{\text{nav}}}_{j=1}^{N_{\text{fr}}}$. To this end, the following assumption, often
met in manifold-learning approaches~\cite{Saul.Roweis.03}, imposes structure on
$\Set{\vect{y}_j^{\text{nav}}}_{j=1}^{N_{\text{fr}}}$.

\begin{assumption}\label{as:ManifoldData}
  Data $\Set{\vect{y}_j^{\text{nav}}}_{j=1}^{N_{\text{fr}}}$ lie on a smooth low-dimensional
  manifold $\mathpzc{M}$~\cite{Tu.book.08} embedded in the high-dimensional Euclidean space
  $\Complex^{\nu N_{\text{f}}}$ (\cf~Fig.~\ref{fig:kspace.LandmarkPoints}).
\end{assumption}

\noindent For example, the most well-known case of $\mathpzc{M}$ is a linear
subspace, {which is the main
  hypothesis behind principal component analysis (PCA). PCA assumes that the
  data lie close to or onto a linear subspace in a low-dimensional space. Along
  the same lines,} based on As.~\ref{as:ManifoldData} and the concept of the
tangent space of a smooth manifold, it is conceivable that neighbouring landmark
points cooperate affinely to describe vector $\vect{y}_j^{\text{nav}}$ (the
gray-colored area in Fig.~\ref{fig:kspace.LandmarkPoints} depicts \textit{all}\/
possible affine combinations of
$\Set{\bm{\ell}_{k_1}, \bm{\ell}_{k_2}, \bm{\ell}_{k_3}}$). {In other words, every temporal frame is approximated by a combination of landmark points, where these landmark points
  can be viewed as a representative of a particular group of frames sharing
  similar phase and motion characteristics.}  Upon defining the
$\nu N_{\text{f}}\times N_{\ell}$ matrix
$\bm{\Lambda} := [\bm{\ell}_1, \bm{\ell}_2, \ldots, \bm{\ell}_{N_{\ell}}]$,
{the above notion can be established
  by the existence of} an $N_{\ell}\times 1$ vector $\vect{b}_j$
that renders the approximation error
$\norm{\vect{y}_j^{\text{nav}} - \bm{\Lambda}\vect{b}_j}$ small, where
$\norm{\cdot}$ denotes the standard Euclidean norm of space
$\Complex^{\nu N_{\text{f}}}$. Since affine combinations are desirable,
$\vect{b}_j$ is constrained to satisfy
$\vect{1}_{N_{\ell}}^{\intercal} \vect{b}_j = 1$, where $\vect{1}_{N_{\ell}}$
stands for the all-one $N_{\ell}\times 1$ vector and superscript $\intercal$
denotes vector/matrix transposition. Moreover, motivated by the low-dimensional
nature of $\mathpzc{M}$ (As.~\ref{as:ManifoldData}), only a few landmark points
are considered to cooperate to represent $\vect{y}_j^{\text{nav}}$ in its closed
vicinity, \ie, $\vect{b}_j$ is sparse. The previous arguments are summarized
into the following modeling hypothesis.

\begin{assumption}\label{as:Ynav.and.LandMarkPoints}
  There exist a sparse $N_{\ell}\times N_{\text{fr}}$ matrix $\vect{B}$, with
  $\vect{1}_{N_{\ell}}^{\intercal} \vect{B} = \vect{1}_{N_{\text{fr}}}^{\intercal}$, and a
  $\nu N_{\text{f}} \times N_{\text{fr}}$ matrix $\vect{E}_2$, which gathers approximation errors, such that
  (s.t.)\ $\vect{Y}_{\text{nav}} = \bm{\Lambda}\vect{B} + \vect{E}_2$.
\end{assumption}

\noindent The previous assumption holds true in the prototypical case where
$\mathpzc{M}$ is a linear subspace and $\vect{Y}_{\text{nav}}$ comprises column
vectors which lie close to or onto $\mathpzc{M}$. Any matrix which includes as
column vectors any basis of $\mathpzc{M}$ may serve as $\bm{\Lambda}$. Since the
affine hull~\cite{Rockafellar.convex.analysis} of the columns of $\bm{\Lambda}$
coincides with $\mathpzc{M}$, there exists surely a coefficient matrix
$\vect{B}$ that satisfies the affine constraints
$\vect{1}_{N_{\ell}}^{\intercal} \vect{B} =
\vect{1}_{N_{\text{fr}}}^{\intercal}$ and
$\vect{Y}_{\text{nav}} = \bm{\Lambda} \vect{B}$. {It is not now difficult to see that
  As.~\ref{as:Ynav.and.LandMarkPoints} covers also the case where $\mathpzc{M}$
  is a union $\cup_{q=1}^Q \mathpzc{M}_q$ of linear subspaces
  $\Set{\mathpzc{M}_q}_{q=1}^Q$, with $\vect{Y}_{\text{nav}}$ comprising columns
  vectors which lie close to or onto $\Set{\mathpzc{M}_q}_{q=1}^Q$. Hence,
  As.~\ref{as:Ynav.and.LandMarkPoints} lays the foundations for more general
  cases where data vectors lie close to or onto a union of smooth manifolds,
  with different dimensions. In such a way, As.~\ref{as:Ynav.and.LandMarkPoints}
  offers theoretical support and justification in cases where data comprise
  disparate groups, where intra-group dependencies can be captured by a single
  smooth manifold, while inter-group relations can be only viewed via the union
  of those manifolds, \eg, the union of data drawn from a static gray-colored
  background and data describing the movement of a beating heart.}

Although several strategies may be implemented to identify the landmark points
$\bm{\Lambda}$, a greedy optimization methodology, introduced in
\cite{Landmark.MinMax}, is adopted here. In short, at every step of the
algorithm, a landmark point is selected from
$\Set{\vect{y}_j^{\text{nav}}}_{j=1}^{N_{\text{fr}}}$ that maximizes, over all
\textit{un-selected}\/ $\Set{\vect{y}_j^{\text{nav}}}_{j=1}^{N_{\text{fr}}}$,
the minimum distance to the landmark points which have been already selected up
to the previous step of the algorithm. The algorithm of \cite{Landmark.MinMax}
scores a computational complexity of order
$\mathcal{O}(N_{\ell} N_{\text{fr}})$, which is naturally heavier than that of a
procedure that selects $\Set{\bm{\ell}_k}_{k=1}^{N_{\ell}}$ randomly from
$\Set{\vect{y}_j^{\text{nav}}}_{j=1}^{N_{\text{fr}}}$.

Still, the landmark points (columns of $\bm{\Lambda}$) are high dimensional. To
meet restrictions imposed by finite computational resources, it is desirable to
reduce the dimensionality of $\bm{\Lambda}$. To this end, the methodology of
\cite{RSE.13}, which is motivated by \cite{Saul.Roweis.03,
  Elhamifar.Vidal.nips.11}, is employed. The approach comprises of two steps:

\begin{subequations}\label{LLE}
  \begin{enumerate}[topsep=0pt,leftmargin=*]

  \item Given $\bm{\Lambda}$ and a user-defined $\lambda_W>0$, solve
    \begin{align}
      \min_{\vect{W} \in \Complex^{N_{\ell}\times
      N_{\ell}}}
      & \norm*{\bm{\Lambda} - \bm{\Lambda}
        \vect{W}}_{\text{F}}^2 + \lambda_W \norm{\vect{W}}_1 \notag
      \\
      \text{s.to}\
      & \vect{1}_{N_{\ell}}^{\intercal} \vect{W} =
        \vect{1}_{N_{\ell}}^{\intercal}\ \text{and}\ \diag(\vect{W})
        = \vect{0}\,, \label{LLE.step1}
    \end{align}
    where $\norm{\cdot}_{\text{F}}$ stands for the Frobenius norm of a matrix. Since $\Set{\bm{\ell}_{k}}_{k=1}^{N_{\ell}}$ lie on
    the manifold $\mathpzc{M}$, then according to As.~\ref{as:ManifoldData} and
    Fig.~\ref{fig:kspace.LandmarkPoints}, any
    point taken from $\Set{\bm{\ell}_{k}}_{k=1}^{N_{\ell}}$ may be faithfully approximated by an
    affine combination of the rest of the landmark points. In other words, there exists a matrix
    $\vect{W}$ s.t.\ $\bm{\Lambda} \approx \bm{\Lambda} \vect{W}$. With
    $\vect{1}_{N_{\ell}}^{\intercal} \vect{W} = \vect{1}_{N_{\ell}}^{\intercal}$ manifesting
    the previous desire for affine combinations, the constraint $\diag(\vect{W}) = \vect{0}$ is used
    to exclude the trivial solution of the identity matrix $\vect{I}_{N_{\ell}}$ for
    $\vect{W}$. Task \eqref{LLE.step1} is an affinely constrained composite convex minimization
    task, and, hence, the framework of \cite{slavakis.FMHSDM} can be employed to solve it, due to
    the flexibility by which \cite{slavakis.FMHSDM} deals with affine constraints when compared with
    state-of-the-art convex optimization techniques.

  \item Once $\vect{W}$ is obtained from the previous step and for a user-defined integer number
    $d \leq \min\{N_{\ell}, \nu N_{\text{f}}\}$, solve
    \begin{align}
      \min_{\check{\bm{\Lambda}}\in \Complex^{d\times
      N_{\ell}}}
      & \norm{\check{\bm{\Lambda}} -
        \check{\bm{\Lambda}} \vect{W}}_{\text{F}}^2\ 
        \text{s.to}\
        \check{\bm{\Lambda}} \check{\bm{\Lambda}}^{\hermconj} = \vect{I}_d\,, \label{LLE.step2}
    \end{align}
    where the constraint $\check{\bm{\Lambda}} \check{\bm{\Lambda}}^{\hermconj} = \vect{I}_d$ is
    used to exclude the trivial solution of $\check{\bm{\Lambda}} = \vect{0}$, and the superscript
    $\hermconj$ denotes the Hermitian transpose of a matrix. The solution of the previous task is
    nothing but the complex conjugate transpose of the matrix which comprises the $d$ minimal
    eigenvectors of
    $(\vect{I}_{N_{\ell}} - \vect{W}) (\vect{I}_{N_{\ell}} - \vect{W})^{\hermconj}$.
  \end{enumerate}
\end{subequations}

{To this point, the identified landmark points were used to capture a latent structure of the MR data cloud and the structure was further used to compress the landmark points to a lower dimension, thus, facilitating the search for affine combinations in a lower dimensional space. However, there arises a need to unfold the points back to the dimensions of the navigator data which leads us to the following model.}  

\begin{assumption}\label{as:linear.model.LandMarkPoints}
  There exist an $N_{\text{k}} \times d$ matrix $\vect{G}_3$ and an
  $N_{\text{k}}\times N_{\ell}$ matrix $\vect{E}_3$, which gathers all approximation
  errors, s.t.\ $\bm{\Lambda} = \vect{G}_3 \check{\bm{\Lambda}} + \vect{E}_3$.
\end{assumption}

\noindent Matrix $\vect{G}_3$ can be viewed as the ``decompression'' operator
which reconstructs the ``full'' $\bm{\Lambda}$ from its low-dimensional
representation $\check{\bm{\Lambda}}$.

{The relationship between the
  undersampled k-space frames and navigator frames can be established, for the
  case of Fig.~\ref{fig:DataCube}, as} 
$\vect{Y}_{\text{nav}} = \bm{\Omega} \vect{Y}$, where $\bm{\Omega}$ is a matrix
with binary entries $\{0, 1\}$ that select entries of $\vect{Y}$. The inverse
problem of recovering $\vect{Y}$ from its partial $\vect{Y}_{\text{nav}}$ is
viable under errors, \eg, an approximation of $\vect{Y}$ can be obtained via
$\bm{\Omega}^{\dagger} \vect{Y}_{\text{nav}}$, where the decompressor
$\bm{\Omega}^{\dagger}$ stands for the Moore-Penrose pseudo-inverse of
$\bm{\Omega}$. The following modeling hypothesis generalizes this argument.

\begin{assumption}\label{as:Y.and.Ynav}
  There exist an $N_{\text{k}} \times \nu N_{\text{f}}$ matrix $\vect{G}_1$ and an
  $N_{\text{k}} \times N_{\text{fr}}$ matrix $\vect{E}_1$, which gathers all
  approximation errors, s.t. $\vect{Y} = \vect{G}_1 \vect{Y}_{\text{nav}} +
  \vect{E}_1$.
\end{assumption}

\noindent A similar modeling assumption can be found in the classical principal component analysis
(PCA)~\cite{friedman2001elements}, where $\vect{G}_1$ serves as the ``decompression'' operator
(usually an orthogonal matrix) that maps the ``compressed'' $\vect{Y}_{\text{nav}}$ back to the
original data $\vect{Y}$ under the approximation error $\vect{E}_1$. However, PCA identifies
best-linear-fit approximations to data, which might be an inappropriate modeling assumption for
several types of data~\cite{Saul.Roweis.03}. The following discussion departs from PCA and
establishes modeling assumptions that allow non-linear data geometries.

Putting modeling assumptions \ref{as:Ynav.and.LandMarkPoints},
\ref{as:linear.model.LandMarkPoints} and \ref{as:Y.and.Ynav} together, it can be verified that there exist matrices
$\vect{G}$ and $\vect{E}$ s.t.\ $\vect{Y} = \vect{G} \check{\bm{\Lambda}} \vect{B} + \vect{E}$. Upon
defining $\vect{U}:= \mathcal{F}^{-1}(\vect{G})$, and since
$\vect{G} \check{\bm{\Lambda}} \vect{B} = \mathcal{F}(\vect{U}) \check{\bm{\Lambda}} \vect{B} =
\mathcal{F}(\vect{U} \check{\bm{\Lambda}} \vect{B})$, by virtue of the linearity of $\mathcal{F}$,
the following bi-linear\/ model between $\vect{Y}$ and the unknowns $(\vect{U}, \vect{B})$ is
established:
\begin{align}
  \vect{Y} = \mathcal{F}(\vect{U} \check{\bm{\Lambda}} \vect{B})
  + \vect{E} \,. \label{bi.linear.model}
\end{align}
Bi-linearity means that if $\vect{U}$ (or $\vect{B}$) is fixed to a
specific value, then $\vect{Y}$ is linear with respect to $\vect{B}$ (or $\vect{U}$), modulo the error $\vect{E}$ term. Interestingly, the linearity of $\mathcal{F}^{-1}$ suggests that
the previous modeling hypothesis holds true also in the image domain:
$\mathcal{F}^{-1}(\vect{Y}) = \vect{U} \check{\bm{\Lambda}} \vect{B} + \mathcal{F}^{-1}(\vect{E})$.

\section{The Bi-Linear Recovery task and its Algorithmic Solution}\label{sec:Task.Algo} 

In practice, only few (k,t)-space data are known. To explicitly take account of the limited number
of data, a (linear) sampling/masking operator $\mathcal{S}(\cdot)$ is introduced, where
$\mathcal{S}(\vect{Y})$ leaves the entries of $\vect{Y}$ as they are at sampled or observed
positions of the k-space domain while nullifying all the rest. The sampling operator is capable
of mimicking any sampling strategy, such as Cartesian, radial, spiral, \etc. It is also often in
dMRI that image frames capture a periodic process, \eg, heart movement, other than the static
background. In other words, it is reasonable to assume that in \eqref{bi.linear.model}, the
one-dimensional Fourier transform $\mathcal{F}_t$ of the $1\times N_{\text{fr}}$ time profile of
every one of the $N_{\text{k}}$ pixels, \ie, every row of the matrix
$\mathcal{F}_t(\vect{U} \check{\bm{\Lambda}} \vect{B})$, is a sparse vector.

All of the previous modeling assumptions are incorporated in the following recovery task: Given
the positive real-valued parameters $\lambda_1, \lambda_2, \lambda_3, C_U$, solve
\begin{align}
  \min_{(\vect{U}, \vect{B}, \vect{Z})} 
  &\ \overbrace{\tfrac{1}{2} \norm*{\mathcal{S}(\vect{Y}) -
    \mathcal{S} \mathcal{F}(\vect{U} 
    \check{\bm{\Lambda}} \vect{B})}_{\text{F}}^2}^{\text{T1}} 
    + \overbrace{\tfrac{\lambda_1}{2} \norm*{\vect{Z} -
    \mathcal{F}_t(\vect{U} \check{\bm{\Lambda}}
    \vect{B})}_{\text{F}}^2}^{\text{T2}} \notag\\
  &\ + \underbrace{\lambda_2 \norm*{\vect{Z}}_1}_{\text{T3}} + \underbrace{ \lambda_3
    \norm*{\vect{B}}_1}_{\text{T4}} \notag\\
  \text{s.to}\
  &\ \underbrace{\norm*{\vect{Ue}_i} \leq C_U,\ \forall i\in\Set{1, \ldots,
    d}}_{\text{C1}};~\underbrace{\vect{1}_{N_{\ell}}^{\intercal} \vect{B} = 
    \vect{1}_{N_{\text{fr}}}^{\intercal}}_{\text{C2}}; \notag\\
  & \vect{U}\in \Complex^{N_{\text{k}} \times d}; \vect{B}\in \Complex^{N_{\ell}
    \times N_{\text{fr}}}; \vect{Z}\in\Complex^{N_{\text{k}} \times
    N_{\text{fr}}} , \label{recovery.task} 
\end{align}
where $\vect{e}_i$ denotes the $i$th column of the identity matrix
$\vect{I}_d$. Elaborating more on the recovery task~\eqref{recovery.task}, T1
corresponds to the data-fit term, while T2 and T3 introduce the auxiliary
variable $\vect{Z}$, used to impose a sparsity constraint on
$\mathcal{F}_t(\vect{U} \check{\bm{\Lambda}} \vect{B})$. T4 imposes a sparsity
constraint on $\vect{B}$, following the discussion on
As.~\ref{as:Ynav.and.LandMarkPoints}. Bound $C_U$ in C1 is used to prevent
unbounded solutions for $\vect{U}$ due to the scaling ambiguity in the bi-linear
term $\vect{U} \check{\bm{\Lambda}} \vect{B}$. C2 adds the affine constraint
discussed in As.~\ref{as:Ynav.and.LandMarkPoints}. {
  Moreover, it is worth noticing here that the reduction of dimensionality,
  achieved via $\check{\bm{\Lambda}}$, reduces also the number of columns and
  unknowns, \ie, degrees of freedom, of $\vect{U}$.}

\begin{algorithm}[!t]
  \begin{algorithmic}[1]
    \renewcommand{\algorithmicindent}{1em}
    \addtolength\abovedisplayskip{-0.5\baselineskip}%
    \addtolength\belowdisplayskip{-0.5\baselineskip}%

    \Require{Available are data $\mathcal{S}(\vect{Y})$, including the navigator
      $\vect{Y}_{\text{nav}}$ ones. Choose parameters
      $\lambda_1, \lambda_2, \lambda_3, C_U, \tau_U, \tau_B >0$, as well as
      $\zeta\in (0,1)$ and $\gamma_0\in (0,1]$.}

    \Ensure{Extract the limit points $\vect{U}_*$ and $\vect{B}_*$ of sequences
      $(\vect{U}_n)_n$ and $(\vect{B}_n)_n$, respectively, and recover the dMRI data
      by the estimate $\hat{\vect{X}} := \vect{U}_* \check{\bm{\Lambda}}\vect{B}_*$.}

    \State\parbox[t]{\dimexpr\linewidth-\algorithmicindent}%
    {Identify landmark points $\bm{\Lambda}$ from the columns of
      $\vect{Y}_{\text{nav}}$ according to~\cite{Landmark.MinMax} (\cf~Sec.~\ref{sec:assume}).}

    \State\parbox[t]{\dimexpr\linewidth-\algorithmicindent}%
    {Compute the ``compressed'' $\check{\bm{\Lambda}}$ according to \eqref{LLE}.}

    \State\parbox[t]{\dimexpr\linewidth-\algorithmicindent}%
    {Arbitrarily fix $(\vect{U}_0, \vect{B}_0, \vect{Z}_0)$ and set $n=0$.}

    \While{$n\geq 0$}\label{alg.step:resume}

    \State\parbox[t]{\dimexpr\linewidth-\algorithmicindent}%
    {Available are $(\vect{U}_n, \vect{B}_n, \vect{Z}_n)$ and
      $\gamma_n$.}

    \State\parbox[t]{\dimexpr\linewidth-\algorithmicindent}%
    {Let $\gamma_{n+1} := \gamma_n (1-\zeta\gamma_n)$.}
    
    \State\parbox[t]{\dimexpr\linewidth-\algorithmicindent}%
    {Obtain $\hat{\vect{U}}_n$ of \eqref{task:min.for.U} and $\hat{\vect{B}}_n$ of
      \eqref{task:min.for.B} via Alg.~\ref{alg:algorithm.Un.Bn}, and the $(i,j)$th entry of
      $\hat{\vect{Z}}_n$, $\forall (i,j)$, via the following soft-thresholding rule:
      \begin{align*}
        [\hat{\vect{Z}}_n]_{ij}\ \mathbin{:=}\
        & [\mathcal{F}_t(\vect{U}_n \check{\bm{\Lambda}} \vect{B}_n) ]_{ij} \cdot 
        \Biggl(1 - \\
        & \frac{\lambda_2/\lambda_1}{\max \left\{\lambda_2/ \lambda_1,
        \left\lvert [\mathcal{F}_t(\vect{U}_n \check{\bm{\Lambda}} \vect{B}_n)]_{ij} \right\rvert
        \right\}} \Biggr) \,.
      \end{align*}}\label{alg.step:convex.tasks}

    \State\parbox[t]{\dimexpr\linewidth-\algorithmicindent}%
    {Update
      \begin{align*}
        (\vect{U}_{n+1}, \vect{B}_{n+1}, \vect{Z}_{n+1})\
        \mathbin{:=}\
        & (1-\gamma_{n+1}) (\vect{U}_n, \vect{B}_n,
          \vect{Z}_n) \\
        & + \gamma_{n+1} (\hat{\vect{U}}_{n},
        \hat{\vect{B}}_{n}, \hat{\vect{Z}}_{n}) \,.
      \end{align*}
    }

    \State\parbox[t]{\dimexpr\linewidth-\algorithmicindent}%
    {Set $n$ equal to $n+1$ and go to step~\ref{alg.step:resume}.}

    \EndWhile\label{alg.step:endwhile}
    
  \end{algorithmic}
  \caption{Recovering the dMRI data}\label{alg:recovery}
\end{algorithm}

{The proposed
  scheme shares with PS-Sparse~\cite{zhao2012pssparse},
  SToRM~\cite{poddar2016dynamic}, (and its other variants bandlimited-SToRM~\cite{poddar2018free}, navigator-less SToRM~\cite{ahmed2019free}) and MLS~\cite{nakarmi2018mls} similarities in
  terms of using navigator lines to learn the underlying data manifold. Instead
  of using the eigen vectors of the co-variance (PS-Sparse) and Laplacian (SToRM and its variants) matrices as the temporal basis, BiLMDM relies on robust sparse
  embedding~\cite{RSE.13} to estimate a low dimensional latent structure. SToRM
  and BiLMDM both aim at learning the structure of ``smooth'' data
  manifolds. Smoothness in SToRM is established globally over the data cloud by
  penalizing the recovery task via several operator norms of a graph-Laplacian
  matrix. On the other hand, BiLMDM enforces smoothness locally over the
  manifold via neighborhoods and affine approximations (patches) of the tangent
  spaces (\cf Fig.~\ref{fig:kspace.LandmarkPoints}), and the data manifold can
  be viewed as an approximation of the union of all those patches (as discussed
  in As.~\ref{as:Ynav.and.LandMarkPoints}). PCA-inspired schemes, such as
  PS-Sparse, search also for local affine/linear patches to
  manifolds. Nevertheless, BiLMDM does not impose any orthogonality constraints
  on any of its factors $\vect{U}$ and $\vect{B}$, as PCA via the singular
  value decomposition (SVD) does. Identification of landmark points to learn an underlying manifold is a novel contribution of the proposed work, given that all of the prior-art methods rely on the entire data cloud for learning. The incorporation of the landmark points in the bi-linear factorization model $\vect{U} \check{\bm{\Lambda}} \vect{B}$
  serves also as a natural encapsulation of the underlying data geometry into
  the optimization task, extending thus the usual way that PCA interprets its
  factors as an orthogonal/decorrelating basis and coefficients. Given that MLS
  employs the same approach of robust sparse embedding, the key feature of
  BiLMDM that sets itself apart from MLS, as well as other manifold-learning
  methods, is the bi-linear factorization model used to represent the MR data
  cloud. This bi-linear factorization model helps to capture the geometry of the
  point cloud locally by imposing a sparsity-promoting penalty on the affine
  combinations, thus, restricting the sharing of data among frames that show
  common phase and structural characteristics; local exploitation of
  dependencies is promoted over any global one. Such a bi-linear model differs also from LASSI~\cite{ravishankar2017.lassi}, which adopts the popular approach of viewing the data matrix as the superposition of a low-rank and a sparse component. The low-rank component describes the static image background, while the sparse component, modeled via a classical bi-linear dictionary-learning term, describes the dynamic foreground. In contrast to LASSI, BiLMDM uses bi-linear modeling to describe the dynamic foreground and the static background together. BiLMDM does not also require any
  pre-processing procedures, like sorting the acquired data into cardiac and
  respiratory phases, as in the motion resolved strategies of
  XD-GRASP~\cite{feng2016xd}, before solving the reconstruction
  task.}

The successive-convex-approximation framework of~\cite{facchinei2015parallel} is
employed to solve \eqref{recovery.task} and is presented in a concise form in
steps~\ref{alg.step:resume}--\ref{alg.step:endwhile} of
Alg.~\ref{alg:recovery}. Convergence to a stationary solution of
\eqref{recovery.task} is
guaranteed~\cite{facchinei2015parallel}. Step~\ref{alg.step:convex.tasks} of
Alg.~\ref{alg:recovery} comprises convex minimization sub-tasks. More
specifically, at every step of the algorithm, given
$(\vect{U}_n, \vect{B}_n, \vect{Z}_n)$, the following estimates are required
(for $\tau_U, \tau_B>0$):
\begin{subequations}\label{solve.for.U.and.B}
  \begin{align}
    \hat{\vect{U}}_n\in
    \Argmin_{\vect{U}}\,
    & \tfrac{1}{2} \norm*{\mathcal{S}(\vect{Y}) -
      \mathcal{S} 
      \mathcal{F}(\vect{U} \check{\bm{\Lambda}}
      \vect{B}_n)}_{\text{F}}^2 + \tfrac{\tau_U}{2}
      \norm*{\vect{U} - \vect{U}_n}_{\text{F}}^2 \notag\\
    & + \tfrac{\lambda_1}{2}
      \norm*{\vect{Z}_n - \mathcal{F}_t(\vect{U} \check{\bm{\Lambda}} 
      \vect{B}_n)}_{\text{F}}^2 \notag\\
    \text{s.to}\ &\ \norm*{\vect{Ue}_i} \leq C_U,\
                   \forall i\in\Set{1, \ldots, d}\,. \label{task:min.for.U} \\
    \hat{\vect{B}}_n\in \Argmin_{\vect{B}} 
    &\ \tfrac{1}{2} \norm*{\mathcal{S}(\vect{Y}) - \mathcal{S} \mathcal{F}(\vect{U}_n
      \check{\bm{\Lambda}} \vect{B})}_{\text{F}}^2 + \tfrac{\tau_B}{2} \norm*{\vect{B}-
      \vect{B}_n}_{\text{F}}^2 \notag\\ 
    &\ + \tfrac{\lambda_1}{2} \norm*{\vect{Z}_n - \mathcal{F}_t(\vect{U}_n \check{\bm{\Lambda}}
      \vect{B})}_{\text{F}}^2 + \lambda_3\norm*{\vect{B}}_1
      \notag\\
    \text{s.to}\
    &\ \vect{1}_{N_{\ell}}^{\intercal} \vect{B} =
      \vect{1}_{N_{\text{fr}}}^{\intercal} \,. \label{task:min.for.B}
  \end{align}
\end{subequations}
Both tasks in \eqref{solve.for.U.and.B} can be viewed as affinely constrained composite convex
minimization tasks, hence allowing the use of~\cite{slavakis.FMHSDM}, as described in
Alg.~\ref{alg:algorithm.Un.Bn}. From a computational complexity perspective, it is worth pointing
out that the proposed scheme relies on minimization sub-tasks, and computational complexities depend
thus on the solver adopted for solving those sub-tasks. Here, \cite{slavakis.FMHSDM} employs only
first-order information (gradients) and proximal mappings, \eg, soft-thresholding rules and
projection mappings. The implementation of \cite{slavakis.FMHSDM} for the specific tasks
\eqref{solve.for.U.and.B} is presented in Alg.~\ref{alg:algorithm.Un.Bn}, and details are deferred
to the appendix section of this manuscript. Notice also that the previous minimization sub-tasks
can be solved in parallel. Furthermore, problems~\ref{solve.for.U.and.B} can be solved inexactly at each iteration (with increasing precision, as described in~\cite{facchinei2015parallel}), which contributes to reducing the computation cost of each iteration; we refer to~\cite{facchinei2015parallel} for more details.

\begin{algorithm}[!t]
  \begin{algorithmic}[1]
    \renewcommand{\algorithmicindent}{1em}
    \addtolength\abovedisplayskip{-0.5\baselineskip}%
    \addtolength\belowdisplayskip{-0.5\baselineskip}%
    
    \Require{$\vect{D}$ is either $\vect{U}_n$ or $\vect{B}_n$ in step~\ref{alg.step:convex.tasks} of
      Alg.~\ref{alg:recovery}. Choose parameter $K_0,~\alpha \in [0.5, 1)$.}

    \Ensure{$\hat{\vect{U}}_n$ and $\hat{\vect{B}}_n$ are set equal to the limit $\vect{H}_{K_0}$ of the
      sequence $(\vect{H}_k)_k$.}

    \State\parbox[t]{\dimexpr\linewidth-\algorithmicindent}%
    {Compute the Lipschitz coefficient $L$ via \eqref{supp:lipschitz.U} or \eqref{supp:lipschitz.B},
      and choose $\lambda \in (0,2(1-\alpha)/L)$.}
    
    \State\parbox[t]{\dimexpr\linewidth-\algorithmicindent}%
    {$\vect{H}_0 := \vect{D}$.}
    
    \State\parbox[t]{\dimexpr\linewidth-\algorithmicindent}%
    {Letting $T$ be either \eqref{supp:T.for.U} or \eqref{supp:T.for.B}, define $T_{\alpha} :=
      \alpha T + (1-\alpha)\Id$, where $\Id$ denotes the identity operator.}

    \State\parbox[t]{\dimexpr\linewidth-\algorithmicindent}%
    {Set $\vect{H}_{1/2} := T_{\alpha}(\vect{H}_{0}) - \lambda \nabla g_1(\vect{H}_{0})$, where the
      gradient $\nabla g_1(\vect{H}_{0})$ takes the form of either \eqref{supp:gradient.U} or
      \eqref{supp:gradient.B}.}
    
    \State\parbox[t]{\dimexpr\linewidth-\algorithmicindent}%
    {Set $\vect{H}_{1} := \prox_{\lambda g_2}(\vect{H}_{1/2})$, where the proximal operator takes
      the form of either \eqref{supp:proximal.U} or \eqref{supp:proximal.B}.}
    
    \While{$k \leq K_0$}\label{alg:resume}

    \State\parbox[t]{\dimexpr\linewidth-\algorithmicindent}%
    {$\vect{H}_{k+3/2} := \vect{H}_{k+1/2} + T(\vect{H}_{k+1})
        - \lambda \nabla g_1(\vect{H}_{k+1}) - T_{\alpha}(\vect{H}_k) + \lambda
        \nabla g_1(\vect{H}_k)$.}\label{alg:step1.iteration}
    
    \State\parbox[t]{\dimexpr\linewidth-\algorithmicindent}%
    {$\vect{H}_{k+2} := \prox_{\lambda g_2}(\vect{H}_{k+3/2})$.}

    \State\parbox[t]{\dimexpr\linewidth-\algorithmicindent}%
    {Set $k$ equal to $k+1$ and go to step~\ref{alg:resume}.}

    \EndWhile
    
  \end{algorithmic}
  \caption{Computing $\hat{\vect{U}}_n$ of \eqref{task:min.for.U} and $\hat{\vect{B}}_n$ of
    \eqref{task:min.for.B}} \label{alg:algorithm.Un.Bn}
\end{algorithm}

\section{Numerical Results}\label{sec:results}
\pgfplotstableread{
  x       bi          error       PS          UK          ST          LASSI     XDGRASP
  4x      0.0438      0.00034     0.0437      0.0437      0.0526      0.0539    0.049
  8x      0.0454      0.00031     0.0449      0.0449      0.064       0.0552    0.0505
  12x     0.0461      0.00084     0.0462      0.0463      0.0762      0.0630    0.0521
  16x     0.0478      0.00079     0.048       0.048       0.085       0.0716    0.0591
  20x     0.0488      0.00025     0.055       0.0512      0.0937      0.0732    0.0653
  24x     0.0499      0.00083     0.0752      0.072       0.0984      0.0960    0.0693
}{\cardiacCineData}
\pgfplotstableread{
  x       bi          error          PS          UK          ST       LASSI
  5x      0.1033      0.00031        0.2939      0.2680      0.2224   0.2287
  9x      0.2283      0.00029        0.3118      0.3400      0.3447   0.2929
  13x     0.2525      0.00025        0.3274      0.3459      0.3877   0.3543
  18x     0.2931      0.00027        0.3450      0.3589      0.4464   0.3832
  24x     0.3235      0.00021        0.3725      0.3714      0.5012   0.4052
}{\aperiodicPincat}
\pgfplotstableread{
  x       bi          error         PS          UK          ST        LASSI     XDGRASP
  6x      0.0488      0.00011        0.0607      0.0632      0.0530    0.0721   0.0535
  12x     0.0507      0.00015        0.0752      0.0791      0.0558    0.0759   0.0561
  17x     0.0527      0.00011        0.0863      0.0892      0.0583    0.0791   0.0587
  21x     0.0592      0.00012        0.0889      0.0932      0.0606    0.0849   0.0631
  34x     0.0636      0.00013        0.0959      0.1029      0.0668    0.0913   0.0674
}{\inVivoMyo}
\begin{figure}[ht!]
  \centering
  \subfloat[\label{fig:mrxcat.nrmse}]{
    \centering
    \begin{tikzpicture}[scale=0.9] {every tick label/.append style={font=\Huge}}
      \begin{axis} [symbolic x coords={4x,8x,12x,16x,20x,24x},xtick=data,
        xlabel= Cartesian {Acceleration Factors}, ylabel= NRMSE,
        ymin=0.04, ymax=0.10, legend style = {nodes = {scale = 0.8, transform shape}, draw = none,
          at = {(0.01,0.95)}, anchor = north west},  
        y tick label style={
          /pgf/number format/.cd,
          fixed,
          fixed zerofill,
          precision=2,
          /tikz/.cd
        }]
        \addplot+[mark = triangle,mark size=3pt, color=blue, error bars/.cd, y dir=both, y explicit]
        table[x=x, y=bi, y error = error] {\cardiacCineData};
        \addlegendentry{BiLMDM}
        \addplot+[color=red, mark=x,mark size=3pt]
        table[x=x,y=PS] {\cardiacCineData};
        \addlegendentry{PS-Sparse}
        \addplot+[color=black, mark=o,mark size=3pt]
        table[x=x,y=UK] {\cardiacCineData};
        \addlegendentry{MLS}
        \addplot+[color=orange, mark=diamond,mark size=3pt]
        table[x=x,y=ST] {\cardiacCineData};
        \addlegendentry{StoRM} 
        \addplot+[color=teal, mark=square,mark size=3pt]
        table[x=x,y=LASSI] {\cardiacCineData};
        \addlegendentry{LASSI}
        \addplot+[style=solid, color=magenta, mark=star,mark size=3pt]
        table[x=x,y=XDGRASP] {\cardiacCineData};
        \addlegendentry{XDGRASP}
      \end{axis}
    \end{tikzpicture}} \\
  \subfloat[\label{fig:perfusion.nrmse}]{
    \centering
    \begin{tikzpicture}[scale=0.9] 
      \begin{axis} [symbolic x coords={6x,12x,17x,21x,34x},xtick=data,
        xlabel = Radial {Acceleration Factors}, ylabel=NRMSE,
        ymin=0.045, ymax=0.11, legend style={nodes={scale = 0.8, transform shape}, draw = none, at =
          {(0.01,0.95)}, anchor = north west}, y tick label style={
          /pgf/number format/.cd,
          fixed,
          fixed zerofill,
          precision=2,
          /tikz/.cd
        }]
        \addplot+[mark = triangle, mark size=3pt, error bars/.cd, y dir=both, y explicit]
        table[x=x,y=bi, y error = error] {\inVivoMyo};
        \addlegendentry{BiLMDM}
        \addplot+[color=red, mark = x,mark size=3pt]
        table[x=x,y=PS] {\inVivoMyo};
        \addlegendentry{PS-Sparse}
        \addplot+[color=black, mark = o,mark size=3pt]
        table[x=x,y=UK] {\inVivoMyo};
        \addlegendentry{MLS}
        \addplot+[color=orange, mark = diamond,mark size=3pt]
        table[x=x,y=ST] {\inVivoMyo};
        \addlegendentry{StoRM}
        \addplot+[color=teal, mark = square,mark size=3pt]
        table[x=x,y=LASSI] {\inVivoMyo};
        \addlegendentry{LASSI}
        \addplot+[style = solid, color=magenta, mark = star,mark size=3pt]
        table[x=x,y=XDGRASP] {\inVivoMyo};
        \addlegendentry{XDGRASP}
      \end{axis}
    \end{tikzpicture}}
  \caption{NRMSE values [\cf~\eqref{def.NRMSE}] computed for~\protect\subref{fig:mrxcat.nrmse} MRXCAT,
    and~\protect\subref{fig:perfusion.nrmse} real cardiac cine data vs.\ acceleration/undersampling
    rates. The NRMSE plot for BiLMDM, given the non-convex nature of the recovery task, is averaged over $25$ independent trials, with different initialization points for each trial. Error bars are also used to indicate the deviation from the sample means due to the random initializations of the non-convex algorithmic scheme. There are points where the error bars are too small to be clearly visible.}
  \label{fig:nrmse.sampling}
\end{figure}
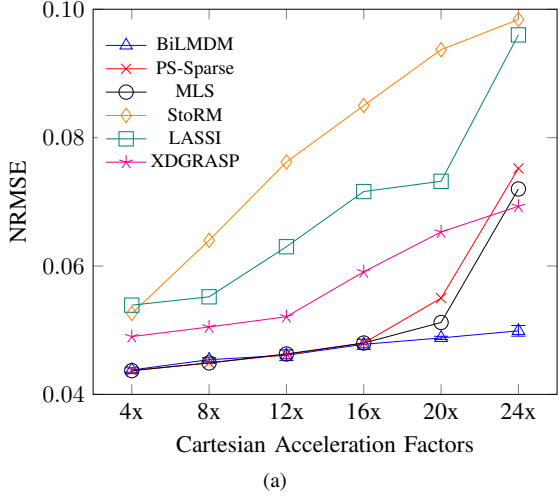
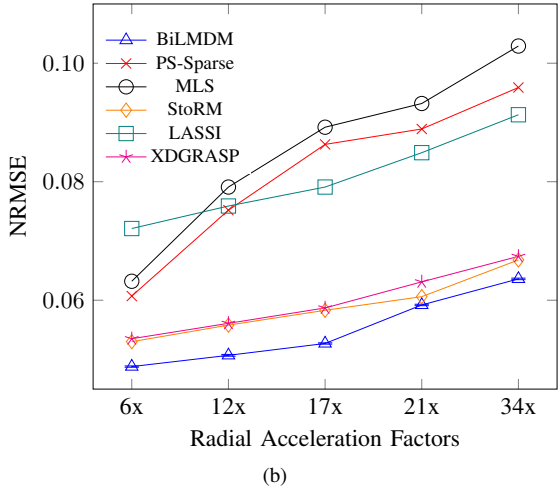
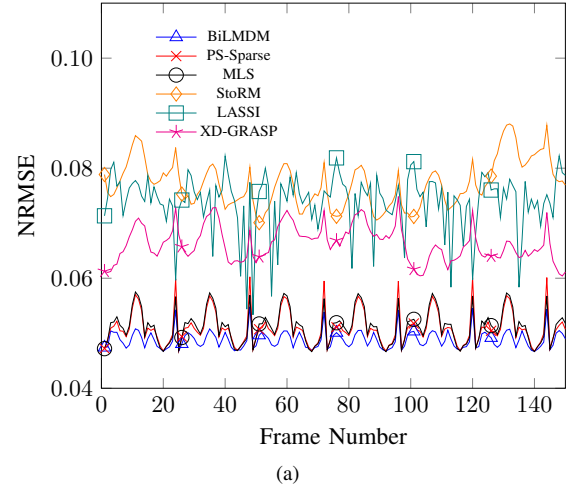
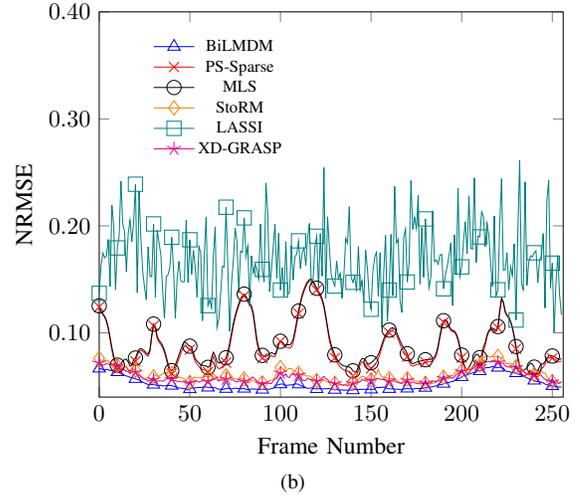
\begin{figure}[ht!]
  \centering
  \subfloat[\label{fig:nrmse.frame.mxrcat}]{
  \centering
  \begin{tikzpicture}[scale = .9] {every tick label/.append style={font=\Huge}}
    \begin{axis} [xlabel= Frame Number, ylabel= NRMSE,
      ymin=0.04, ymax=0.11, xmin=0, xmax =150,
      legend style = {nodes={scale = 0.65, transform shape}, draw = none, at =
          {(0.1,0.95)}, anchor = north west}, mark repeat={25},
      y tick label style={
        /pgf/number format/.cd,
        fixed,
        fixed zerofill,
        precision=2,
        /tikz/.cd
      }]
      \addplot+[color=blue, mark=triangle,mark size=3pt ,select coords between index={1}{150}]
      table[x expr=\coordindex, y=bi] {mrxcat-bi-frame-error.txt};
      \addlegendentry{BiLMDM}
      \addplot+[color=red, mark=x,mark size=3pt, select coords between index={1}{150}]
      table[x expr=\coordindex, y=ps] {mrxcat-ps-frame-error.txt};
      \addlegendentry{PS-Sparse}
      \addplot+[color=black, mark=o,mark size=3pt, select coords between index={1}{150}]
      table[x expr=\coordindex,y=mls] {mrxcat-mls-frame-error.txt};
      \addlegendentry{MLS}
      \addplot+[color=orange, mark=diamond,mark size=3pt,select coords between index={1}{150}]
      table[x expr=\coordindex,y=storm] {mrxcat-storm-frame-error.txt};
      \addlegendentry{StoRM} 
      \addplot+[color=teal, mark=square,mark size=3pt, select coords between index={1}{150}]
      table[x expr=\coordindex,y=lassi] {mrxcat-lassi-frame-error.txt};
      \addlegendentry{LASSI}
      \addplot+[style = solid, color=magenta, mark=star,mark size=3pt, select coords between index={1}{150}]
      table[x expr=\coordindex,y=storm] {mrxcat-xdgrasp-frame-error.txt};
      \addlegendentry{XD-GRASP}
    \end{axis}
  \end{tikzpicture}
  } \\
  \subfloat[\label{fig:nrmse.frame.perfusion}]{
  \centering
  \begin{tikzpicture}[scale = .9] {every tick label/.append style={font=\Huge}}
    \begin{axis} [xlabel= Frame Number, ylabel= NRMSE,
      ymin=0.04, ymax=0.4,
      xmin=0, xmax =256,
      mark repeat={10},
      legend style={nodes={scale = 0.7, transform shape}, draw = none, at =
          {(0.1,0.95)}, anchor = north west}, 
      y tick label style={
        /pgf/number format/.cd,
        fixed,
        fixed zerofill,
        precision=2,
        /tikz/.cd
      }]
      \addplot+[color=blue, mark=triangle,mark size=3pt]
      table[x expr=\coordindex, y=bi] {perfusion-bi-frame-error.txt};
      \addlegendentry{BiLMDM}
      \addplot+[color=red, mark=x,mark size=3pt]
      table[x expr=\coordindex, y=ps] {perfusion-ps-frame-error.txt};
      \addlegendentry{PS-Sparse}
      \addplot+[color=black, mark=o,mark size=3pt]
      table[x expr=\coordindex,y=mls] {perfusion-mls-frame-error.txt};
      \addlegendentry{MLS}
      \addplot+[color=orange, mark=diamond,mark size=3pt]
      table[x expr=\coordindex,y=storm] {perfusion-storm-frame-error.txt};
      \addlegendentry{StoRM} 
      \addplot+[color=teal, mark=square,mark size=3pt]
      table[x expr=\coordindex,y=lassi] {perfusion-lassi-frame-error.txt};
      \addlegendentry{LASSI}
      \addplot+[style=solid, color=magenta, mark=star,mark size=3pt]
      table[x expr=\coordindex,y=xdgrasp] {perfusion-xdgrasp-frame-error.txt};
      \addlegendentry{XD-GRASP}
    \end{axis}
  \end{tikzpicture}
  }
  \caption{Framewise-NRMSE values for \protect\subref{fig:nrmse.frame.mxrcat} MRXCAT cardiac cine data
    (acceleration rate: 12x), and \protect\subref{fig:nrmse.frame.perfusion} real cardiac cine data (acceleration
    rate: 17x) for PS-Sparse [\protect\subref{fig:nrmse.frame.mxrcat}
    $0.054 \pm 3\times 10^{-3}$~\protect\subref{fig:nrmse.frame.perfusion}
    $0.0863 \pm 2.1\times 10^{-2}$], MLS [\protect\subref{fig:nrmse.frame.mxrcat}
    $0.051\pm 3.1\times 10^{-3}$~\protect\subref{fig:nrmse.frame.perfusion}
    $0.0892 \pm 2\times 10^{-2}$], SToRM [\protect\subref{fig:nrmse.frame.mxrcat}
    $0.0844\pm 2.6\times 10^{-2}$~\protect\subref{fig:nrmse.frame.perfusion}
    $0.0583 \pm 6.6\times 10^{-3}$], LASSI [\protect\subref{fig:nrmse.frame.mxrcat}
    $0.0732\pm 6\times 10^{-3}$~\protect\subref{fig:nrmse.frame.perfusion}
    $0.0791 \pm 3.1\times 10^{-2}$]{XD-GRASP [\protect\subref{fig:nrmse.frame.mxrcat}
    $0.065 \pm 2.5 \times 10^{-2}$~\protect\subref{fig:nrmse.frame.perfusion} 
    $0.0587 \pm 2.7 \times 10^{-2}$]} and BiLMDM [\protect\subref{fig:nrmse.frame.mxrcat}
    $0.0488 \pm 1.6\times 10^{-3}$,~\protect\subref{fig:nrmse.frame.perfusion} 
    $0.0527 \pm 6.4\times 10^{-3}$]. The previous numerical
    values demonstrate the (sample mean over all $N_{\text{fr}}$ frames) $\pm$ (standard deviation
    from the sample mean). Consistent with simulation results, BiLMDM exhibits less
    fluctuations in error across all the frames over the rest of the methods.}
  \label{fig:nrmse.frame}
\end{figure}

\begin{figure*}[ht!]
  \centering
  \subfloat{\begin{annotate}
                {\includegraphics[height=2.27cm, width=0.11\textwidth]{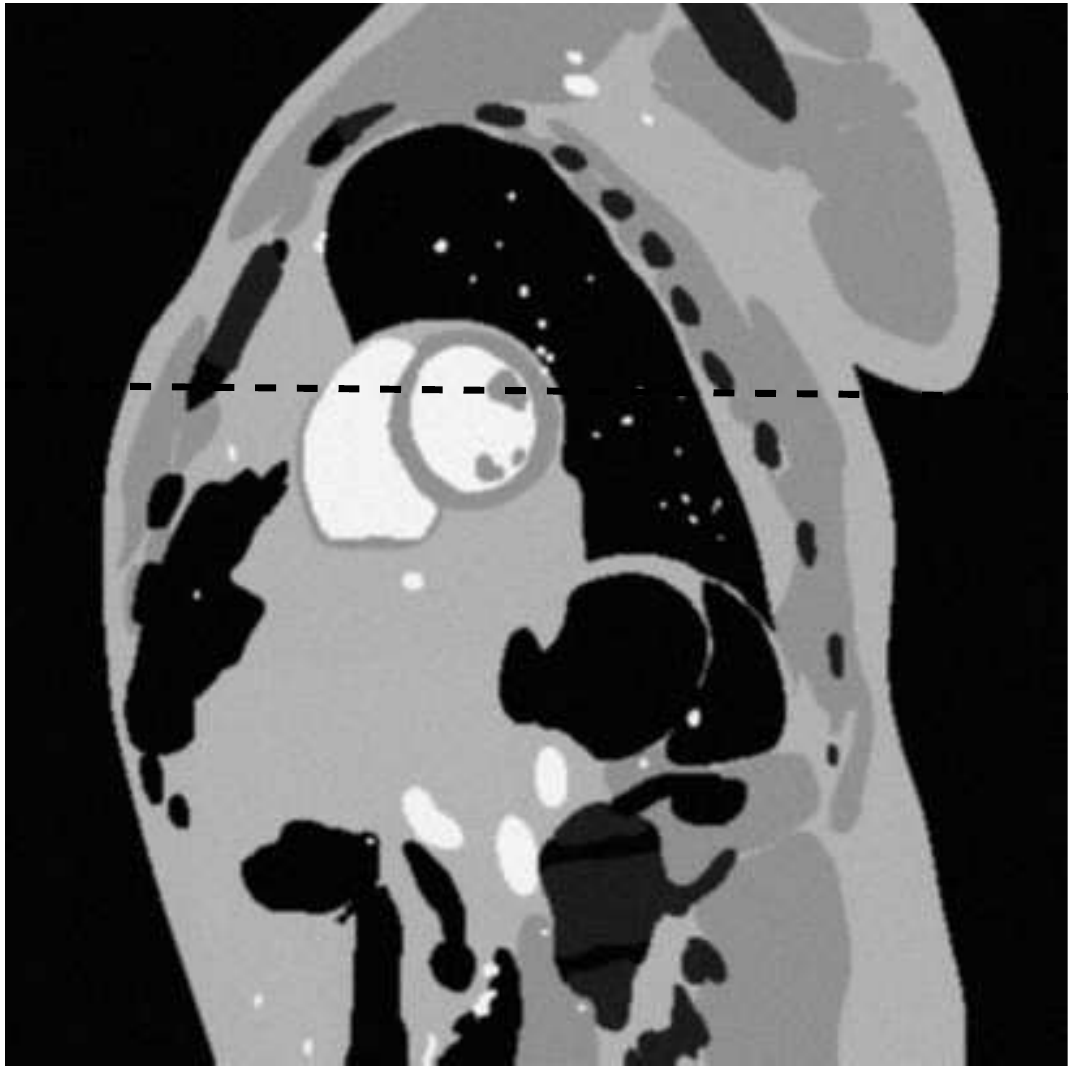}}{0.5}
                \draw[yellow, thick] (-1,1) rectangle (0.2, -0.1); 
                \draw[yellow, thick, dashed] (-2, 0.6) -- (2, 0.6);
            \end{annotate}}
  \hspace{-0.55cm}
  \subfloat{\begin{annotate}
                {\includegraphics[width = 0.11\textwidth]{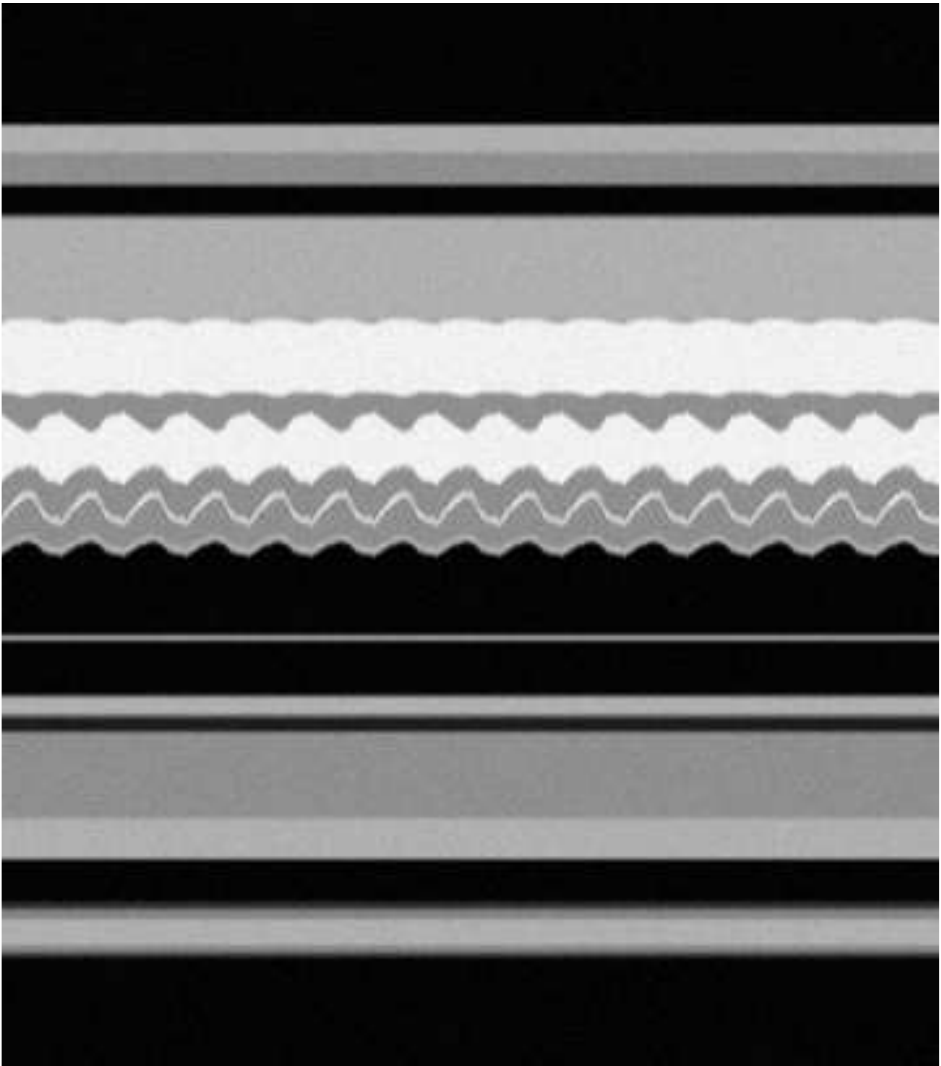}}{0.5}
                \draw[yellow, thick, ->] (0,0.6) -- (0.5,0.1);
                \draw[yellow, thick, ->] (-1,1.4) -- (-0.5,0.9);
                \draw[yellow, thick, ->] (0,1.7) -- (0.5,1.2);
            \end{annotate}}
  \hspace{-0.55cm} 
  \subfloat{\begin{annotate}
                {\includegraphics[width = 0.11\textwidth]{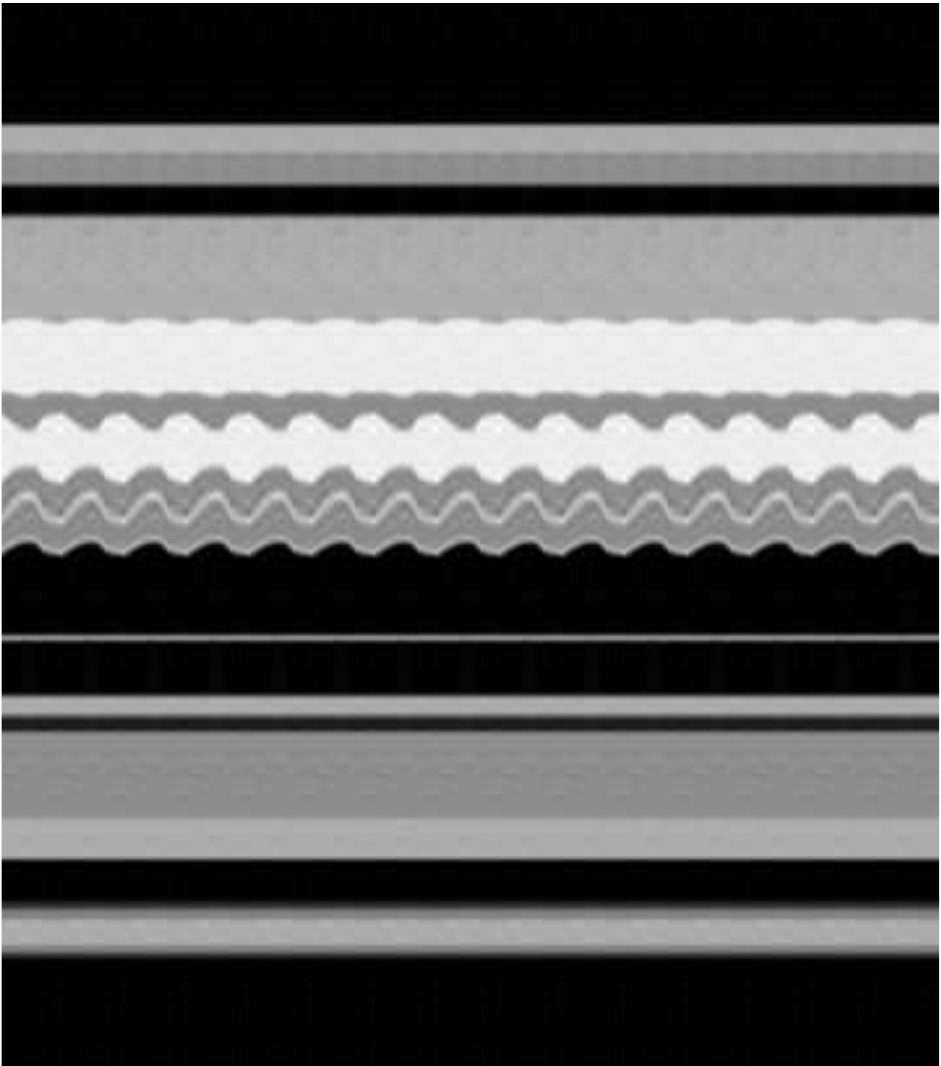}}{0.5}
                \draw[red, thick, ->] (0,0.6) -- (0.5,0.1);
                \draw[red, thick, ->] (-1,1.4) -- (-0.5,0.9);
                \draw[red, thick, ->] (0,1.7) -- (0.5,1.2);
            \end{annotate}}
  \hspace{-0.55cm} 
  \subfloat{\begin{annotate}
                {\includegraphics[width = 0.11\textwidth]{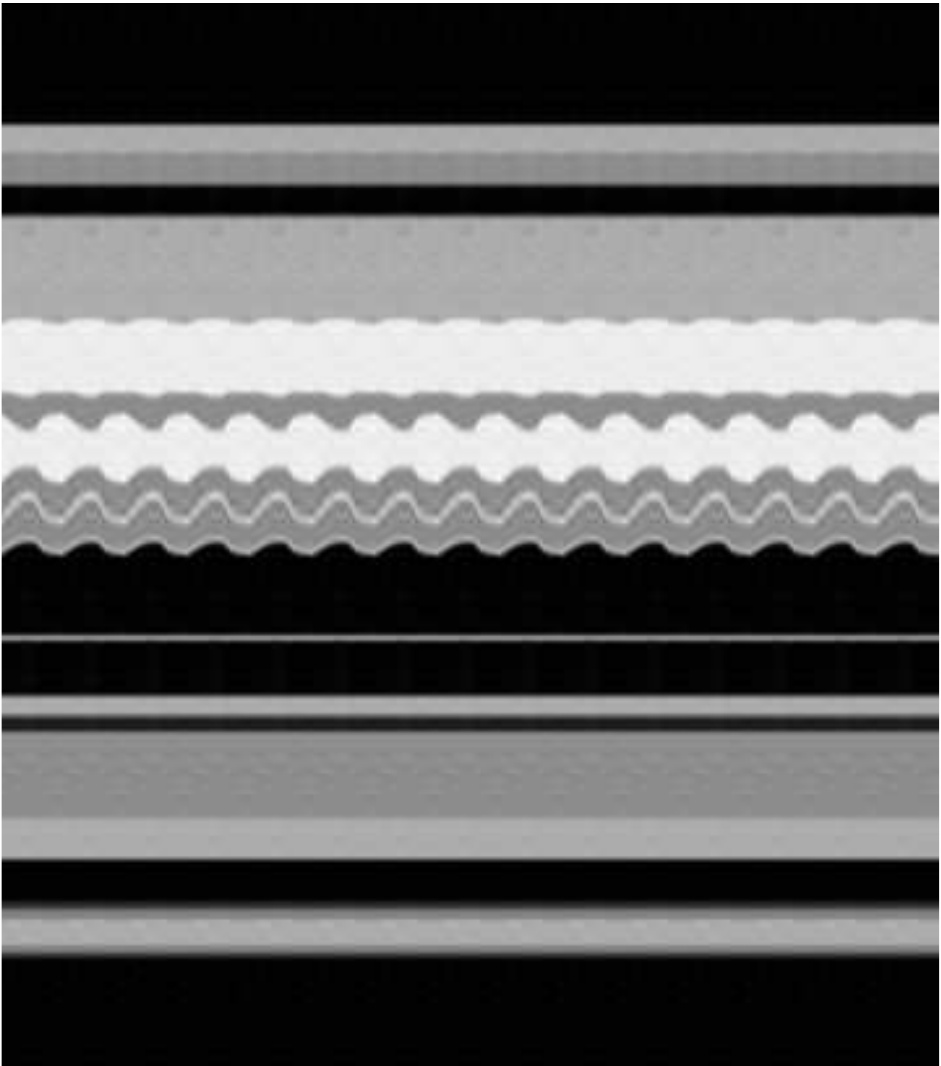}}{0.5}
                \draw[red, thick, ->] (0,0.6) -- (0.5,0.1);
                \draw[red, thick, ->] (-1,1.4) -- (-0.5,0.9);
                \draw[red, thick, ->] (0,1.7) -- (0.5,1.2);
            \end{annotate}}
  \hspace{-0.55cm} 
  \subfloat{\begin{annotate}
                {\includegraphics[width = 0.11\textwidth]{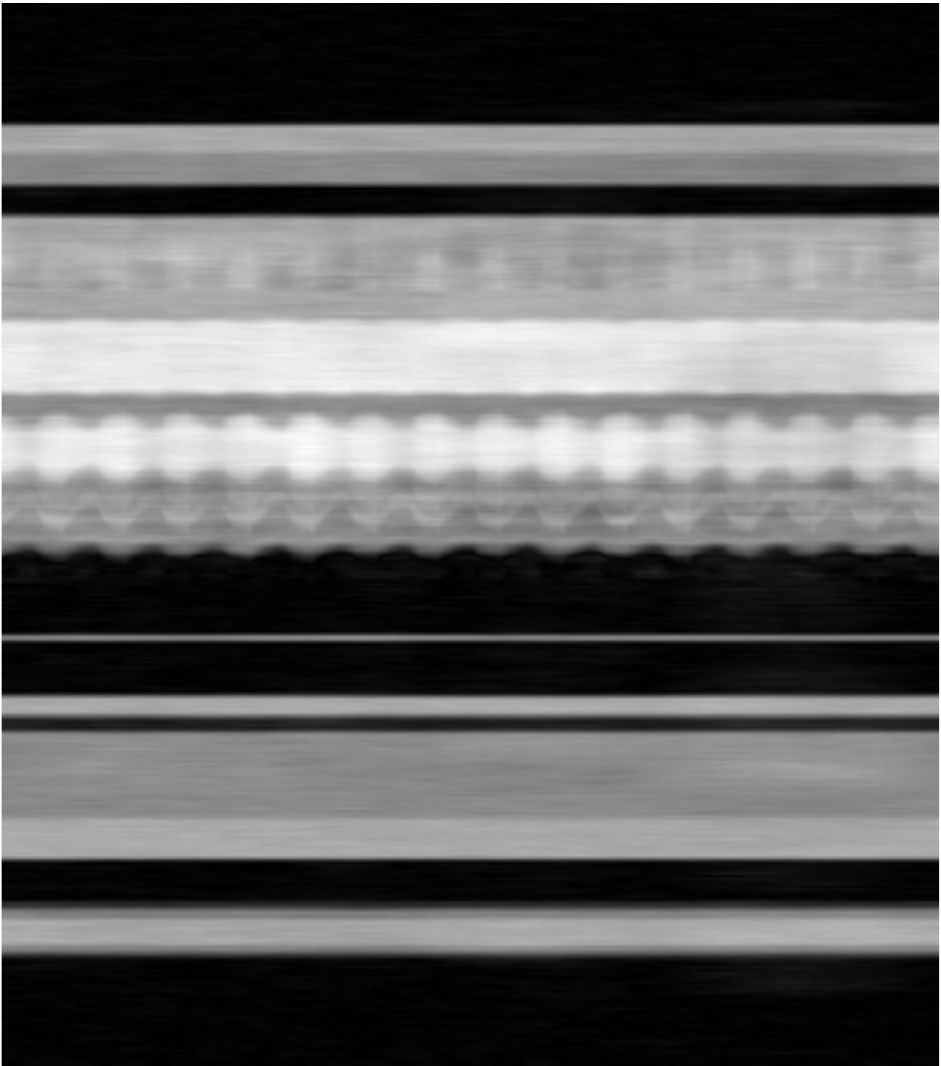}}{0.5}
                \arrow{-1,1.4}{-0.5,0.9}
                \arrow{0,1.7}{0.5,1.2}
                \arrow{0,0.5}{0.5,0}
            \end{annotate}}
  \hspace{-0.55cm} 
  \subfloat{\begin{annotate}
                {\includegraphics[width = 0.11\textwidth]{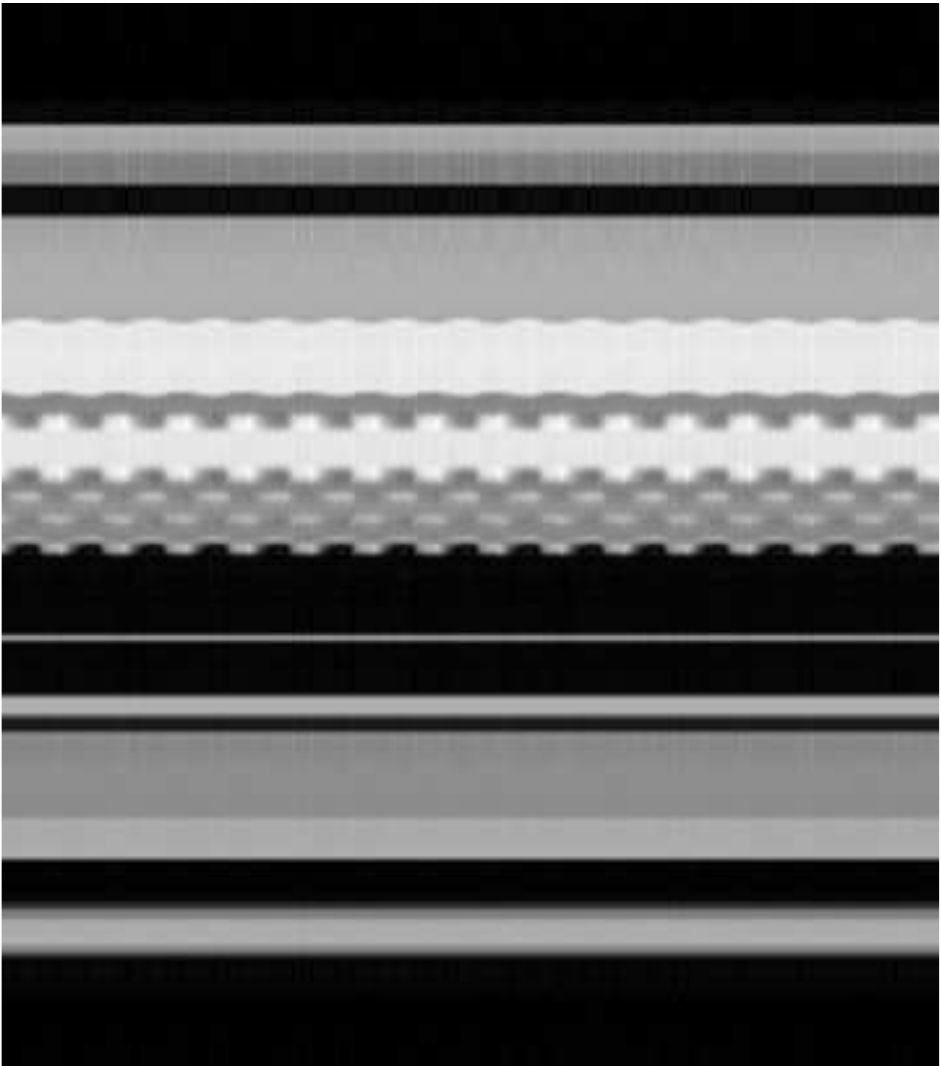}}{0.5}
                \arrow{-1,1.4}{-0.5,0.9}
                \arrow{0,0.6}{0.5,0.1}
            \end{annotate}}
  \hspace{-0.55cm}
  \subfloat{\begin{annotate}
                {\includegraphics[width = 0.11\textwidth]{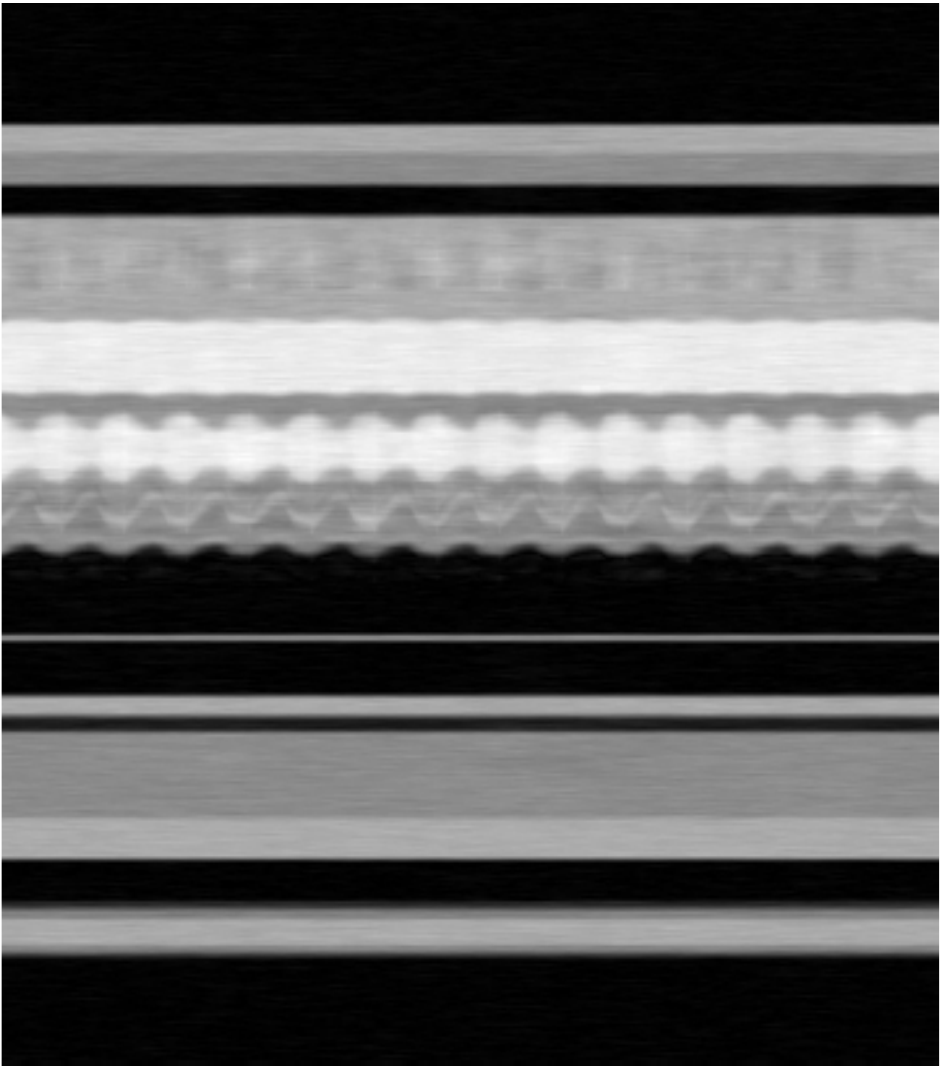}}{0.5}
                \arrow{-1,1.4}{-0.5,0.9}
                \arrow{0,1.7}{0.5,1.2}
                \arrow{0,0.5}{0.5,0}
            \end{annotate}}
  \hspace{-0.55cm}
  \subfloat{\begin{annotate}
                {\includegraphics[width = 0.11\textwidth]{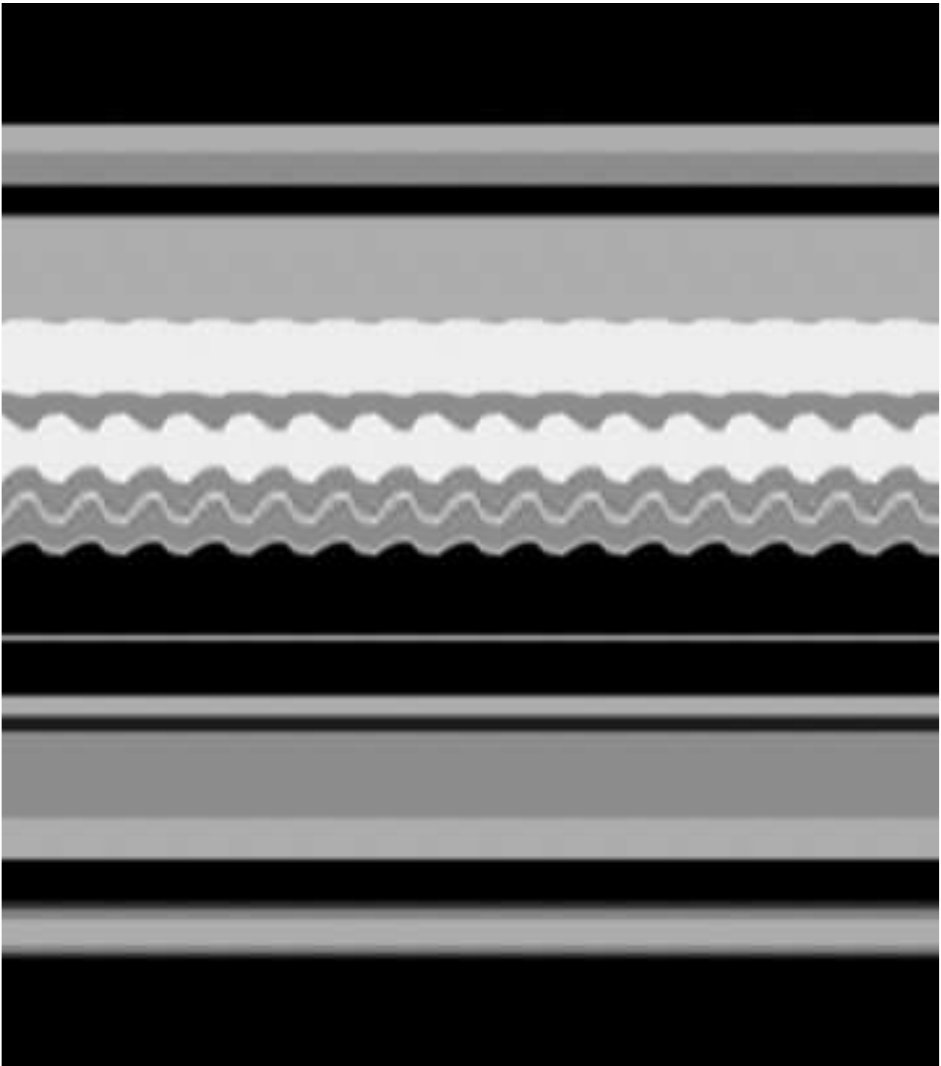}}{0.5}
                \draw[red, thick, ->] (0,0.6) -- (0.5,0.1);
                \draw[red, thick, ->] (-1,1.4) -- (-0.5,0.9);
                \draw[red, thick, ->] (0,1.7) -- (0.5,1.2);
            \end{annotate}}\\
  \vspace{-0.21cm} 
  \hspace{4cm}
  \subfloat{\begin{annotate}
                {\includegraphics[width=0.11\textwidth]{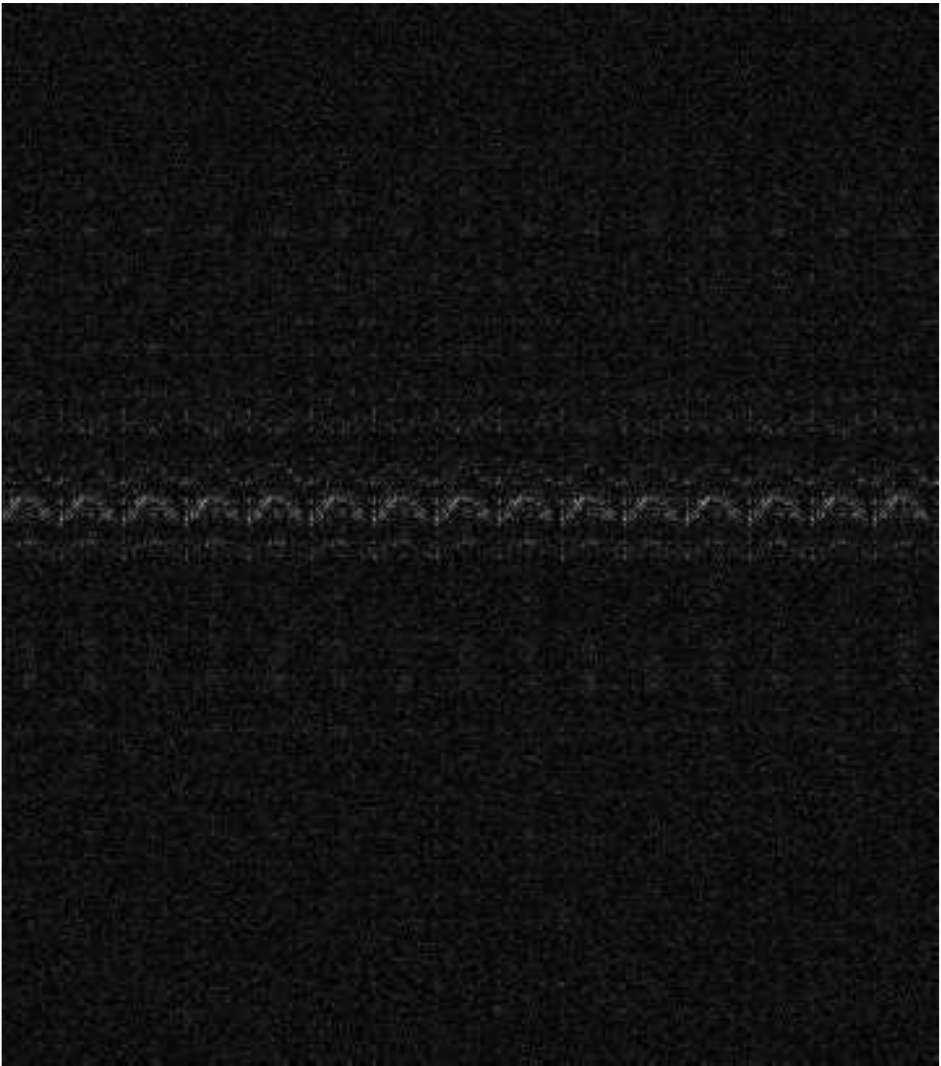}}{0.5}
            \end{annotate}}
  \hspace{-0.55cm} 
  \subfloat{\begin{annotate}
                {\includegraphics[width = 0.11\textwidth]{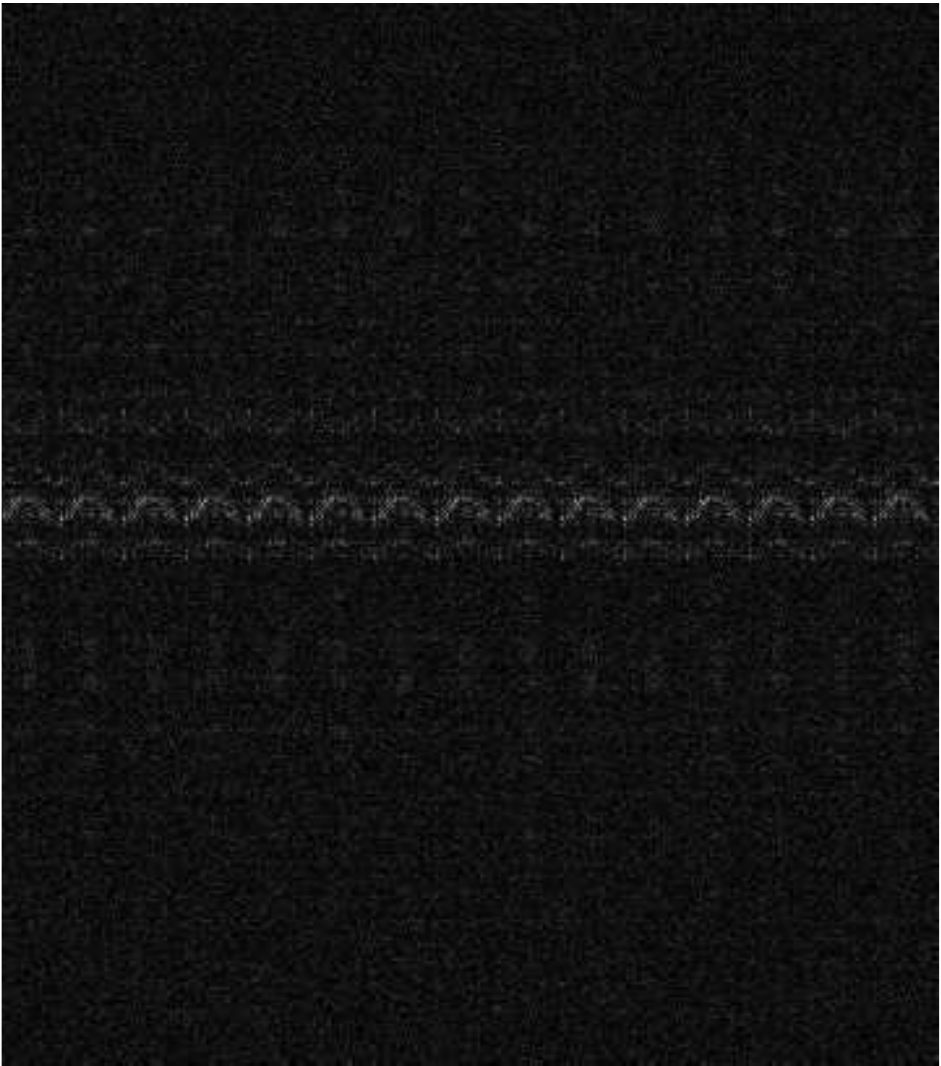}}{0.5}
            \end{annotate}}
  \hspace{-0.55cm} 
  \subfloat{\begin{annotate}
                {\includegraphics[width = 0.11\textwidth]{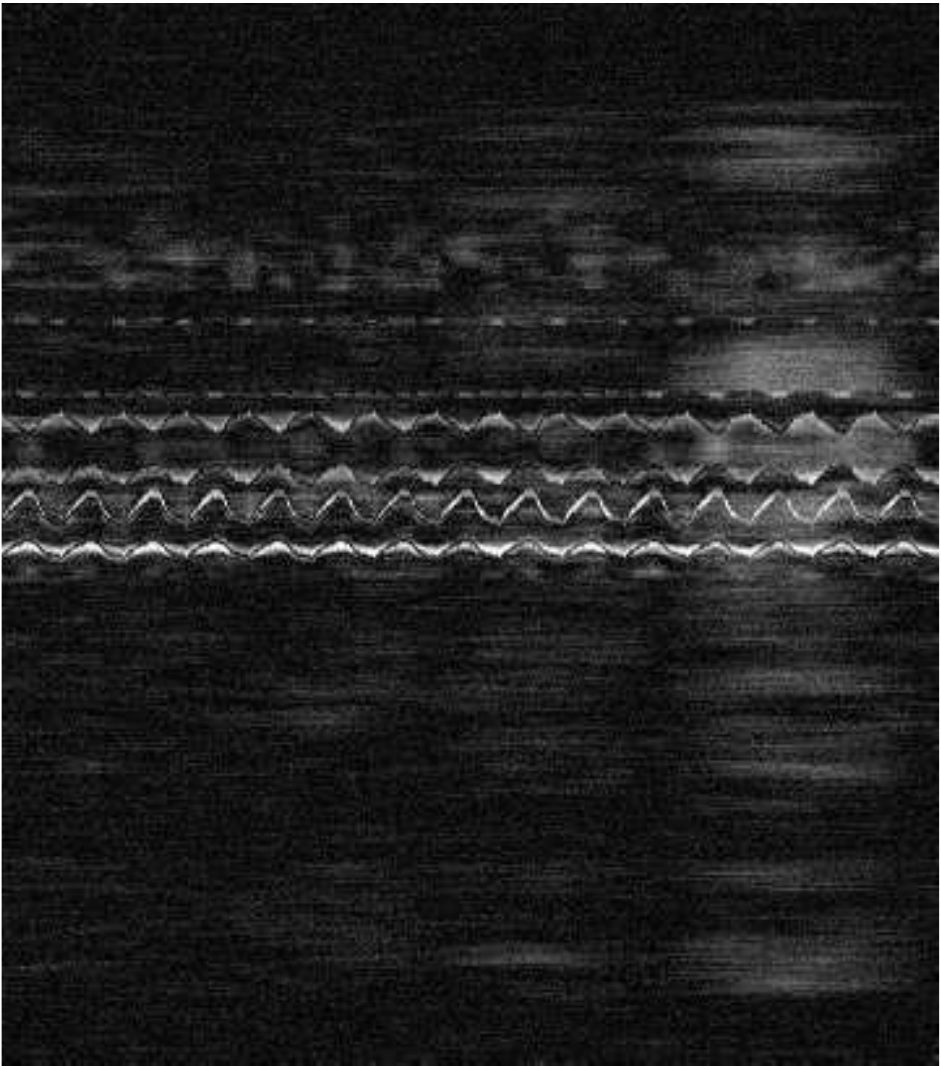}}{0.5}
            \end{annotate}}
  \hspace{-0.55cm} 
  \subfloat{\begin{annotate}
                {\includegraphics[width = 0.11\textwidth]{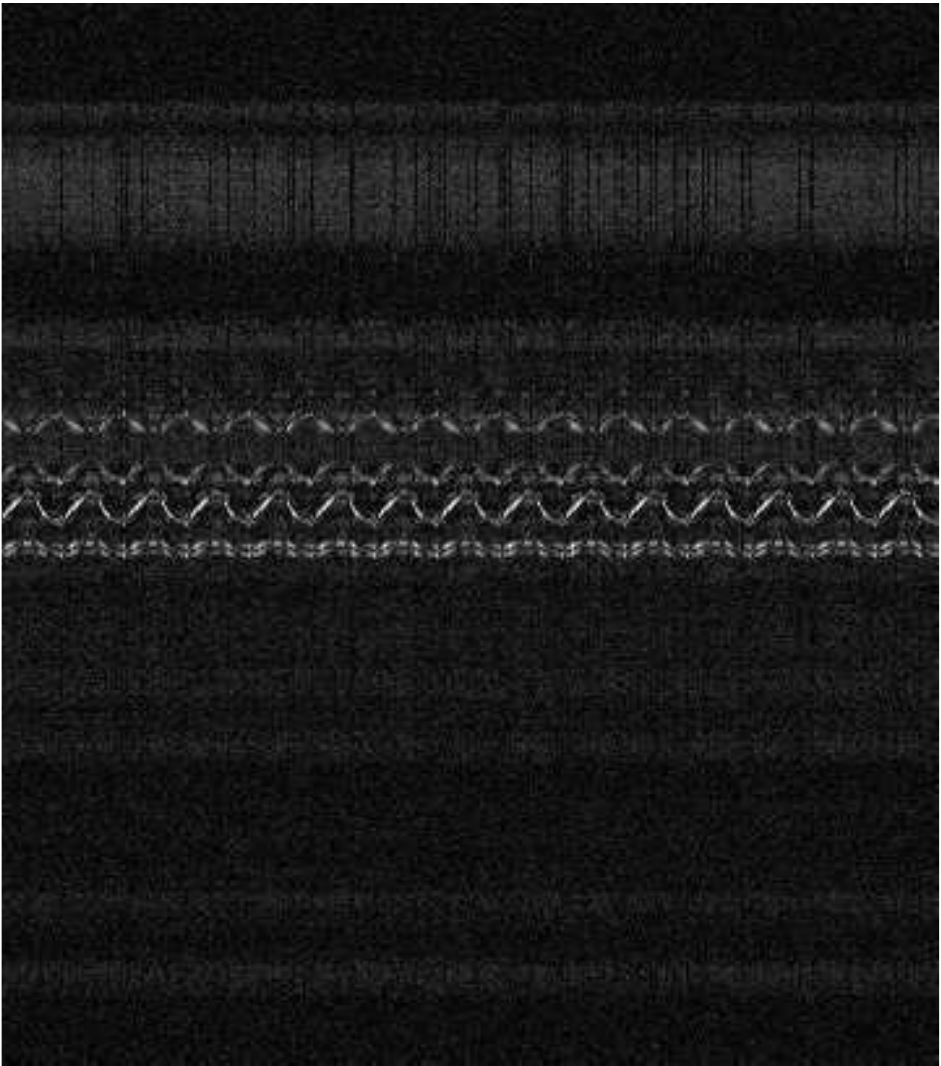}}{0.5}
            \end{annotate}}
  \hspace{-0.55cm}
  \subfloat{\begin{annotate}
                {\includegraphics[width = 0.11\textwidth]{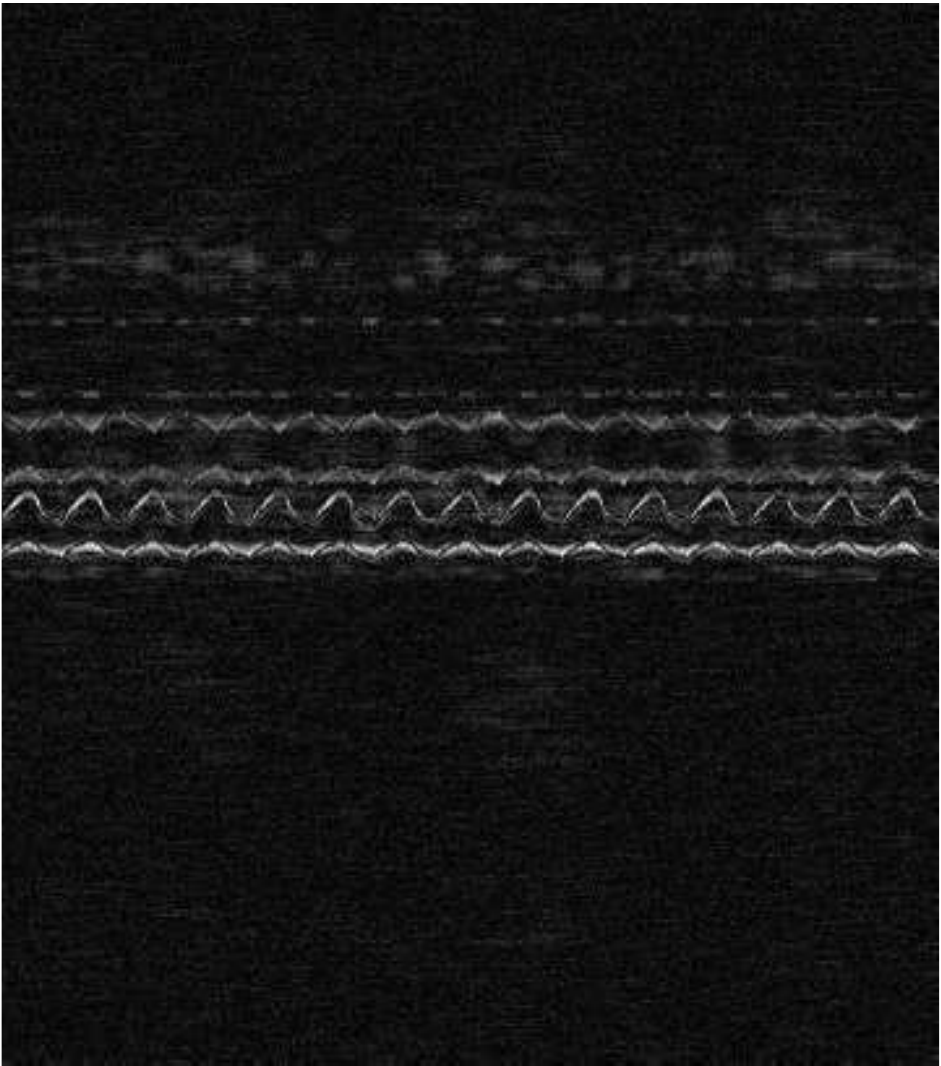}}{0.5}
            \end{annotate}}
  \hspace{-0.55cm}
  \subfloat{\begin{annotate}
                {\includegraphics[width = 0.11\textwidth]{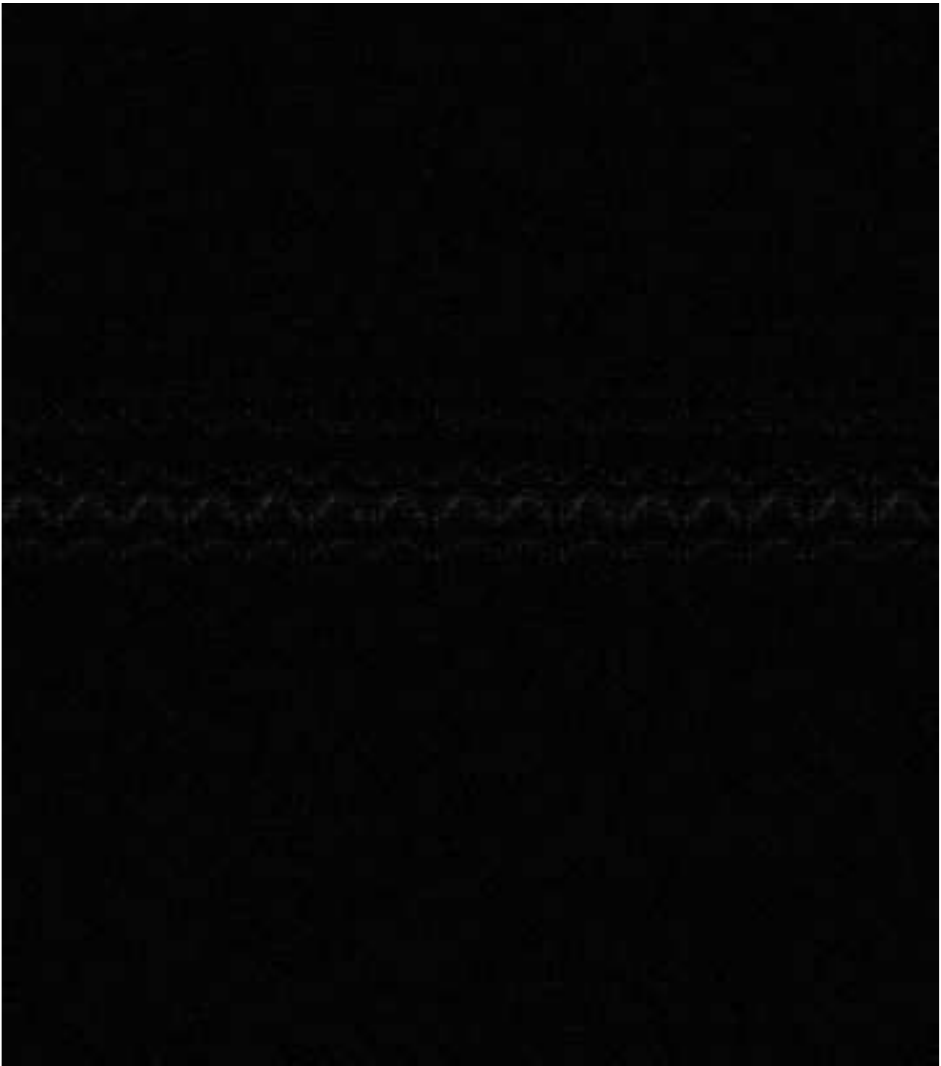}}{0.5}
            \end{annotate}}
  \caption{Temporal cross-sections for MRXCAT cardiac cine (acceleration rate: 20x). Left to right:
    Gold standard (spatial frame), gold standard (temporal cross-section), PS-Sparse ($0.055$), MLS
    ($0.0512$), SToRM ($0.0844$), LASSI ($0.0732$){, XD-GRASP ($0.0653$)} and BiLMDM
    ($0.0488 \pm 2.5 \times 10^{-4}$). The previous numerical values indicate the NRMSE for the complete dataset, in addition to the standard deviation (for BiLMDM only) obtained after running the non-convex algorithmic scheme for $25$ independent trials. Top to bottom: Temporal cross section and error maps. The temporal
    location of the frames is indicated by the yellow dotted line in the gold-standard (spatial-frame)
    image. }
  \label{fig:mrxcat.temporal}
\end{figure*}
The proposed framework is tested and validated on three datasets, used also
in~\cite{zhao2012pssparse, lingala2011ktslr, nakarmi2017m, nakarmi2018mls, candes2013unbiased}:
\begin{enumerate*}[label={\bfseries\roman*)}]
\item Magnetic-resonance extended cardiac-torso (MRXCAT) cine phantom~\cite{wissmann2014mrxcat};
\item {cardiac phantom generated from real MR scans~\cite{zhao2012pssparse}}
\item {prospectively undersampled free breathing, real time, cardiac cine data.}
\end{enumerate*} 
All experiments were conducted on a $12$-core Intel(R) $2.40$ GHz Linux-based system with $48$GB
RAM, with implementations realized in MATLAB~\cite{MATLAB.2018b}. The proposed method is compared
with: PS-Sparse~\cite{zhao2012pssparse}, MLS~\cite{nakarmi2018mls}, SToRM~\cite{poddar2016dynamic}, LASSI~\cite{ravishankar2017.lassi} {and XD-GRASP}~\cite{feng2016xd}.

The publicly available MATLAB implementations of SToRM~\cite{code.storm},
LASSI~\cite{code.lassi} {and XD-GRASP}~\cite{code.xdgrasp} were
employed. MATLAB code was also written to realize the algorithmic solutions
of~\cite{slavakis.FMHSDM} to the convex-optimization problems posed in PS-Sparse
and MLS. Parameters were tuned to produce the least NRMSE for each algorithm at
each sampling ratio and on every dataset. {The proposed framework currently underperforms
  in terms of computation times versus the competing algorithms and their
  optimized software implementations. Since the present manuscript serves as a
  proof of concept of BiLMDM, an optimized and time-efficient C/C++ version of
  the developed MATLAB code falls beyond the scope of this work and it will be
  presented in the near future via an open-source software-distribution venue.}

{Tasks~\eqref{recovery.task},
  \eqref{task:min.for.U} and~\eqref{task:min.for.B} require parameter tuning to
  produce the desirable MR images from the scanner acquired data. These
  parameters were chosen empirically to minimize the reconstruction error for
  the datasets where the gold standard was known. A range of values for every
  parameter was realized, and as long as these parameters are within that range,
  the NRMSEs were not so sensitive to the parameters. This empirical research
  into the behaviour of the parameter values revealed the following: for hyper
  parameters penalizing the periodicity along the time series ($\lambda_1$,
  $\lambda_2$), higher the ratio $\lambda_2/\lambda_1$, severe temporal bleeding
  was observed; hence, it should be low enough to avoid temporal bleeding and at
  the same time exploit the periodicity along the temporal axis.  The parameter
  which penalizes the sparsity on $\vect{B}$, $\lambda_3$, controls the size of
  the neighborhood used for approximation of the time series image. Higher the
  value of $\lambda_3$, lower the size of the neighborhood, allowing thus frames
  with very similar characteristics to share data among themselves. However,
  decreasing the value excessively would result in dissimilar frames influencing each other. Parameter $\text{N}_\ell$ determines the number
  of landmark points which act as a representative of a collection of frames
  with similar phase and motion characteristics. The value for $\text{N}_\ell$
  should be high enough to describe the data cloud concisely. For the datasets
  used in this work, desirable values range between $15\%-20\%$ of
  $N_\text{fr}$. Any increase $\text{N}_\ell$ beyond that didn't yield any
  significant changes in the image quality. Bounding constant $C_{U}$ set
  to 1 usually yields adequate estimates for $\vect{U}$, although it is worth
  pointing out that the estimates for $\vect{U}$ are not sensitive to a change
  in $C_{U}$. The step sizes for convergence of Alg.~\ref{alg:recovery}
  ($\gamma_0=0.9$, $\zeta_0=0.001$) and the step size for
  Alg.~\ref{alg:algorithm.Un.Bn} ($\alpha = 0.5$) ensures desirable
  results. These values were reached by observing the effects of these
  parameters on the NRMSE values for the reconstruction of the two synthetically
  generated cardiac cine dataset.}

Quality of reconstruction is evaluated by the normalized-root-mean-square error (NRMSE), defined as
\begin{align}
    \text{NRMSE} := \tfrac{\norm{\vect{X} - \hat{\vect{X}}}_\text{F}}{\norm{\vect{X}}_\text{F}}
  \,, \label{def.NRMSE} 
\end{align}
where $\vect{X}$ represents the fully sampled, high-fidelity and original
image-domain data, while $\hat{\vect{X}}$ represents an estimate of $\vect{X}$
computed by the reconstruction schemes. {In
  addition to the NRMSE, the effectiveness of the algorithms are also inspected
  along the edges using the high-frequency error norm (HFEN) and sharpness
  measures. HFEN is defined as
\begin{align}
    \text{HFEN} = \tfrac{\norm{\text{LoG}(\vect{X}) -
  \text{LoG}(\hat{\vect{X}})}_\text{2}}{\norm{\text{LoG}(\vect{X})}_\text{2}}
  \,,
\end{align}
where LoG is a rotationally symmetric Laplacian of Gaussian filter. As
in~\cite{ravishankar2011mr}, a filter kernel of size $15 \times 15$ with
standard deviation of 1.5 pixels is employed. Two sharpness measures are used:
M1 (intensity variance based) and M2 (energy of the image gradient based), as
described in~\cite{subbarao1993focusing}. Better insights into the structural
information are extracted using the Structural Similarity (SSIM)
index~\cite{wang2004image} which investigates the local patterns in the pixel
intensities after normalizing for luminance and contrast. The metrics (except
for sharpness measure) can be used for assessment only when the ground truth
images are present and hence are used for the two phantom datasets and not the
prospectively undersampled data.}

The proposed framework is validated over a range of
{under}sampling/acceleration rates, defined by
$N_{\text{k}} N_{\text{fr}} / (\text{\# of acquired voxels})$. 1-D Cartesian
(Fig.~\ref{fig:sampling.strategy.cartesian}) as well as radial
(Fig.~\ref{fig:sampling.strategy.radial}) sampling were applied to 
{both the phantom datasets}. To save space, only the 1-D
Cartesian-sampling results are demonstrated for the MRXCAT dataset, while
radial-sampling ones are shown for the phantom generated
using real MR scans. {Nevertheless, BiLMDM's performance against the
  competing reconstruction algorithms follows a similar trend also for sampling
  strategies not included in the manuscript due to space limitations.}

\begin{figure*}[ht]
  \centering
  \subfloat{\begin{annotate}
        {\includegraphics[width = 0.12\textwidth]{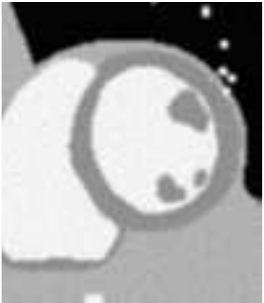}}{0.5}
        \end{annotate}}\hspace{-0.5cm}
  \subfloat{\begin{annotate}
        {\includegraphics[width = 0.12\textwidth]{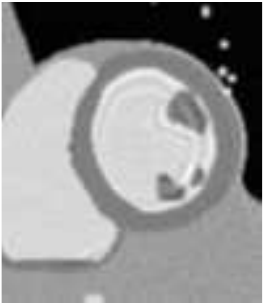}}{0.5}
        \arrow{-1,1.8}{-0.2, 1}
        \end{annotate}}\hspace{-0.5cm}
  \subfloat{\begin{annotate}
        {\includegraphics[width = 0.12\textwidth]{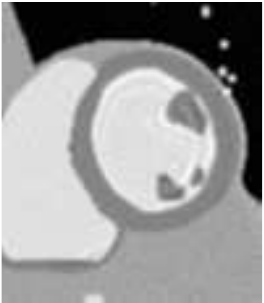}}{0.5}
        \arrow{-1,1.8}{-0.2, 1}
        \end{annotate}}\hspace{-0.5cm}
  \subfloat{\begin{annotate}
        {\includegraphics[width = 0.12\textwidth]{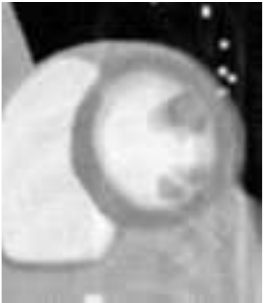}}{0.5}
        \draw[red, thick] (0.2,1.1) rectangle (1.5,0);
        \draw[red, thick] (1,-1) rectangle (2,-2);
        \end{annotate}}\hspace{-0.5cm}
  \subfloat{\begin{annotate}
        {\includegraphics[width = 0.12\textwidth]{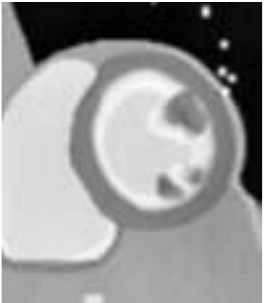}}{0.5}
        \arrow{-1,1.8}{0, 0.8}
        \end{annotate}}\hspace{-0.5cm}
  \subfloat{\begin{annotate}
        {\includegraphics[width = 0.12\textwidth]{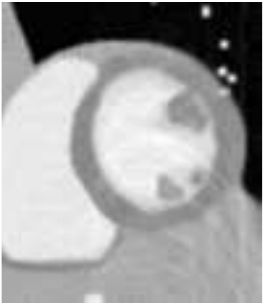}}{0.5}
        \draw[red, thick] (1,-1) rectangle (2,-2);
        \end{annotate}}\hspace{-0.5cm}    
  \subfloat{\begin{annotate}
        {\includegraphics[width = 0.12\textwidth]{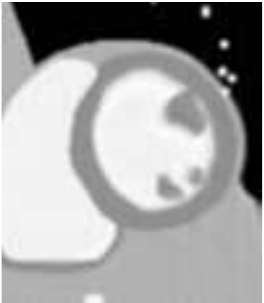}}{0.5} 
        \end{annotate}}\\ 
  \vspace{-0.5cm} 
  \subfloat{\begin{annotate}
        {\includegraphics[width = 0.12\textwidth]{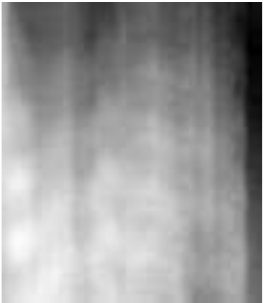}}{0.5}
        \end{annotate}}\hspace{-0.5cm}
  \subfloat{\begin{annotate}
        {\includegraphics[width = 0.12\textwidth]{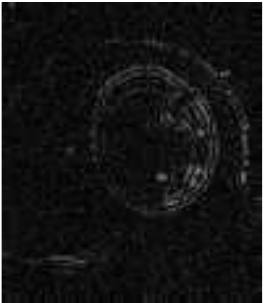}}{0.5}
        \end{annotate}}\hspace{-0.5cm}
  \subfloat{\begin{annotate}
        {\includegraphics[width = 0.12\textwidth]{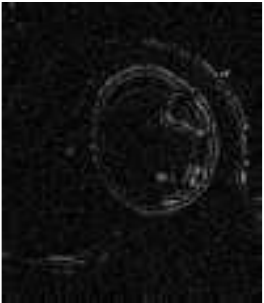}}{0.5}
        \end{annotate}}\hspace{-0.5cm}
  \subfloat{\begin{annotate}
        {\includegraphics[width = 0.12\textwidth]{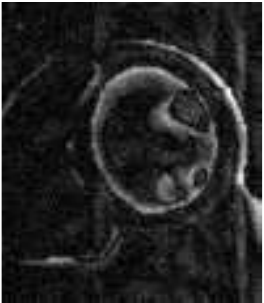}}{0.5}
        \end{annotate}}\hspace{-0.5cm}
  \subfloat{\begin{annotate}
        {\includegraphics[width = 0.12\textwidth]{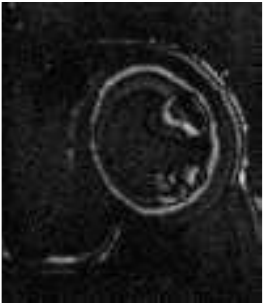}}{0.5}
        \end{annotate}}\hspace{-0.5cm}
  \subfloat{\begin{annotate}
        {\includegraphics[width = 0.12\textwidth]{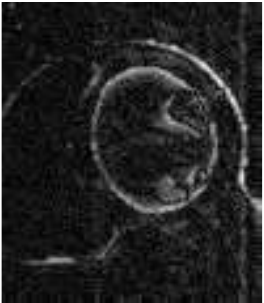}}{0.5}
        \end{annotate}}\hspace{-0.5cm}
  \subfloat{\begin{annotate}
        {\includegraphics[width = 0.12\textwidth]{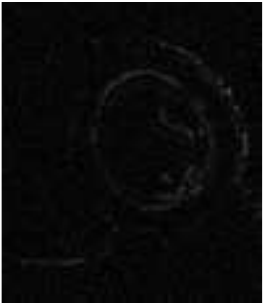}}{0.5}
        \end{annotate}}\\ 
  \vspace{-0.5cm}
  \subfloat{\begin{annotate}
        {\includegraphics[width = 0.12\textwidth]{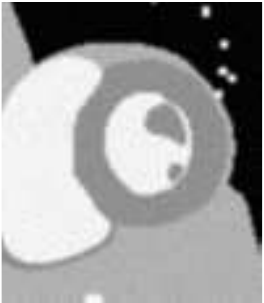}}{0.5}
        \end{annotate}}\hspace{-0.5cm}
  \subfloat{\begin{annotate}
        {\includegraphics[width = 0.12\textwidth]{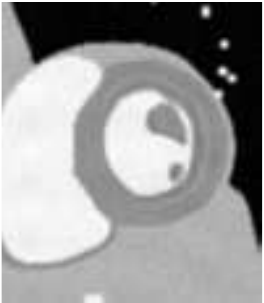}}{0.5}
        \arrow{0,0}{1,-0.5}
        \end{annotate}}\hspace{-0.5cm}
  \subfloat{\begin{annotate}
        {\includegraphics[width = 0.12\textwidth]{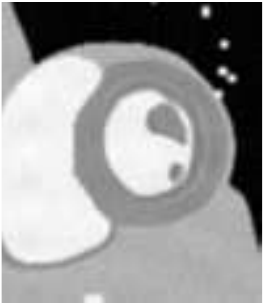}}{0.5}
        \arrow{0,0}{1,-0.5}
        \end{annotate}}\hspace{-0.5cm}
  \subfloat{\begin{annotate}
        {\includegraphics[width = 0.12\textwidth]{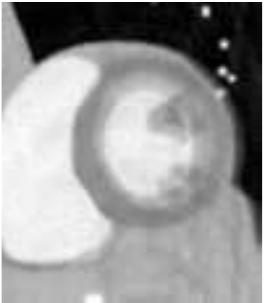}}{0.5}
        \draw[red, thick] (0.2,1) rectangle (1.2,0.2);
        \end{annotate}}\hspace{-0.5cm}
  \subfloat{\begin{annotate}
        {\includegraphics[width = 0.12\textwidth]{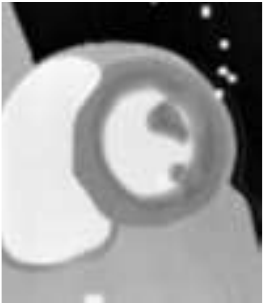}}{0.5}
        \arrow{-1,2}{0,1.2}
        \end{annotate}}\hspace{-0.5cm}
  \subfloat{\begin{annotate}
        {\includegraphics[width = 0.12\textwidth]{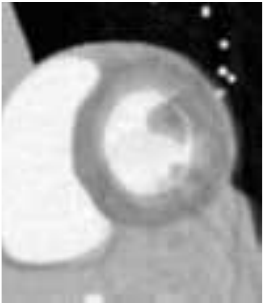}}{0.5}
        \draw[red, thick] (0.2,1) rectangle (1.2,0);
        \end{annotate}}\hspace{-0.5cm}
  \subfloat{\begin{annotate}
        {\includegraphics[width = 0.12\textwidth]{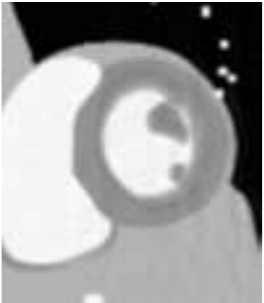}}{0.5}
        \end{annotate}}\\ 
  \vspace{-0.5cm}
  \subfloat{\begin{annotate}
        {\includegraphics[width = 0.12\textwidth]{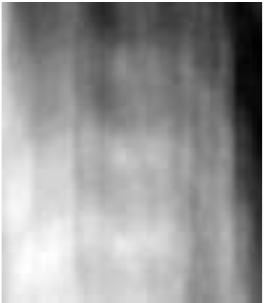}}{0.5}
        \end{annotate}}\hspace{-0.5cm}
  \subfloat{\begin{annotate}
        {\includegraphics[width = 0.12\textwidth]{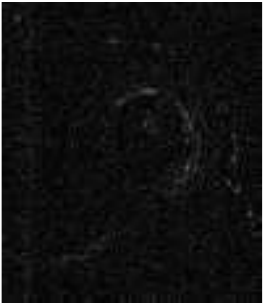}}{0.5}
        \end{annotate}}\hspace{-0.5cm}
  \subfloat{\begin{annotate}
        {\includegraphics[width = 0.12\textwidth]{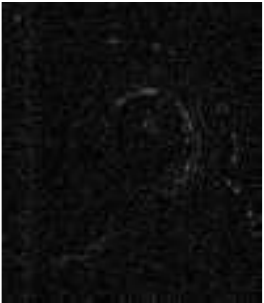}}{0.5}
        \end{annotate}}\hspace{-0.5cm}
  \subfloat{\begin{annotate}
        {\includegraphics[width = 0.12\textwidth]{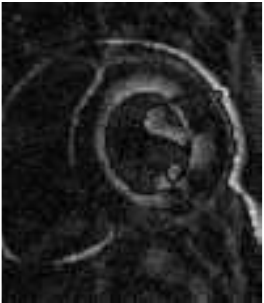}}{0.5}
        \end{annotate}}\hspace{-0.5cm}
  \subfloat{\begin{annotate}
        {\includegraphics[width = 0.12\textwidth]{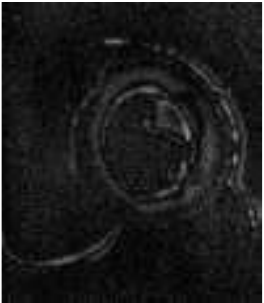}}{0.5}
        \end{annotate}}\hspace{-0.5cm}
  \subfloat{\begin{annotate}
        {\includegraphics[width = 0.12\textwidth]{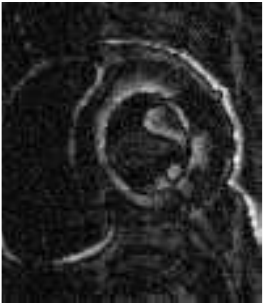}}{0.5}
        \end{annotate}}\hspace{-0.5cm}
  \subfloat{\begin{annotate}
        {\includegraphics[width = 0.12\textwidth]{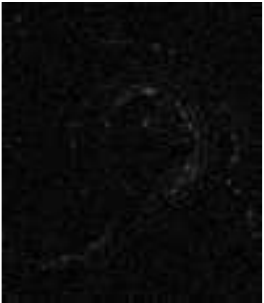}}{0.5}
        \end{annotate}}
  \caption{Spatial results for {region of interest in (marked with a yellow rectangle in the gold standard spatial-frame image of Fig.~\ref{fig:mrxcat.temporal})} MRXCAT cardiac cine (acceleration rate: 20x). Left to right: Gold standard, PS-Sparse ($0.055$), MLS ($0.0512$), SToRM ($0.0844$), LASSI ($0.0732$), {XD-GRASP ($0.0653$)} and BiLMDM ($0.0488 \pm 2.5 \times 10^{-4}$). The previous numerical values indicate the NRMSE for the complete dataset, in addition to the standard deviation (for BiLMDM only) obtained after running the non-convex algorithmic scheme for $25$ independent trials. Top to bottom: Diastole phase (frame $1$ of the time series), the under-sampled image followed by error maps, systole phase (frame $12$ of the time
    series) and the under-sampled image followed by error maps.}
  \label{fig:mrxcat.error}
\end{figure*}

\subsection{MRXCAT phantom}\label{subsec:mrxcat.results}

The extended cardiac torso (XCAT) framework was used to generate the MRXCAT
phantom~\cite{wissmann2014mrxcat}; a breath-hold cardiac cine data of spatial size
$(\text{N}_\text{p},\text{N}_\text{f}) = (408, 408)$ corresponding to a spatial resolution of
$1.56 \times 1.56~\text{mm}^2$ for a FOV of $400 \times 400~\text{mm}^2$. The
cardiac cine phantom generated is spread across $\text{N}_\text{fr}=360$ time frames in the temporal
direction, consisting of 15 cardiac cycles and 24 cardiac phases. The MRXCAT-phantom dataset is
characterized by its limited inter-frame variations and periodic nature along the temporal
direction, unlike the other two datasets.

On the quantitative front, Fig.~\ref{fig:mrxcat.nrmse} provides quantitative comparisons in terms of NRMSE for a range of acceleration factors; while { Fig.~\ref{fig:nrmse.frame.mxrcat} provides the fluctuations in the NRMSE across the time frames in addition to Tab.~\ref{tab:mrxcat.quant} which summarizes all the performance metrics for an acceleration factor of 20x.} It is evident from Fig.~\ref{fig:mrxcat.nrmse}
that the proposed BiLMDM consistently outperforms the state-of-the-art schemes over the entire range
of acceleration rates. It is worth noticing here that BiLMDM scores similar NRMSE values to
PS-Sparse and MLS, at specific acceleration rates, even though it uses a small subset of the
navigator data (landmark points) in this case $N_{\ell} = 50,~d = 4$, in contrast to PS-Sparse and MLS that utilize the whole set of navigator data. BiLMDM's performance is consistent for every frame in the time series as displayed in Fig.~\ref{fig:nrmse.frame.mxrcat}, and exhibits the least NRMSE and least NRMSE deviation for every frame, unlike {XD-GRASP,} LASSI and SToRM. {Tab.~\ref{tab:mrxcat.quant} provides evidence that BiLMDM not only produces result closest to the ground truth, as exhibited by both the NRMSE and SSIM values, but also produces the sharpest image of all indicated by the smallest HFEN value and the largest M1 measure supported by a high M2 measure.}

{Shifting focus towards the qualitative aspect, }Fig.~\ref{fig:mrxcat.error} focuses on a dynamic region of interest (ROI) from the end-diastolic and
end-systolic phases. The proposed framework exhibits reconstruction of a high-resolution image with
contrast close to the gold standard, as opposed to the alias-infected reconstruction of XD-GRASP and SToRM or the
noisy reconstructions of PS-Sparse, MLS and LASSI. The error maps in Figs.~\ref{fig:mrxcat.temporal}
and~\ref{fig:mrxcat.error} provide visual proof of the improvements in reconstruction. BiLMDM
shows significant improvements at the image edges, as visualized in the error maps for both the
diastolic and systolic phases {and further supported by HFEN, M1 and M2 values}. Moreover, PS-Sparse, MLS and LASSI exhibit faint-right artifacts {(marked by red arrows in Fig.~\ref{fig:mrxcat.error})},
unlike BiLMDM. The temporal cross sections in Fig.~\ref{fig:mrxcat.temporal} are consistent with the
spatial results in Fig.~\ref{fig:mrxcat.error}. {Reconstructions from XD-GRASP and SToRM exhibit temporal blurring, while LASSI reconstructions show presence of noisy artifacts in Fig.~\ref{fig:mrxcat.temporal}. SToRM, LASSI and XD-GRASP also witness breaks in the periodicity of the cardiac phases as pointed by red arrows in Fig.~\ref{fig:mrxcat.temporal}.}

\begin{table}[]
\caption{Quantitative Performance Analysis for MRXCAT Phantom (Acceleration Rate: 20x)}
\centering
\resizebox{\columnwidth}{!}{%
{
\begin{tabular}{|l|l|l|l|l|l|}
\hline
          & NRMSE       & SSIM      & HFEN      & M1                & M2         \\ \hline
PS-Sparse & $0.055$     & $0.8913$  & $0.1548$  & $4856.8$   & $1.7\times10^7$     \\ \hline
MLS       & $0.0512$    & $0.8905$  & $0.1558$  & $4855.1$   & $1.7\times10^7$     \\ \hline
SToRM     & $0.0937$    & $0.776$   & $0.2882$  & $4594$     & $\bm{1.9\times10^7}$     \\ \hline
LASSI     & $0.0732$    & $0.7954$  & $0.1785$  & $4616.1$   & $1.6\times10^7$     \\ \hline
XD-GRASP  & $0.0653$    & $0.8205$  & $0.2478$  & $4764.3$   & $1.8\times10^7$           \\ \hline
BiLMDM    & $\bm{0.0488}$    & $\bm{0.9218}$  & $\bm{0.1314}$  & $\bm{4958.1}$   & $1.8\times10^7$     \\ \hline
\end{tabular}%
}
}\label{tab:mrxcat.quant}
\end{table}

\begin{table}[]
\caption{Quantitative Performance Analysis for cardiac phantom generated from real MR scans (Acceleration Rate:16x)}
\centering
\resizebox{\columnwidth}{!}{%
{
\begin{tabular}{|l|l|l|l|l|l|}
\hline
          & NRMSE    & SSIM      & HFEN     & M1           & M2         \\ \hline
PS-Sparse & $0.0863$ & $0.9216$  & $0.1375$ & $2.6\times10^{-6}$     & $175.6$     \\ \hline
MLS       & $0.0892$ & $0.9134$  & $0.2684$ & $2.6\times10^{-6}$     & $176.6$     \\ \hline
SToRM     & $0.0583$ & $0.9336$  & $0.1375$ & $2.6\times10^{-6}$     & $191.9$     \\ \hline
LASSI     & $0.0791$ & $0.7425$  & $0.368$  & $2.2\times10^{-6}$     & $180.7$     \\ \hline
XD GRASP  & $0.0587$ & $0.9348$  & $0.1193$ & $2.6\times10^{-6}$     & $194.4$                     \\ \hline
BiLMDM    & $\bm{0.0526}$ & $\bm{0.946}$   & $\bm{0.089}$  & $\bm{2.7\times10^{-6}}$     & $\bm{194.6}$     \\ \hline
\end{tabular}%
}
}\label{tab:real.quant}
\end{table}

\subsection{Cardiac phantom generated from real MR scan}\label{subsec:real.results}

Besides synthetically generated MR images, real human-cardiac cine MR data were also
acquired~\cite{zhao2012pssparse} with the following parameters: spatial size
$(\text{N}_\text{p},\text{N}_\text{f}) = (200,256)$, $\text{FOV} = 273 \times 350~\text{mm}^2$,
resulting in a spatial resolution of $1.36 \times 1.36~\text{mm}^2$. The data were acquired during a
single breath hold. Multiple time-wraps (introducing temporal variations) and quasi-periodic spatial
deformation~\cite{lin2000thin} (to model respiration) were then used to generate a temporal sequence
with $\text{N}_{\text{fr}} = 256$ frames. The results in Fig.~\ref{fig:perfusion.nrmse}, similar to
those on the previous cine dataset, show that BiLMDM outperforms the rest of the techniques with
NRMSE values ranging from $0.05$ to $0.061$ for $N_{\ell}=35,~d=12$. {Fig.~\ref{fig:nrmse.frame.perfusion}, further supports the ability of BiLMDM to create reconstructions with the least NRMSE fluctuations across the many frames. As seen in Tab.~\ref{tab:real.quant}, BiLMDM exhibits the least HFEN, M1 and M2 measure values which are testament to the fact that reconstructions obtained are sharper and are precise around the edges. The high similarity index proves that the BiLMDM is capable of producing results closest to the ground truth.}  

\begin{figure*}[ht]
    \centering
        \subfloat{\begin{annotate}
            {\includegraphics[height = 1.8cm, width = 0.1\textwidth, angle=90,origin=c]{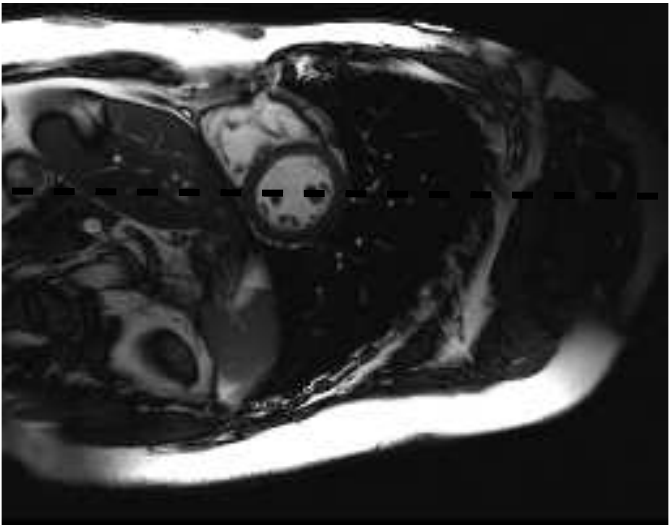}}{0.5}
            \draw[yellow, thick] (-1.2, 0) rectangle (-0.2,-0.8);
            \draw[yellow, thick, dashed] (-0.5, -1.75) -- (-0.5, 1.75);
            \end{annotate}}\hspace{-0.55cm} 
        \subfloat{\begin{annotate}
            {\includegraphics[ width = 0.1\textwidth]{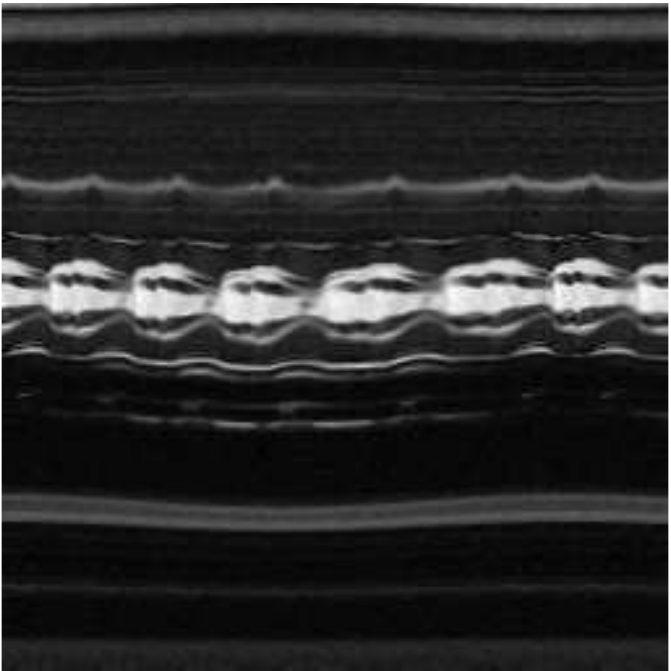}}{0.5}
            \draw[yellow, ->, thick] (0.5, 1.5) -- (1, 0.85);
            \draw[yellow, thick] (-1,0.6) rectangle (0,-0.2);
            \end{annotate}}\hspace{-0.55cm} 
        \subfloat{\begin{annotate}
            {\includegraphics[width = 0.1\textwidth]{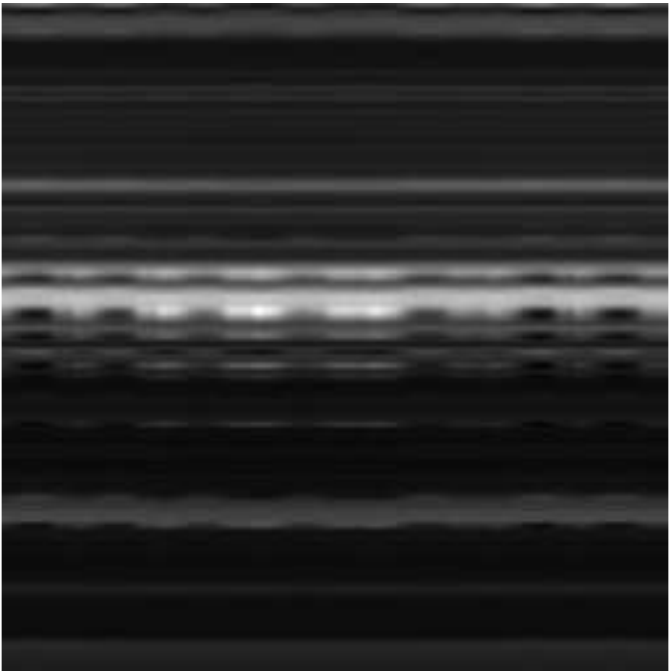}}{0.5}
            \draw[red, ->, thick] (0.5, 1.5) -- (1, 0.85);
            \draw[red, thick] (-1,0.6) rectangle (0,-0.2);
            \end{annotate}}\hspace{-0.55cm} 
        \subfloat{\begin{annotate}
            {\includegraphics[width = 0.1\textwidth]{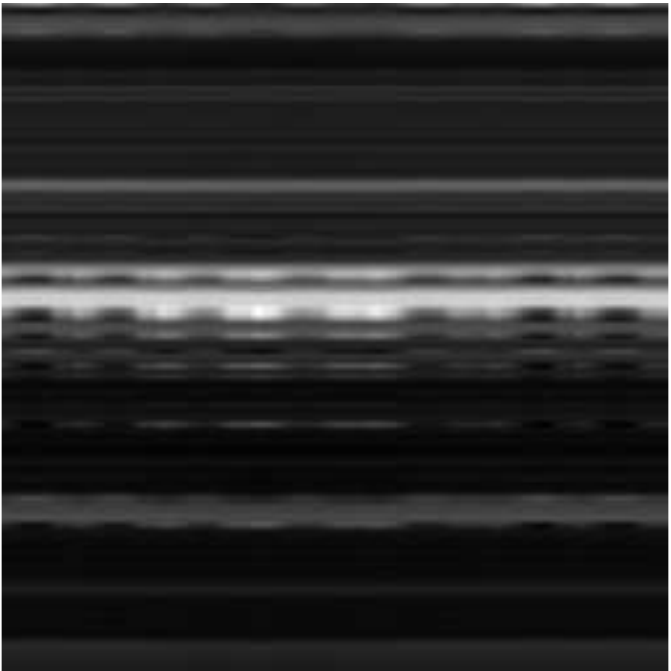}}{0.5}
            \draw[red, ->, thick] (0.5, 1.5) -- (1, 0.85);
            \draw[red, thick] (-1,0.6) rectangle (0,-0.2);
            \end{annotate}}\hspace{-0.55cm} 
        \subfloat{\begin{annotate}
            {\includegraphics[width = 0.1\textwidth]{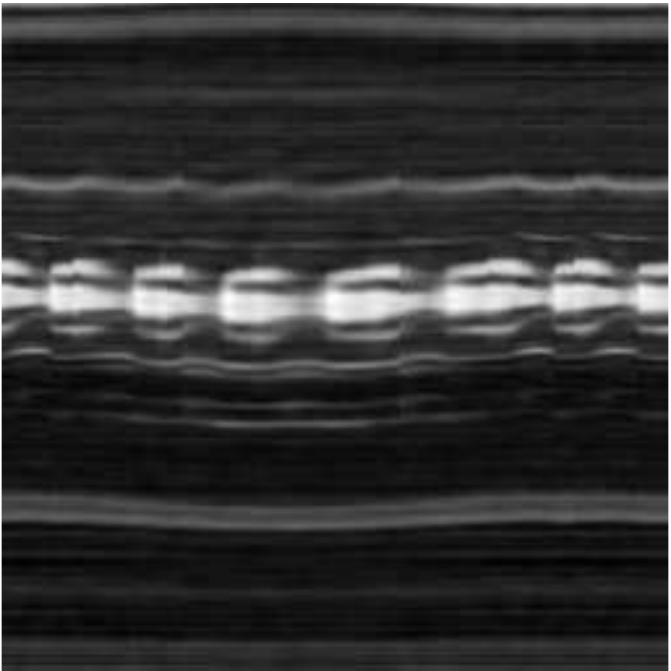}}{0.5}
            \draw[red, ->, thick] (0.5, 1.5) -- (1, 0.85);
            \draw[red, thick] (-1,0.6) rectangle (0,-0.2);
            \end{annotate}}\hspace{-0.55cm} 
        \subfloat{\begin{annotate}
            {\includegraphics[width = 0.1\textwidth]{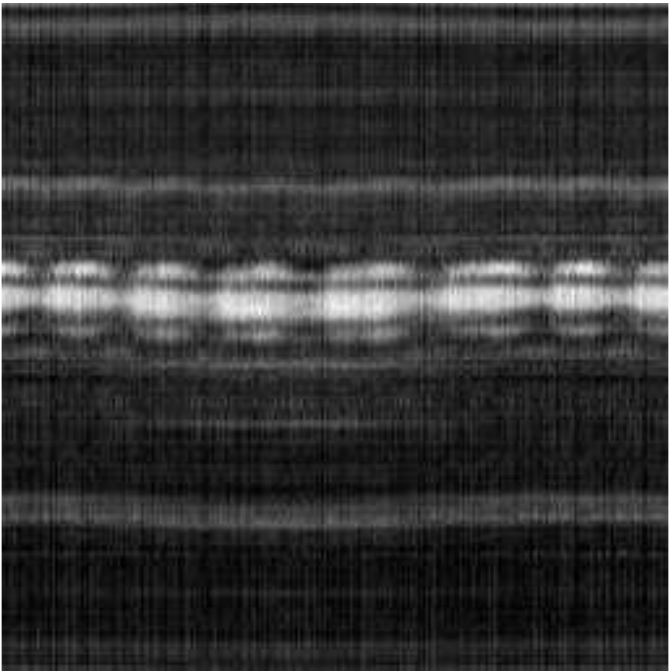}}{0.5}
            \draw[red, ->, thick] (0.5, 1.5) -- (1, 0.85);
            \draw[red, thick] (-1,0.6) rectangle (0,-0.2);
            \end{annotate}}\hspace{-0.55cm}
        \subfloat{\begin{annotate}
            {\includegraphics[width = 0.1\textwidth]{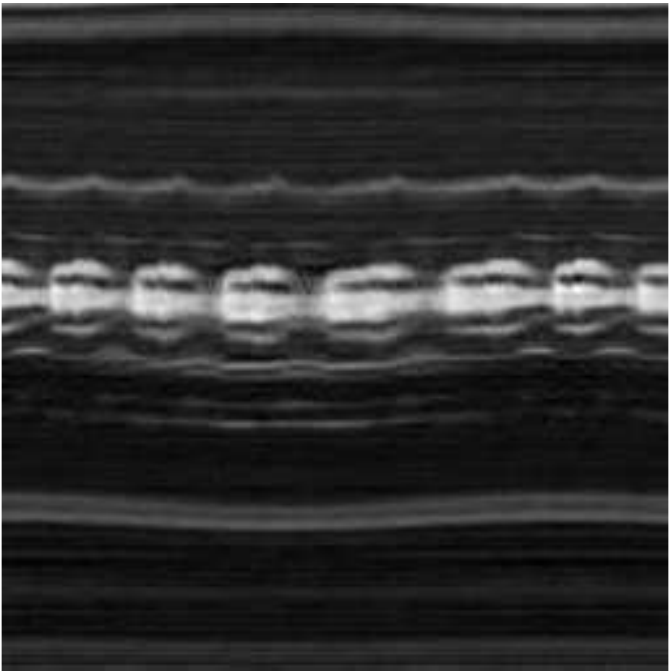}}{0.5}
            \draw[red, ->, thick] (0.5, 1.5) -- (1, 0.85);
            \draw[red, thick] (-1,0.6) rectangle (0,-0.2);
            \end{annotate}}\hspace{-0.55cm}   
        \subfloat{\begin{annotate}
            {\includegraphics[width = 0.1\textwidth]{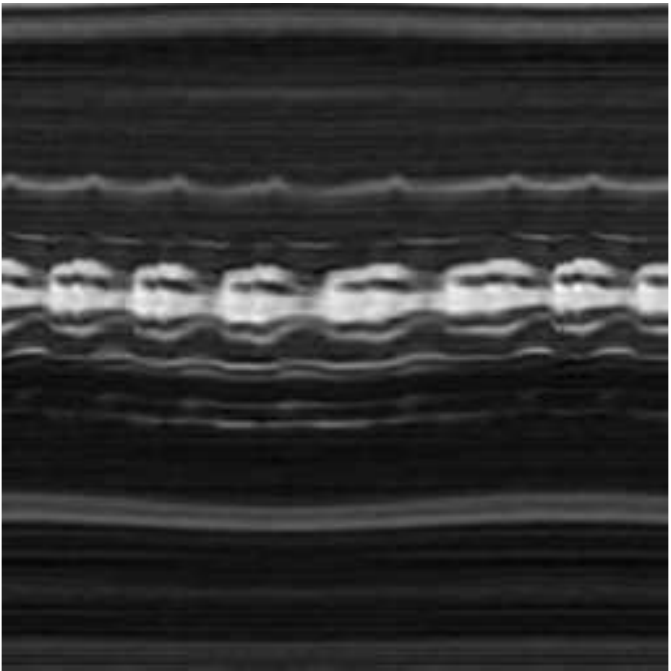}}{0.5}
            \draw[red, ->, thick] (0.5, 1.5) -- (1, 0.85);
            \draw[red, thick] (-1,0.6) rectangle (0,-0.2);
            \end{annotate}}\\  
        \vspace{-0.21cm} 
        \hspace{3.6cm}
        \subfloat{\begin{annotate}
            {\includegraphics[width = 0.1\textwidth]{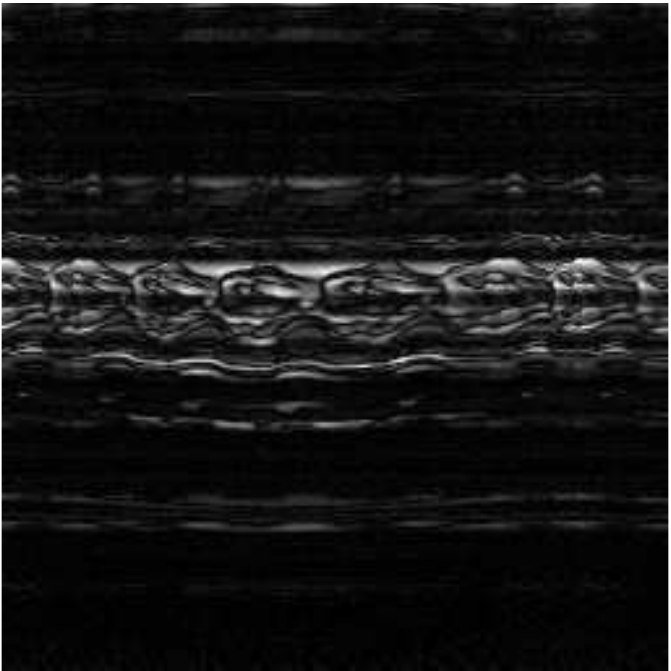}}{0.5}
            \end{annotate}}\hspace{-0.55cm} 
        \subfloat{\begin{annotate}
            {\includegraphics[width = 0.1\textwidth]{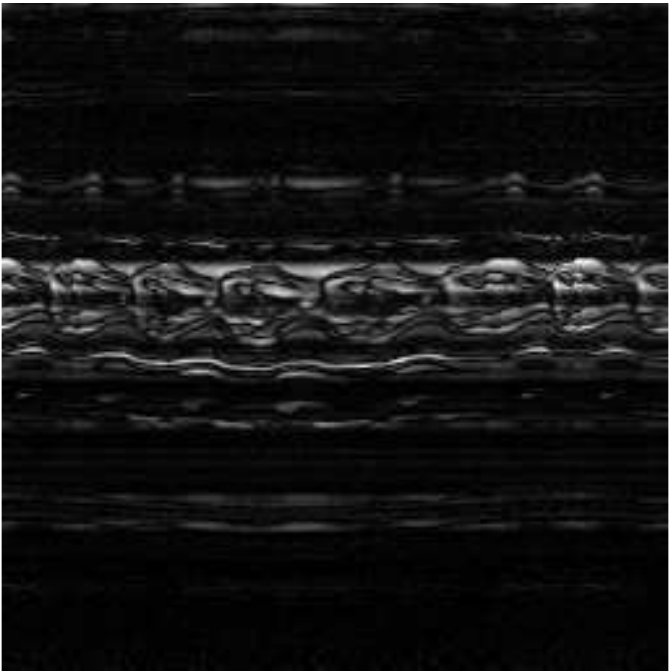}}{0.5}
            \end{annotate}}\hspace{-0.55cm} 
        \subfloat{\begin{annotate}
            {\includegraphics[width = 0.1\textwidth]{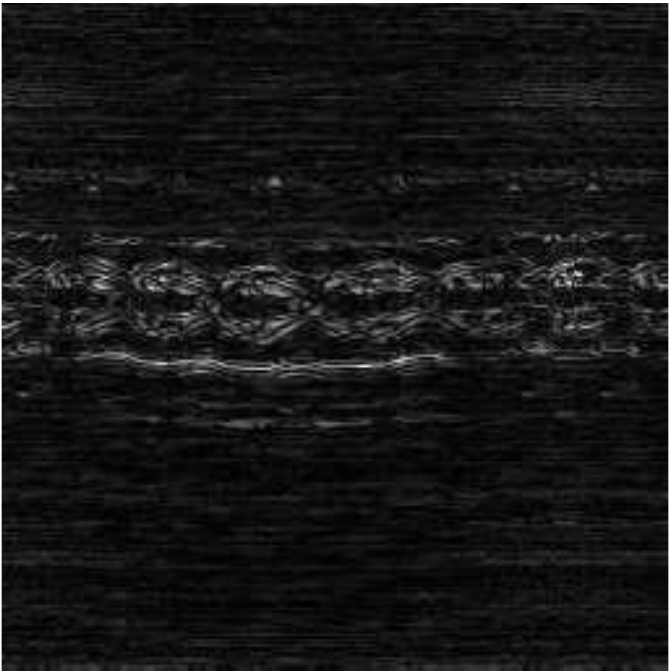}}{0.5}
            \end{annotate}}\hspace{-0.55cm} 
        \subfloat{\begin{annotate}
            {\includegraphics[width = 0.1\textwidth]{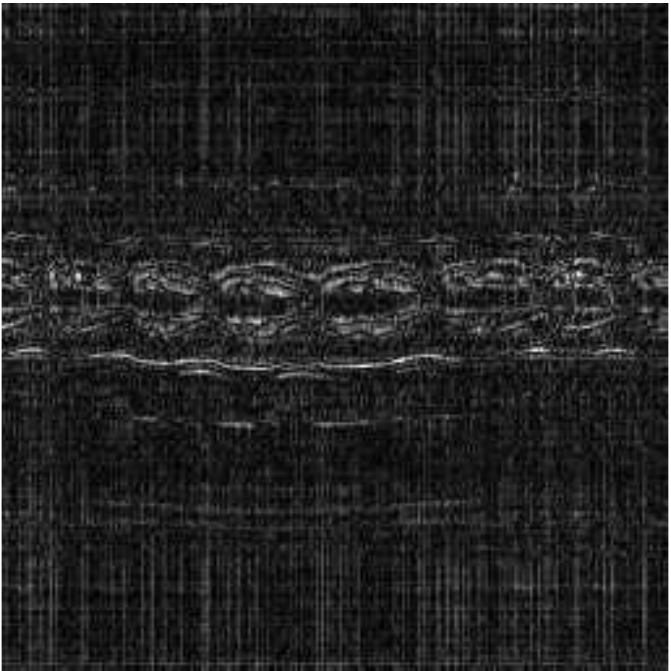}}{0.5}
            \end{annotate}}\hspace{-0.55cm}
        \subfloat{\begin{annotate}
            {\includegraphics[width = 0.1\textwidth]{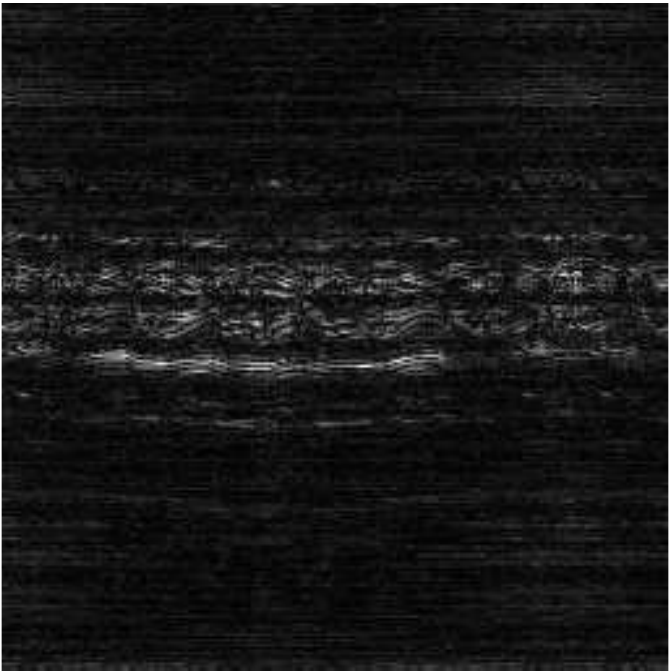}}{0.5}
            \end{annotate}}\hspace{-0.55cm}  
        \subfloat{\begin{annotate}
            {\includegraphics[width = 0.1\textwidth]{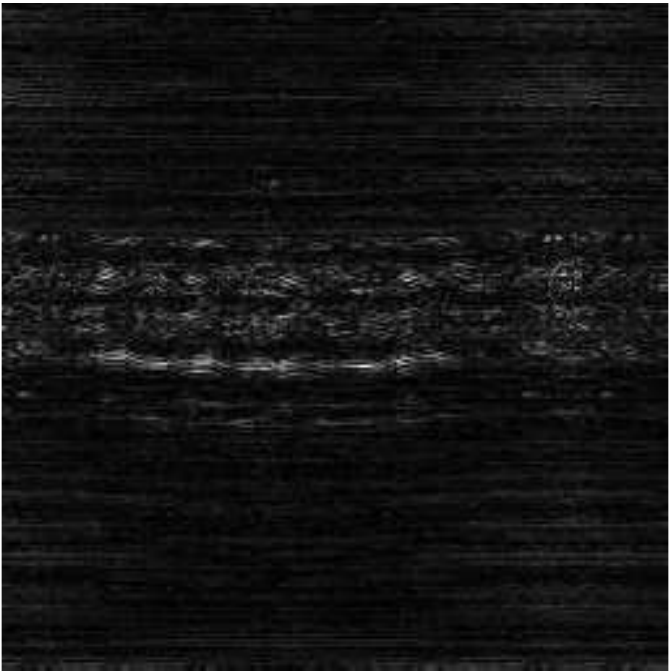}}{0.5}
            \end{annotate}}
        \caption{Temporal cross-sections for real cardiac cine data (acceleration rate:
          17x). Left to right: Gold standard (spatial frame), gold standard (temporal
          cross-section), PS-Sparse ($0.0863$), MLS ($0.0892$), SToRM ($0.0583$), LASSI ($0.0791$){, XD-GRASP ($0.0587$) }
          and BiLMDM ($0.0527 \pm 1.1 \times 10^{-4}$). The previous numerical values indicate the NRMSE for the complete dataset, in addition to the standard deviation (for the proposed scheme only) obtained after running the non-convex algorithmic scheme for $25$ independent trials. Top to bottom: Temporal cross
          section and error maps. The temporal location of the frames is indicated by the yellow dotted line in the gold-standard (spatial-frame) image.}
    \label{fig:perfusion.temporal}
\end{figure*}

\begin{figure*}[ht]
  \centering
  \subfloat{\begin{annotate}
        {\includegraphics[width = 0.12\textwidth, trim={0 0.8cm 1.9cm 0},clip, angle=90,origin=c]{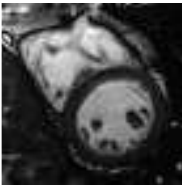}}{0.5} 
        \end{annotate}}\hspace{-0.55cm}
  \subfloat{\begin{annotate}
        {\includegraphics[width = 0.12\textwidth, trim={0 0.8cm 1.9cm 0},clip, angle=90,origin=c]{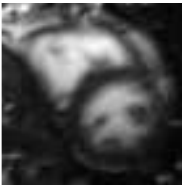}}{0.5} 
        \arrow{-1,1}{-0.3,1}
        \arrow{-0.5, -1.2}{-1,-0.8}
        \arrow{1.3,-0.3}{1.3,0.2}
        \end{annotate}}\hspace{-0.55cm} 
  \subfloat{\begin{annotate}
        {\includegraphics[width = 0.12\textwidth, trim={0 0.8cm 1.9cm 0},clip, angle=90,origin=c]{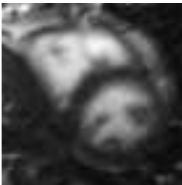}}{0.5}
        \arrow{-1,1}{-0.3,1}
        \arrow{-0.5, -1.2}{-1,-0.8}
        \arrow{1.3,-0.3}{1.3,0.2}
        \end{annotate}}\hspace{-0.55cm} 
  \subfloat{\begin{annotate}
        {\includegraphics[width = 0.12\textwidth, trim={0 0.8cm 1.9cm 0},clip, angle=90,origin=c]{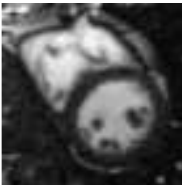}}{0.5}
        \arrow{-1,1}{-0.3,1}
        \arrow{-0.5, -1.2}{-1,-0.8}
        \arrow{1.3,-0.3}{1.3,0.2}
        \end{annotate}}\hspace{-0.55cm} 
  \subfloat{\begin{annotate}
        {\includegraphics[width = 0.12\textwidth, trim={0 0.8cm 1.9cm 0},clip, angle=90,origin=c]{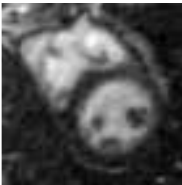}}{0.5}
        \arrow{-1,1}{-0.3,1}
        \arrow{-0.5, -1.2}{-1,-0.8}
        \arrow{1.3,-0.3}{1.3,0.2}
        \end{annotate}}\hspace{-0.55cm}
  \subfloat{\begin{annotate}
        {\includegraphics[width = 0.12\textwidth, trim={0 0.8cm 1.9cm 0},clip, angle=90,origin=c]{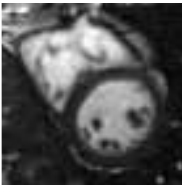}}{0.5}
        \arrow{-1,1}{-0.3,1}
        \arrow{-0.5, -1.2}{-1,-0.8}
        \arrow{1.3,-0.3}{1.3,0.2}
        \end{annotate}}\hspace{-0.55cm}    
  \subfloat{\begin{annotate}
        {\includegraphics[width = 0.12\textwidth, trim={0 0.8cm 1.9cm 0},clip, angle=90,origin=c]{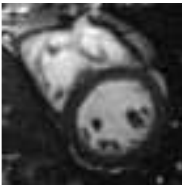}}{0.5}
        \arrow{-1,1}{-0.3,1}
        \arrow{-0.5, -1.2}{-1,-0.8}
        \arrow{1.3,-0.3}{1.3,0.2}
        \end{annotate}}\\  
  \vspace{-0.55cm} 
  \subfloat{\begin{annotate}
        {\includegraphics[width = 0.12\textwidth, trim={0 0.8cm 1.9cm 0},clip, angle=90,origin=c]{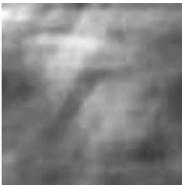}}{0.5}
        \end{annotate}}\hspace{-0.55cm}
  \subfloat{\begin{annotate}
        {\includegraphics[width = 0.12\textwidth, trim={0 0.8cm 1.9cm 0},clip, angle=90,origin=c]{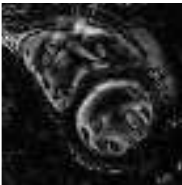}}{0.5}
        \end{annotate}}\hspace{-0.55cm} 
  \subfloat{\begin{annotate}
        {\includegraphics[width = 0.12\textwidth, trim={0 0.8cm 1.9cm 0},clip, angle=90,origin=c]{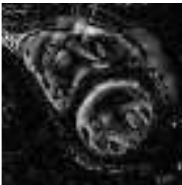}}{0.5}
        \end{annotate}}\hspace{-0.55cm} 
  \subfloat{\begin{annotate}
        {\includegraphics[width = 0.12\textwidth, trim={0 0.8cm 1.9cm 0},clip, angle=90,origin=c]{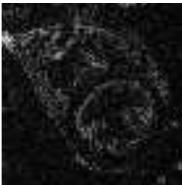}}{0.5}
        \end{annotate}}\hspace{-0.55cm} 
  \subfloat{\begin{annotate}
        {\includegraphics[width = 0.12\textwidth, trim={0 0.8cm 1.9cm 0},clip, angle=90,origin=c]{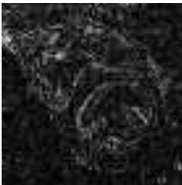}}{0.5}
        \end{annotate}}\hspace{-0.55cm}
  \subfloat{\begin{annotate}
        {\includegraphics[width = 0.12\textwidth, trim={0 0.8cm 1.9cm 0},clip, angle=90,origin=c]{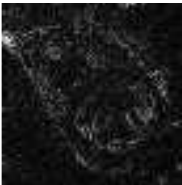}}{0.5}
        \end{annotate}}\hspace{-0.55cm}
  \subfloat{\begin{annotate}
        {\includegraphics[width = 0.12\textwidth, trim={0 0.8cm 1.9cm 0},clip, angle=90,origin=c]{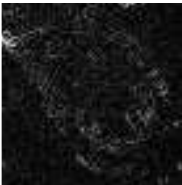}}{0.5}
        \end{annotate}}\\ 
  \vspace{-0.55cm}
  \subfloat{\begin{annotate}
        {\includegraphics[width = 0.12\textwidth, trim={0 0.8cm 1.9cm 0},clip, angle=90,origin=c]{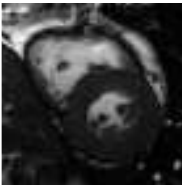}}{0.5}
        \end{annotate}}\hspace{-0.55cm} 
  \subfloat{\begin{annotate}
        {\includegraphics[width = 0.12\textwidth, trim={0 0.8cm 1.9cm 0},clip, angle=90,origin=c]{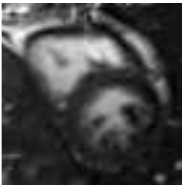}}{0.5}
        \draw[red, thick] (-0.2,1.5) rectangle (1.7,-0.4);
        \arrow{0,-1.5}{-0.2,-1}
        \end{annotate}}\hspace{-0.55cm} 
  \subfloat{\begin{annotate}
        {\includegraphics[width = 0.12\textwidth, trim={0 0.8cm 1.9cm 0},clip, angle=90,origin=c]{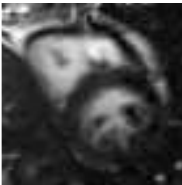}}{0.5}
        \draw[red, thick] (-0.2,1.5) rectangle (1.7,-0.4);
        \arrow{0,-1.5}{-0.2,-1}
        \end{annotate}}\hspace{-0.55cm} 
  \subfloat{\begin{annotate}
        {\includegraphics[width = 0.12\textwidth, trim={0 0.8cm 1.9cm 0},clip, angle=90,origin=c]{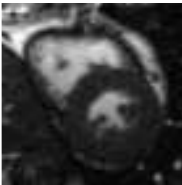}}{0.5}
        \draw[red, thick] (-0.1,1) rectangle (1.1,-0.2);
        \arrow{0,-1.5}{-0.2,-1}
        \end{annotate}}\hspace{-0.55cm} 
  \subfloat{\begin{annotate}
        {\includegraphics[width = 0.12\textwidth, trim={0 0.8cm 1.9cm 0},clip, angle=90,origin=c]{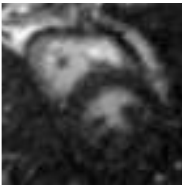}}{0.5}
        \draw[red, thick] (-0.2,1.5) rectangle (1.7,-0.4);
        \arrow{0,-1.5}{-0.2,-1}
        \end{annotate}}\hspace{-0.55cm}
  \subfloat{\begin{annotate}
        {\includegraphics[width = 0.12\textwidth, trim={0 0.8cm 1.9cm 0},clip, angle=90,origin=c]{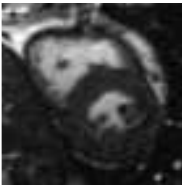}}{0.5}
        \draw[red, thick] (-0.1,1) rectangle (1.1,-0.2);
        \arrow{0,-1.5}{-0.2,-1}
        \end{annotate}}\hspace{-0.55cm}    
  \subfloat{\begin{annotate}
        {\includegraphics[width = 0.12\textwidth, trim={0 0.8cm 1.9cm 0},clip, angle=90,origin=c]{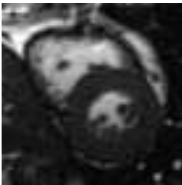}}{0.5}
        \draw[red, thick] (-0.1,1) rectangle (1.1,-0.2);
        \arrow{0,-1.5}{-0.2,-1} 
        \end{annotate}}\\  
  \vspace{-0.55cm}
  \subfloat{\begin{annotate}
        {\includegraphics[width = 0.12\textwidth, trim={0 0.8cm 1.9cm 0},clip, angle=90,origin=c]{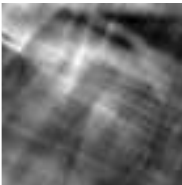}}{0.5}
        \end{annotate}}\hspace{-0.55cm}
  \subfloat{\begin{annotate}
        {\includegraphics[width = 0.12\textwidth, trim={0 0.8cm 1.9cm 0},clip, angle=90,origin=c]{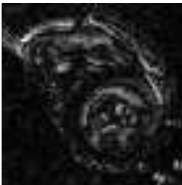}}{0.5}
        \end{annotate}}\hspace{-0.55cm} 
  \subfloat{\begin{annotate}
        {\includegraphics[width = 0.12\textwidth, trim={0 0.8cm 1.9cm 0},clip, angle=90,origin=c]{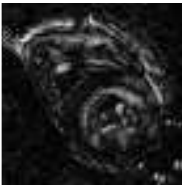}}{0.5}
        \end{annotate}}\hspace{-0.55cm} 
  \subfloat{\begin{annotate}
        {\includegraphics[width = 0.12\textwidth, trim={0 0.8cm 1.9cm 0},clip, angle=90,origin=c]{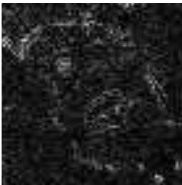}}{0.5}
        \end{annotate}}\hspace{-0.55cm} 
  \subfloat{\begin{annotate}
        {\includegraphics[width = 0.12\textwidth, trim={0 0.8cm 1.9cm 0},clip, angle=90,origin=c]{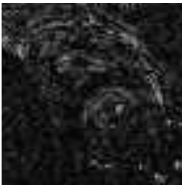}}{0.5}
        \end{annotate}}\hspace{-0.55cm}
  \subfloat{\begin{annotate}
        {\includegraphics[width = 0.12\textwidth, trim={0 0.8cm 1.9cm 0},clip, angle=90,origin=c]{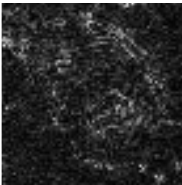}}{0.5}
        \end{annotate}}\hspace{-0.55cm}    
  \subfloat{\begin{annotate}
        {\includegraphics[width = 0.12\textwidth, trim={0 0.8cm 1.9cm 0},clip, angle=90,origin=c]{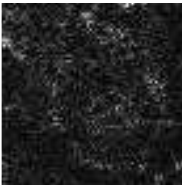}}{0.5}
        \end{annotate}}
  \caption{Spatial results for {region of interest marked with a yellow rectangle in the gold standard (spatial-frame) image of Fig.~\ref{fig:perfusion.temporal}} real cardiac cine data (acceleration rate: 17x). Left to right: Gold standard, PS-Sparse ($0.0863$), MLS ($0.0892$), SToRM ($0.0583$), LASSI ($0.0791$){, XD-GRASP ($0.0587$)} and
    BiLMDM ($0.0527 \pm 1.1 \times 10^{-4}$). The previous numerical values indicate the NRMSE for the complete dataset, in addition to the standard deviation (for BiLMDM only) obtained after running the non-convex algorithmic scheme for $25$ independent trials. Top to bottom: Diastole phase (frame $1$ of the
    time series), the under-sampled image followed by error maps, systole phase (frame $16$ of the
    time series) and the under-sampled image followed by error maps.}
  \label{fig:perfusion.error}
\end{figure*}

With regards to Figs.~\ref{fig:perfusion.temporal} and \ref{fig:perfusion.error}, significant
blurring and deformations can be observed for PS-Sparse, MLS and LASSI, in contrast to the sharper
images produced by BiLMDM for both the end-diastolic and end-systolic phases. There are slight
improvements over SToRM {and XD-GRASP}, which are illustrated via the error maps. At the same time, it is worth noting that in the end-systolic phase, PS-Sparse, MLS and LASSI produce an enlargement (deformation) of the area {marked with a red rectangle} on the image. The temporal cross sections in Fig.~\ref{fig:perfusion.temporal}
indicate significant motion blurring for PS-Sparse, MLS and LASSI. Moreover, it can be verified that
a large number of artifacts are present in the temporal cross section produced by LASSI. {Significant temporal bleeding is observed in results from PS-Sparse and MLS which is consistent with the undesirable enlargement in the end-systolic phase discussed earlier. Even though SToRM and XD-GRASP provide reconstructions almost as good as BiLMDM, it can be seen that there are signs of temporal blurring in the marked areas in Fig.~\ref{fig:perfusion.temporal} for the corresponding methods.} On the
contrary, BiLMDM provides a temporal cross-section very close to the gold standard.

\subsection{Prospectively Undersampled Cardiac Cine Data}\label{subsec:prospective}

{The prospectively undersampled real-time free breathing cardiac cine data was acquired using a FLASH sequence from a volunteer breathing normally under the following acquisition parameters: TR/TE = $5.8/4$ ms, FOV $284\times350$ mm, spatial resolution = $1.8$ mm. A 12 channel scanner was used to continuously acquire 4500 phase encoding lines under 1-D Cartesian trajectory. These phase encoded lines were divided in groups of 15 (including 5 navigator lines) to form 300 frames, resulting in a data matrix of size $156\times192\times300$ for each channel. The above acquisition corresponds to a undersampling factor of nearly 10x. For fairness in comparison, the multi-channel data was reconstructed coil by coil for all methods and then combined with the sum-of-squares strategy.}

\begin{figure*}[ht]
  \centering
  \subfloat{\begin{annotate}
        {\includegraphics[width = 0.16\textwidth,clip, angle=90,origin=c]{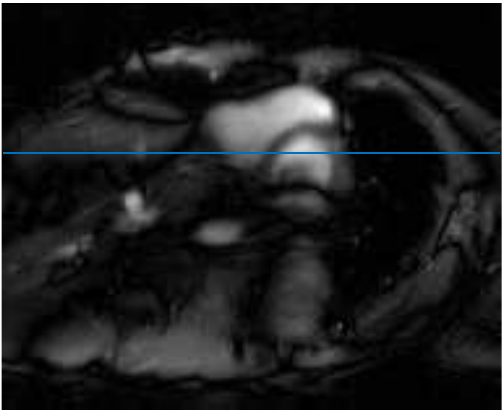}}{0.5}
        \draw[yellow, thick] (-1.4, 1.2) rectangle (0.2, -0.5);
        \draw[yellow,thick, dashed] (-0.6, 2.75) -- (-0.6, -2.75);
        \draw[red, thick, ->] (-0,-1.2) -- (0.5,-1.4);
        \draw[red, thick, ->] (0.5,0) -- (1,0);
        \arrow{-0.2,-2}{0,-2.5}
        \end{annotate}}\hspace{-0.55cm} 
  \subfloat{\begin{annotate}
        {\includegraphics[width = 0.16\textwidth,clip, angle=90,origin=c]{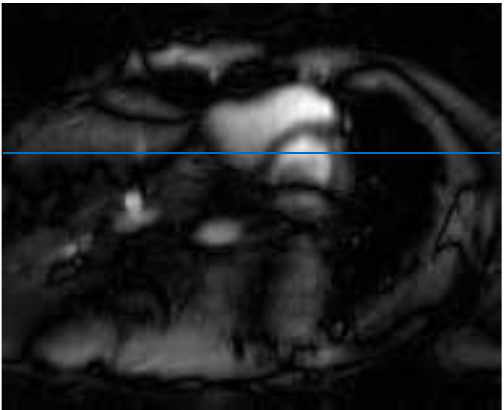}}{0.5}
        \draw[yellow, thick] (-1.4, 1.2) rectangle (0.2, -0.5);
        \draw[yellow,thick, dashed] (-0.6, 2.75) -- (-0.6, -2.75);
        \draw[red, thick, ->] (-0,-1.2) -- (0.5,-1.4);
        \draw[red, thick, ->] (0.5,0) -- (1,0);
        \arrow{-0.2,-2}{0,-2.5}
        \end{annotate}}\hspace{-0.55cm} 
  \subfloat{\begin{annotate}
        {\includegraphics[width = 0.16\textwidth,clip, angle=90,origin=c]{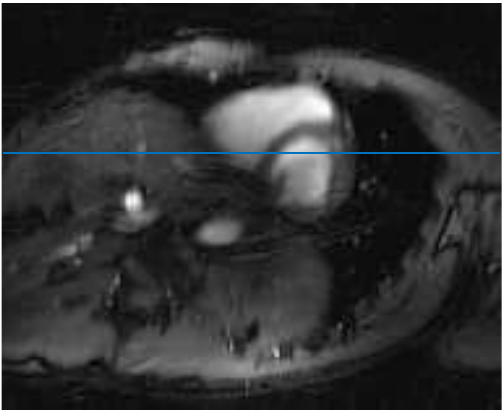}}{0.5}
        \draw[yellow, thick] (-1.4, 1.2) rectangle (0.2, -0.5);
        \draw[yellow,thick, dashed] (-0.6, 2.75) -- (-0.6, -2.75);
        \draw[red, thick, ->] (-0,-1.2) -- (0.5,-1.4);
        \draw[red, thick, ->] (-0.2,-2) -- (0,-2.5);
        \draw[red, thick, ->] (0.5,0) -- (1,0);
        \end{annotate}}\hspace{-0.55cm} 
  \subfloat{\begin{annotate}
        {\includegraphics[width = 0.16\textwidth,clip, angle=90,origin=c]{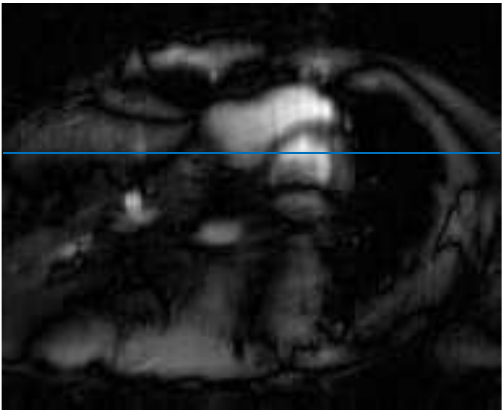}}{0.5}
        \draw[yellow, thick] (-1.4, 1.2) rectangle (0.2, -0.5);
        \draw[yellow,thick, dashed] (-0.6, 2.75) -- (-0.6, -2.75);
        \draw[red, thick, ->] (-0,-1.2) -- (0.5,-1.4);
        \draw[red, thick, ->] (0.5,0) -- (1,0);
        \arrow{-0.2,-2}{0,-2.5}
        \end{annotate}}\hspace{-0.55cm}
  \subfloat{\begin{annotate}
        {\includegraphics[width = 0.16\textwidth,clip, angle=90,origin=c]{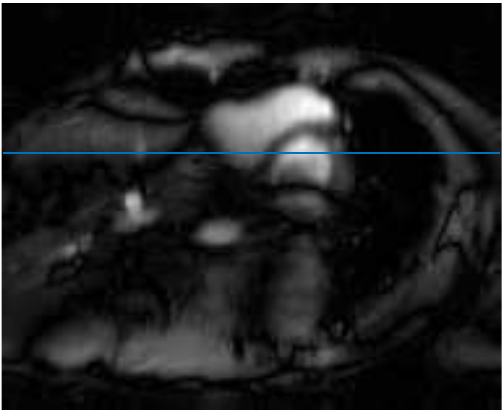}}{0.5}
        \draw[yellow, thick] (-1.4, 1.2) rectangle (0.2, -0.5);
        \draw[yellow,thick, dashed] (-0.6, 2.75) -- (-0.6, -2.75);
        \draw[red, thick, ->] (-0,-1.2) -- (0.5,-1.4);
        \draw[red, thick, ->] (0.5,0) -- (1,0);
        \arrow{-0.2,-2}{0,-2.5}
        \end{annotate}}\hspace{-0.55cm}    
  \subfloat{\begin{annotate}
        {\includegraphics[width = 0.16\textwidth,clip, angle=90,origin=c]{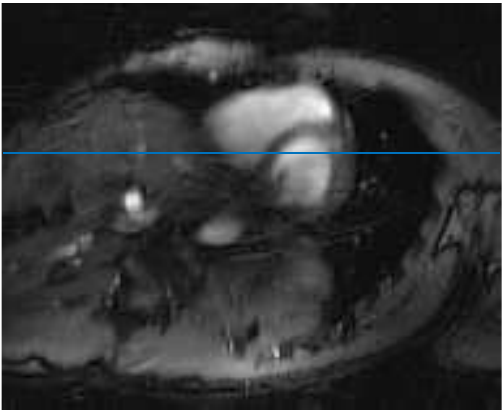}} {0.5}
        \draw[yellow, thick] (-1.4, 1.2) rectangle (0.2, -0.5);
        \draw[yellow,thick, dashed] (-0.6, 2.75) -- (-0.6, -2.75);
        \draw[red, thick, ->] (-0,-1.2) -- (0.5,-1.4);
        \draw[red, thick, ->] (-0.2,-2) -- (0,-2.5);
        \draw[red, thick, ->] (0.5,0) -- (1,0);
        \end{annotate}}\\ 
  \vspace{-0.58cm}
  \subfloat{\begin{annotate}
        {\includegraphics[width = 2.37cm,clip, angle=90,origin=c]{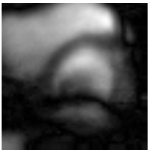}}{0.5}
        \arrow{0.3,-0.6}{0.8,-0.2}
        \end{annotate}}\hspace{-0.55cm} 
  \subfloat{\begin{annotate}
        {\includegraphics[width = 2.37cm,clip, angle=90,origin=c]{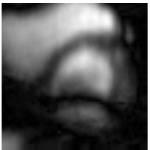}}{0.5}
        \arrow{0.3,-0.6}{0.8,-0.2}
        \end{annotate}}\hspace{-0.55cm} 
  \subfloat{\begin{annotate}
        {\includegraphics[width = 2.37cm,clip, angle=90,origin=c]{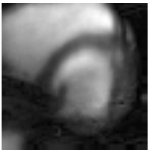}}{0.5}
        \draw[red, thick,->] (0.3,-0.8) -- (0.8,-0.5);
        \end{annotate}}\hspace{-0.55cm} 
  \subfloat{\begin{annotate}
        {\includegraphics[width = 2.37cm,clip, angle=90,origin=c]{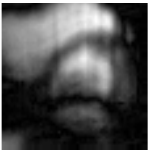}}{0.5}
        \arrow{0.3,-0.6}{0.8,-0.2}
        \arrow{-0.5,0.3}{0.2,0.3}
        \end{annotate}}\hspace{-0.55cm}
  \subfloat{\begin{annotate}
        {\includegraphics[width = 2.37cm,clip, angle=90,origin=c]{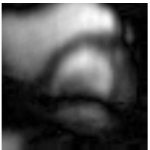}}{0.5}
        \arrow{0.3,-0.6}{0.8,-0.2}
        \end{annotate}}\hspace{-0.55cm}    
  \subfloat{\begin{annotate}
        {\includegraphics[width = 2.37cm,clip, angle=90,origin=c]{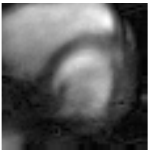}}{0.5} 
        \draw[red, thick,->] (0.3,-0.8) -- (0.8,-0.5);
        \end{annotate}}\\ 
  \vspace{-0.55cm}
  \subfloat{\begin{annotate}
        {\includegraphics[width = 2.36cm,clip, angle=180,origin=c]{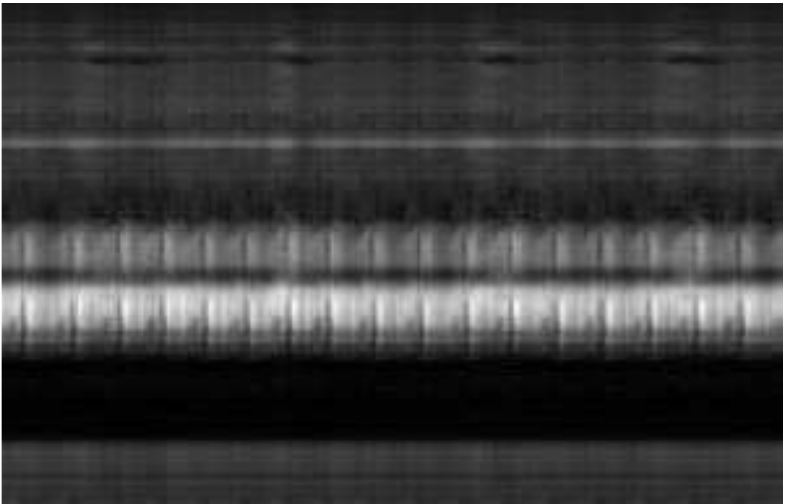}}{0.5}
        \draw[red, thick] (0,0.7) rectangle (1,-0.4);
        \end{annotate}}\hspace{-0.55cm} 
  \subfloat{\begin{annotate}
        {\includegraphics[width = 2.36cm,clip, angle=180,origin=c]{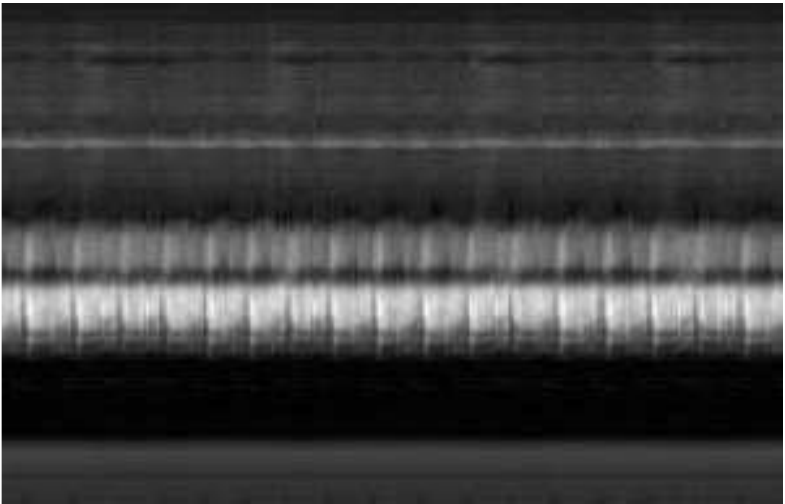}}{0.5}
        \draw[red, thick] (0,0.7) rectangle (1,-0.4);
        \end{annotate}}\hspace{-0.55cm} 
  \subfloat{\begin{annotate}
        {\includegraphics[width = 2.36cm,clip, angle=180,origin=c]{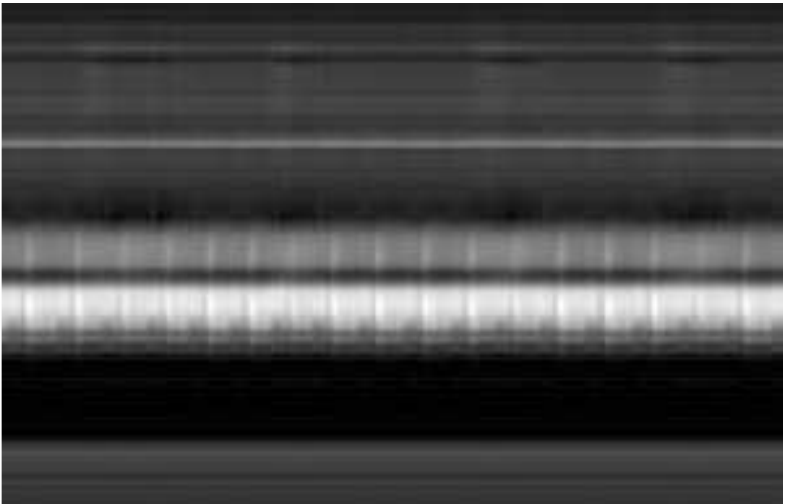}}{0.5}
        \draw[red, thick] (0,0.7) rectangle (1,-0.4);
        \end{annotate}}\hspace{-0.55cm} 
  \subfloat{\begin{annotate}
        {\includegraphics[width = 2.36cm,clip, angle=180,origin=c]{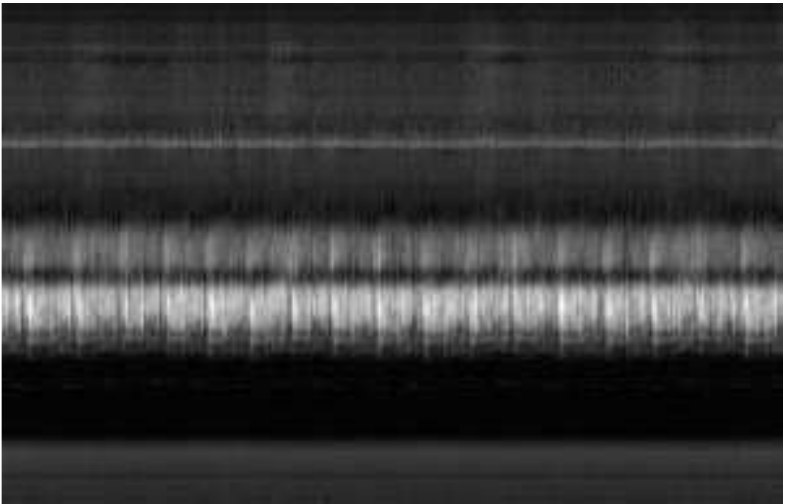}}{0.5}
        \draw[red, thick] (0,0.7) rectangle (1,-0.4);
        \end{annotate}}\hspace{-0.55cm}
  \subfloat{\begin{annotate}
        {\includegraphics[width = 2.36cm,clip, angle=180,origin=c]{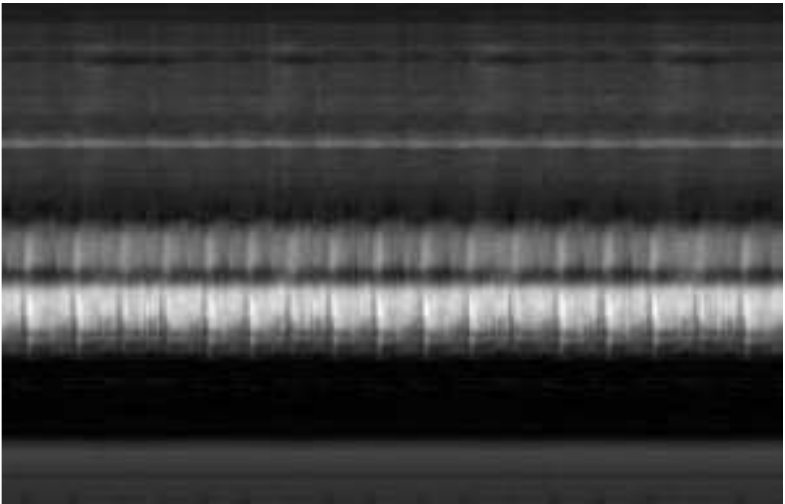}}{0.5}
        \draw[red, thick] (0,0.7) rectangle (1,-0.4);
        \end{annotate}}\hspace{-0.55cm}    
  \subfloat{\begin{annotate}
        {\includegraphics[width = 2.36cm,clip, angle=180,origin=c]{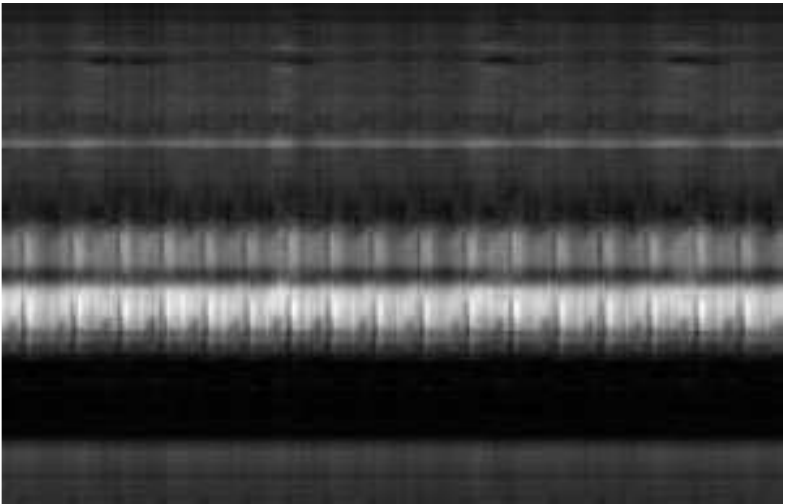}}{0.5} 
        \draw[red, thick] (0,0.7) rectangle (1,-0.4);
        \end{annotate}}
  \caption{Spatial results for prospectively undersampled cardiac cine data (acceleration rate: 10x). Left to right: PS-Sparse, MLS, SToRM, LASSI, XD-GRASP and BiLMDM. Top to bottom: Reconstructed frame $1$ of the time series, the region of interest (cardiac region marked with a yellow rectangle in the images of the topmost row), temporal cross-section of the reconstructed time series along the yellow dotted line in the images of the topmost row.}
  \label{fig:prospetive.error}
\end{figure*}

{Fig.~\ref{fig:prospetive.error} shows the reconstruction results from the proposed and state-of-the methods. The validation skips most of the performance metrics except for the sharpness measure due to the absence of a ground truth. Hence the validation is solely dependent on the visual quality as shown in Fig.~\ref{fig:prospetive.error} and the sharpness measures. The full FOV images exhibit that BiLMDM along with SToRM, unlike other methods show well-defined structures and result in a sharper image. LASSI exhibits some aliasing effects, while PS-Sparse and MLS observes some blurring in the spatial frame. This is further supported by sharpness measures (M1 followed by M2): PS-Sparse ($8.0\times10^{-8}$, $81.7$), MLS ($8.2\times10^{-8}$, $79.9$), SToRM ($8.6\times10^{-7}$, $\bm{91.8}$), LASSI ($8.2\times10^{-8}$, $83.6$), XD-GRASP ($8.2\times10^{-7}$, $80.6$) and BiLMDM ($\bm{8.8\times10^{-7}}$, $91.7$). Comparing the temporal cross sections for the reconstructed images, even though the FOV images look pretty sharp for SToRM and BiLMDM, SToRM suffers from more temporal blurring than BiLMDM, while LASSI and PS-Sparse show grainy artifacts in the temporal cross section.}


\section{Discussion}\label{sec:discussion}

{The proposed BiLMDM scheme relies on data
  sharing among different frames by approximating each frame by an affine
  combination of its neighboring frames on the manifold that describes locally
  the MR data cloud. This manifold and sparse-approximation based recovery model
  enables the reconstruction of desirable, good quality and artifact free MR
  images from severely undersampled k-space measurements. The quality of the
  reconstructed images and the corresponding performance in numbers discussed
  under Sec.~\ref{sec:results} is a testament to the fact that the BiLMDM
  outperforms the competing state-of-the-art methods. BiLMDM has consistently
  provided better results for both synthetically generated, retrospectively
  undersampled as well as experimentally acquired, prospectively undersampled
  cardiac cine data.

  Methods like PS-Sparse and MLS have reconstructed desirable MR images in the
  case of the MRXCAT phantom, under cartesian sampling trajectories; however,
  they suffered severe motion blurring and deformations (enlargements) in the
  case of the cardiac cine data retrospectively sampled via radial
  trajectories. In contrast, SToRM produced good quality images for the data
  using radial sampling but suffered severe spatial blurring and temporal
  bleeding in the cases of Cartesian sampling. However, BiLMDM reconstructions
  have been artifact free, without any temporal bleeding irrespective of what
  sampling strategy was used to acquire data. Similar results to the ones
  discussed here hold also for the sampling strategies not included in the
  manuscript, for all employed algorithms. Even in the case of the
  experimentally acquired data, where techniques like PS-Sparse, MLS, XD-GRASP
  and LASSI suffered from some missing structures and distortions, while SToRM
  suffered temporal blurring, BiLMDM produced images which seem to best preserve
  the features of the cardiac structure spatially and temporally. BiLMDM seems
  capable of reconstructing desirable MR images irrespective of the sampling
  strategy into consideration, which appears to be a dominant factor in the
  competing state-of-the-art methods.}
\section{Conclusions}\label{sec:conclusion}

This paper proposed the novel bi-linear modeling for data manifolds (BiLMDM); a
new framework for data reconstruction using manifold-learning and
sparse-approximation arguments. BiLMDM comprises several modules: Extracting a
set of landmark points from a data cloud helps in learning the latent manifold
geometry while identifying low-dimensional renditions of the landmark points
facilitates efficient means for data storage and computations. Finally, a
bi-linear optimization task is used to achieve data recovery. Quantitative and
qualitative analyses on dynamic MRI data, described in Sec.~\ref{sec:results},
provided evidence that the proposed BiLMDM achieves improvements in dMRI image
reconstruction and artifact suppression over state-of-the-art approaches such as
PS-Sparse, MLS, LASSI{, XD-GRASP} and SToRM in cardiac cine
data. {BiLMDM reconstruction is not limited to cine MR data but has
  also produced reliable reconstructions for perfusion MR data. The efficacy of
  BiLMDM on perfusion data and extensive comparison tests versus
  state-of-the-art schemes are reserved for a future publication.} By
introducing BiLMDM, this work paves the way for devising efficient ways to
utilize fewer data points for reconstruction than state-of-the-art solutions and
opens the door for further advances in geometric data-approximation methods.

\appendices
\section{Mathematical Preliminaries}\label{supp:basic.def}

For positive integers $m, n$, the space of matrices $\Complex^{m\times n}$ is equipped with the
inner product $\innerp{\vect{A}}{\vect{B}} := \trace(\vect{A^{\hermconj}}\vect{B})$,
$\forall \vect{A}, \vect{B}\in \Complex^{m\times n}$, where the superscript $\hermconj$ denotes the
Hermitian transpose of a matrix. It is worth noting that the inner product is not commutative:
$\innerp{\vect{B}}{\vect{A}} = \conj{\innerp{\vect{A}}{\vect{B}}}$, where the overline symbol
denotes the complex conjugate of a number. The induced norm of $\Complex^{m\times n}$ by the
previous inner product coincides with the Frobenius norm of a matrix:
$\norm{\cdot}_{\text{F}} = \innerp{\cdot}{\cdot}^{1/2}$. Moreover, the spectral norm
$\norm{\vect{A}}_2$ of matrix $\vect{A}\in \Complex^{m\times n}$ is defined as
$\lambda_{\max}^{1/2}(\vect{A}^{\hermconj} \vect{A})$, where $\lambda_{\max}(\cdot)$ stands for the
maximum eigenvalue of a symmetric matrix. In the case where $m = n$, then
$\norm{\vect{A}}_2 = \lambda_{\max}(\vect{A})$.

Given the positive integers $m_1, m_2, n_1, n_2$ and the linear mapping
$\mathcal{L}: \Complex^{m_1\times n_1}\to \Complex^{m_2\times n_2}$, the adjoint of $\mathcal{L}$ is
the linear mapping $\mathcal{L}^*: \Complex^{m_2\times n_2}\to \Complex^{m_1\times n_1}$ defined as
$\innerp{\vect{A}}{\mathcal{L}(\vect{B})} = \innerp{\mathcal{L}^*(\vect{A})}{\vect{B}}$,
$\forall \vect{A}\in \Complex^{m_2\times n_2}$, $\forall \vect{B}\in \Complex^{m_1\times n_1}$. For
example, with regards to the sampling mapping $\mathcal{S}(\cdot)$ in Sec.~\ref{sec:Task.Algo}, its
adjoint $\mathcal{S}^*(\cdot) = \mathcal{S}(\cdot)$, \ie, $\mathcal{S}(\cdot)$ is self-adjoint, and
$\mathcal{S}^2(\cdot) = \mathcal{S}(\cdot)$. Moreover, for the MATLAB implementation of the Fourier
transform~\cite{MATLAB.2018b}, $\mathcal{F}^* = N_{\text{k}} \mathcal{F}^{-1}$ and
$\mathcal{F}_t^* = N_{\text{fr}} \mathcal{F}_t^{-1}$. If matrix $\vect{A}\in \Complex^{m\times n}$
is viewed as a linear mapping $\vect{A}: \Complex^n\to \Complex^m$, then
$\vect{A}^* = \vect{A}^{\hermconj}$. Mapping $\mathcal{L}(\cdot)$ is called Lipschitz continuous,
with coefficient $L>0$, if
$\norm{\mathcal{L}(\vect{A}) - \mathcal{L}(\vect{B})}_{\text{F}} \leq L \norm{\vect{A} -
  \vect{B}}_{\text{F}}$, $\forall\vect{A}, \vect{B}$.

Given a convex function $g(\cdot): \Complex^{m\times n}\to \Real$ and a positive real number
$\lambda$, the proximal mapping
$\prox_{\lambda g}(\cdot): \Complex^{m\times n}\to \Complex^{m\times n}$ is defined as
$\prox_{\lambda g}(\vect{A}) := \arg\min_{\vect{B}} \lambda g(\vect{B}) + (1/2)
\norm{\vect{A}-\vect{B}}_{\text{F}}^2$. For example, in the case where $g$ becomes the indicator
function $\iota_{\mathcal{C}}$ with respect to a closed convex set
$\mathcal{C} \subset \Complex^{m\times n}$, \ie, $\iota_{\mathcal{C}}(\vect{A}) := 0$, if
$\vect{A}\in \mathcal{C}$, while $\iota_{\mathcal{C}}(\vect{A}) := +\infty$, if
$\vect{A}\notin \mathcal{C}$, then $\prox_{\lambda\iota_{\mathcal{C}}}$ becomes the metric
projection mapping onto $\mathcal{C}$:
$\prox_{\lambda\iota_{\mathcal{C}}}(\vect{A}) = \arg\min_{\vect{B}\in \mathcal{C}}
\norm{\vect{A}-\vect{B}}_{\text{F}}$. Moreover, in the case where $g$ is the $\ell_1$-norm
$\norm{\cdot}_1$, then the $(i,j)$th entry of $\prox_{\lambda\norm{\cdot}_1}(\vect{A})$ is given by
the soft-thresholding rule~\cite[Lemma~V.I]{maleki2013asymptotic}
\begin{align}
  [\prox_{\lambda\norm{\cdot}_1}(\vect{A})]_{ij} = [\vect{A}]_{ij} \left(1 -
  \tfrac{\lambda}{\max\{\lambda, \lvert[\vect{A}]_{ij}\rvert\}} \right)
  \,. \label{soft.thresholding}
\end{align}

\section{Solving for $\hat{\vect{U}}_n$}\label{app:solve.Un}

This refers to the convex minimization sub-task~\eqref{task:min.for.U} and provides important
details essential to the implementation of Alg.~\ref{alg:algorithm.Un.Bn}. All derivations,
including those in Sec.~\ref{app:solve.Bn}, are performed on the basis of viewing the complex-valued
$\vect{U}$ as $(\Re(\vect{U}), \Im(\vect{U}))$, where $\Re(\cdot)$ and $\Im(\cdot)$ stand for the
real and imaginary parts, respectively, of a complex-valued matrix. Gradients are
\textit{not} considered in the complex-differentiability sense (Cauchy-Riemann
conditions)~\cite{Lang.complex}. Nevertheless, to save space and use compact mathematical
expressions, all subsequent results, including those in Sec.~\ref{app:solve.Bn}, are stated in their
complex-valued form.

Upon defining the convex constraint
$\mathcal{C}_i := \Set{\vect{U}\ \given \norm{\vect{Ue}_i} \leq C_U }, \forall i\in\Set{1, \ldots,
  d}$, \eqref{task:min.for.U} can be expressed as
\begin{align}
  \hat{\vect{U}}_n \in \arg\min\nolimits_{\vect{U}} g_1(\vect{U}) +
  g_2(\vect{U})\,, \label{supp:define.uhat}
\end{align}
where
\begin{alignat}{2}
  g_1(\vect{U}) 
  & {} := {} && \tfrac{1}{2} \norm{\mathcal{S}(\vect{Y})    -
    \mathcal{SF}(\vect{U}\check{\bm\Lambda}\vect{B}_n)}^{2}_{\text{F}}
    +  \tfrac{\tau_U}{2}\norm{\vect{U} - \vect{U}_n}^2_{\text{F}} \notag\\
  &&& + \tfrac{\lambda_1}{2} \norm{{\vect{Z}_n-
      \mathcal{F}_t(\vect{U}\check{\bm\Lambda}\vect{B}_n)}}_{\text{F}}^2\,, \label{supp:originial.g1} \\
  g_2(\vect{U}) 
  & := && \sum\nolimits_{i=1}^d \iota_{\mathcal{C}_i}(\vect{U})\,. \label{supp:original.g2}
\end{alignat}

It can be verified that $\forall\vect{U}$,
\begin{alignat}{2}
  \nabla g_1(\vect{U}) 
  & {} = {} && \left[ N_{\text{k}} \mathcal{F}^{-1}\mathcal{S} \mathcal{F} 
    (\vect{U}\check{\bm\Lambda}\vect{B}_n) + \lambda_1 N_\text{fr}
    \vect{U}\check{\bm\Lambda}\vect{B}_n \right] \vect{B}_n^{\hermconj} \check{\bm\Lambda}^{\hermconj}
  \notag\\
  &&& + \tau_U (\vect{U} - \vect{U}_n) \notag\\
  &&& - \left[ N_{\text{k}} \mathcal{F}^{-1}\mathcal{S}(\vect{Y}) + \lambda_1
  N_\text{fr}\mathcal{F}_t^{-1}(\vect{Z}_n) \right] \vect{B}_n^{\hermconj}
  \check{\bm\Lambda}^{\hermconj}\,. \label{supp:gradient.U}
\end{alignat}
By virtue of the fact $\norm{\mathcal{F}^{-1} \mathcal{S} \mathcal{F} (\vect{U})}_{\text{F}} \leq
\norm{\vect{U}}_{\text{F}}$, $\forall \vect{U}$, it can be also verified that $\forall\vect{U}_1,
\vect{U}_2$,
\begin{align}
  & \norm{\nabla{g_1(\vect{U}_1)} - \nabla{g_1(\vect{U}_2)}}_{\text{F}} \notag \\ 
  & \leq \norm*{(N_{\text{k}} + \lambda_1 N_\text{fr})
    \check{\bm\Lambda}\vect{B}_n \vect{B}_n^{\hermconj} \check{\bm\Lambda}^{\hermconj} +
    \tau_U\vect{I}_d}_2 \cdot\norm{\vect{U}_1 - \vect{U}_2}_{\text{F}} \,,
\end{align}
which yields the Lipschitz coefficient
\begin{align}
    L := (N_{\text{k}} + \lambda_1 N_\text{fr}) \cdot \lambda_{\max}
  \left( \check{\bm\Lambda}\vect{B}_n \vect{B}_n^{\hermconj} \check{\bm\Lambda}^{\hermconj} \right)
  + \tau_U \,. \label{supp:lipschitz.U} 
\end{align}

The $i$th column of $\prox_{\lambda g_2}(\vect{U})$, $\forall \lambda>0$ and $\forall \vect{U}$,
is computed by the (metric) projection mapping onto $\mathcal{C}_i$:
\begin{align}\label{supp:proximal.U}
  \prox_{\lambda g_2}(\vect{U}) \vect{e}_i = \tfrac{C_U}{\max\{C_U, \norm{\vect{U}\vect{e}_i} \}}
  \vect{U}\vect{e}_i \,.
\end{align}
Lastly, since \eqref{supp:define.uhat} is not affinely constrained, \cite{slavakis.FMHSDM} suggests
\begin{align}
  T(\vect{U}) = \vect{U}, \quad \forall\vect{U}\,, \label{supp:T.for.U}
\end{align}
for Alg.~\ref{alg:algorithm.Un.Bn}.

\section{Solving for $\hat{\vect{B}}_n$}\label{app:solve.Bn}

This refers to the convex minimization sub-task~\eqref{task:min.for.B} and provides important
details, essential to the implementation of Alg.~\ref{alg:algorithm.Un.Bn}. Upon defining the affine
constraint
$\mathcal{C}_{\text{aff}} := \Set{ \vect{B} \mid \vect{1}_{N_{\ell}
    }^{\intercal}\vect{B} = \vect{1}_{N_\text{fr}}^{\intercal}}$, \eqref{task:min.for.B}
can be expressed as
\begin{align}
  \hat{\vect{B}}_n \in \arg\min\nolimits_{\vect{B}\in\mathcal{C}_{\text{aff}}}\ g_1(\vect{B}) +
  g_2(\vect{B})\,, \label{supp:define.bhat}
\end{align}
where
\begin{alignat}{2}
  g_1(\vect{B})
  & {} = {} &&
  \tfrac{1}{2} \norm{\mathcal{S}(\vect{Y}) -
    \mathcal{SF}(\vect{U}_n\check{\bm{\Lambda}}\vect{B})}^{2}_\text{F} 
  + \tfrac{\tau_B}{2}\norm{\vect{B} - \vect{B}_n}^2_\text{F}\notag \label{supp:b.g1.term}\\ 
  &&& + \tfrac{\lambda_1}{2} \norm{{\vect{Z}_n -
      \mathcal{F}_t(\vect{U}_n\check{\bm{\Lambda}}\vect{B})}}_\text{F}^2 \,, \\ 
  g_2(\vect{B})
  & = && \lambda_3\norm{\vect{B}}_1 \,. \label{supp:b.g2.term}
\end{alignat}

It can be verified that $\forall\vect{B}$,
\begin{alignat}{2}
  \nabla{g_1}(\vect{B})
  & {} = {} && \check{\bm{\Lambda}}^{\hermconj}\vect{U}_n^{\hermconj}
  [N_{\text{k}} \mathcal{F}^{-1}\mathcal{S}
  \mathcal{F}(\vect{U}_n\check{\bm{\Lambda}}\vect{B}) + \lambda_1 N_{\text{fr}}
  \vect{U}_n\check{\bm{\Lambda}}\vect{B}] \notag \\
  &&& + \tau_B (\vect{B} - \vect{B}_n) \notag\\
  &&& -\check{\bm{\Lambda}}^{\hermconj} \vect{U}_n^{\hermconj}
  [N_{\text{k}} \mathcal{F}^{-1}\mathcal{S}(\vect{Y}) 
  + \lambda_1 N_{\text{fr}} \mathcal{F}_t^{-1}(\vect{Z}_n)]\,, \label{supp:gradient.B}
\end{alignat}
and by steps similar to those in Sec.~\ref{app:solve.Un}, $\nabla{g_1}(\cdot)$ is Lipschitz
continuous with coefficient
\begin{align}
  L = (N_{\text{k}}+ \lambda_1 N_{\text{fr}}) \cdot
  \lambda_{\max} \left(\check{\bm{\Lambda}}^{\hermconj} \vect{U}_n^{\hermconj}
  \vect{U}_n\check{\bm{\Lambda}} \right) + \tau_B \,. \label{supp:lipschitz.B}
\end{align}
According to \eqref{soft.thresholding}, $\prox_{\lambda g_2}(\cdot)$ becomes:
$\forall \vect{B}$,
\begin{align}
  [\prox_{\lambda g_2}]_{ij} = [\vect{B}]_{ij} 
  \left( 1 - \tfrac{\lambda\lambda_3} {\max\{\lambda\lambda_3,{|[\vect{B}]_{ij}|\}}} \right)
  \,. \label{supp:proximal.B}
\end{align}
Moreover, according to~\cite{slavakis.FMHSDM}, mapping $T(\cdot)$, used in
Alg.~\ref{alg:algorithm.Un.Bn}, is set equal to the projection mapping onto
$\mathcal{C}_{\text{aff}}$: $\forall\vect{B}$,
\begin{align}
  T(\vect{B}) = \vect{B} - \tfrac{1}{N_{\ell}} \vect{1}_{N_{\ell}}
  (\vect{1}_{N_{\ell}}^\intercal\vect{B} - \vect{1}_{N_\text{fr}}^\intercal) \,.\label{supp:T.for.B}
\end{align}

\begin{thebibliography}{10}
\providecommand{\url}[1]{#1}
\csname url@samestyle\endcsname
\providecommand{\newblock}{\relax}
\providecommand{\bibinfo}[2]{#2}
\providecommand{\BIBentrySTDinterwordspacing}{\spaceskip=0pt\relax}
\providecommand{\BIBentryALTinterwordstretchfactor}{4}
\providecommand{\BIBentryALTinterwordspacing}{\spaceskip=\fontdimen2\font plus
\BIBentryALTinterwordstretchfactor\fontdimen3\font minus
  \fontdimen4\font\relax}
\providecommand{\BIBforeignlanguage}[2]{{%
\expandafter\ifx\csname l@#1\endcsname\relax
\typeout{** WARNING: IEEEtran.bst: No hyphenation pattern has been}%
\typeout{** loaded for the language `#1'. Using the pattern for}%
\typeout{** the default language instead.}%
\else
\language=\csname l@#1\endcsname
\fi
#2}}
\providecommand{\BIBdecl}{\relax}
\BIBdecl

\bibitem{Liang.Lauterbur.book}
Z.-P. Liang and P.~C. Lauterbur, \emph{Principles of Magnetic Resonance
  Imaging: A Signal Processing Perspective}.\hskip 1em plus 0.5em minus
  0.4em\relax IEEE Press, 2000.

\bibitem{Petersen.Middleton.62}
D.~P. Petersen and D.~Middleton, ``Sampling and reconstruction of
  wave-number-limited functions in {N}-dimensional {E}uclidean spaces,''
  \emph{Information and Control}, vol.~5, no.~4, pp. 279--323, 1962.

\bibitem{Liang.Lauterbur.dMRI.94}
Z.-P. Liang and P.~C. Lauterbur, ``An efficient method for dynamic magnetic
  resonance imaging,'' \emph{IEEE Trans.\ Medical Imag.}, vol.~13, no.~4, pp.
  677--686, 1994.

\bibitem{wang2016accelerating}
S.~Wang, Z.~Su, L.~Ying, X.~Peng, S.~Zhu, F.~Liang, D.~Feng, and D.~Liang,
  ``Accelerating magnetic resonance imaging via deep learning,'' in
  \emph{Proc.\ ISBI}, 2016, pp. 514--517.

\bibitem{chen2017low}
H.~Chen, Y.~Zhang, M.~K. Kalra, F.~Lin, Y.~Chen, P.~Liao, J.~Zhou, and G.~Wang,
  ``Low-dose {CT} with a residual encoder-decoder convolutional neural
  network,'' \emph{IEEE Trans.\ Medical Imag.}, vol.~36, no.~12, pp.
  2524--2535, 2017.

\bibitem{zhu2018image}
B.~Zhu, J.~Z. Liu, S.~F. Cauley, B.~R. Rosen, and M.~S. Rosen, ``Image
  reconstruction by domain-transform manifold learning,'' \emph{Nature}, vol.
  555, no. 7697, p. 487, 2018.

\bibitem{jin2017deep}
K.~H. Jin, M.~T. McCann, E.~Froustey, and M.~Unser, ``Deep convolutional neural
  network for inverse problems in imaging,'' \emph{IEEE Transactions on Image
  Processing}, vol.~26, no.~9, pp. 4509--4522, 2017.

\bibitem{aggarwal2018model}
H.~K. Aggarwal, M.~P. Mani, and M.~Jacob, ``Model based image reconstruction
  using deep learned priors ({MODL}),'' in \emph{2018 IEEE 15th International
  Symposium on Biomedical Imaging (ISBI 2018)}.\hskip 1em plus 0.5em minus
  0.4em\relax IEEE, 2018, pp. 671--674.

\bibitem{schlemper2018deep}
J.~Schlemper, J.~Caballero, J.~V. Hajnal, A.~N. Price, and D.~Rueckert, ``A
  deep cascade of convolutional neural networks for dynamic {MR} image
  reconstruction,'' \emph{IEEE Trans. Medical Imag.}, vol.~37, no.~2, pp.
  491--503, 2018.

\bibitem{hammernik2018learning}
K.~Hammernik, T.~Klatzer, E.~Kobler, M.~P. Recht, D.~K. Sodickson, T.~Pock, and
  F.~Knoll, ``Learning a variational network for reconstruction of accelerated
  {MRI} data,'' \emph{Magnetic Resonance in Medicine}, vol.~79, no.~6, pp.
  3055--3071, 2018.

\bibitem{mardani2017recurrent}
M.~Mardani, E.~Gong, J.~Y. Cheng, J.~Pauly, and L.~Xing, ``Recurrent generative
  adversarial neural networks for compressive imaging,'' in \emph{Proc.\ IEEE
  CAMSAP}, 2017, pp. 1--5.

\bibitem{mardani2017deep}
M.~Mardani, E.~Gong, J.~Y. Cheng, S.~Vasanawala, G.~Zaharchuk, M.~Alley,
  N.~Thakur, S.~Han, W.~Dally, J.~M. Pauly \emph{et~al.}, ``Deep generative
  adversarial networks for compressed sensing automates {MRI},'' \emph{arXiv
  e-print}, 2017, 1706.00051.

\bibitem{lustig2006kt}
M.~Lustig, J.~M. Santos, D.~L. Donoho, and J.~M. Pauly, ``k-t {SPARSE}: {H}igh
  frame rate dynamic {MRI} exploiting spatio-temporal sparsity,'' in
  \emph{Proc.\ ISMRM}, vol. 2420, 2006.

\bibitem{Jung.07}
H.~Jung, J.~C. Ye, and E.~Y. Kim, ``Improved k-t {BLAST} and k-t {SENSE} using
  {FOCUSS},'' \emph{Physics in Medicine and Biology}, vol.~52, no.~11, pp.
  3201--3226, 2007.

\bibitem{otazo2010combination}
R.~Otazo, D.~Kim, L.~Axel, and D.~K. Sodickson, ``Combination of compressed
  sensing and parallel imaging for highly accelerated first-pass cardiac
  perfusion {MRI},'' \emph{Magnetic Resonance in Medicine}, vol.~64, no.~3, pp.
  767--776, 2010.

\bibitem{liang2012k}
D.~Liang, E.~V.~R. DiBella, R.-R. Chen, and L.~Ying, ``k-t {ISD}: {D}ynamic
  cardiac {MR} imaging using compressed sensing with iterative support
  detection,'' \emph{Magnetic Resonance in Medicine}, vol.~68, no.~1, pp.
  41--53, 2012.

\bibitem{lingala2011ktslr}
S.~G. Lingala, Y.~Hu, E.~V.~R. DiBella, and M.~Jacob, ``Accelerated dynamic
  {MRI} exploiting sparsity and low-rank structure: k-t {SLR},'' \emph{IEEE
  Trans.\ Medical Imag.}, vol.~30, no.~5, pp. 1042--1054, 2011.

\bibitem{zhao2012pssparse}
B.~Zhao, J.~P. Haldar, A.~G. Christodoulou, and Z.-P. Liang, ``Image
  reconstruction from highly undersampled (k,t)-space data with joint partial
  separability and sparsity constraints,'' \emph{IEEE Trans.\ Medical Imag.},
  vol.~31, no.~9, pp. 1809--1820, 2012.

\bibitem{block2007undersampled}
K.~T. Block, M.~Uecker, and J.~Frahm, ``Undersampled radial {MRI} with multiple
  coils: {I}terative image reconstruction using a total variation constraint,''
  \emph{Magnetic Resonance in Medicine}, vol.~57, no.~6, pp. 1086--1098, 2007.

\bibitem{knoll2011second}
F.~Knoll, K.~Bredies, T.~Pock, and R.~Stollberger, ``Second order total
  generalized variation ({TGV}) for {MRI},'' \emph{Magnetic Resonance in
  Medicine}, vol.~65, no.~2, pp. 480--491, 2011.

\bibitem{feng2014golden}
L.~Feng, R.~Grimm, K.~T. Block, H.~Chandarana, S.~Kim, J.~Xu, L.~Axel, D.~K.
  Sodickson, and R.~Otazo, ``Golden-angle radial sparse parallel {MRI}:
  combination of compressed sensing, parallel imaging, and golden-angle radial
  sampling for fast and flexible dynamic volumetric {MRI},'' \emph{Magnetic
  Resonance in Medicine}, vol.~72, no.~3, pp. 707--717, 2014.

\bibitem{poddar2016dynamic}
S.~Poddar and M.~Jacob, ``Dynamic {MRI} using smoothness regularization on
  manifolds {(SToRM)},'' \emph{IEEE Trans.\ Medical Imag.}, vol.~35, no.~4, pp.
  1106--1115, 2016.

\bibitem{ravishankar2017.lassi}
S.~Ravishankar, B.~E. Moore, R.~R. Nadakuditi, and J.~A. Fessler, ``Low-rank
  and adaptive sparse signal ({LASSI}) models for highly accelerated dynamic
  imaging,'' \emph{IEEE Trans.\ Medical Imag.}, vol.~36, no.~5, pp. 1116--1128,
  2017.

\bibitem{feng2016xd}
L.~Feng, L.~Axel, H.~Chandarana, K.~T. Block, D.~K. Sodickson, and R.~Otazo,
  ``{XD-GRASP}: {G}olden-angle radial {MRI} with reconstruction of extra
  motion-state dimensions using compressed sensing,'' \emph{Magnetic resonance
  in medicine}, vol.~75, no.~2, pp. 775--788, 2016.

\bibitem{awate2012spatiotemporal}
S.~P. Awate and E.~V.~R. DiBella, ``Spatiotemporal dictionary learning for
  undersampled dynamic {MRI} reconstruction via joint frame-based and
  dictionary-based sparsity,'' in \emph{Proc.\ ISBI}, 2012, pp. 318--321.

\bibitem{wang2014compressed}
Y.~Wang and L.~Ying, ``Compressed sensing dynamic cardiac cine {MRI} using
  learned spatiotemporal dictionary.'' \emph{IEEE Trans. Biomed. Engineering},
  vol.~61, no.~4, pp. 1109--1120, 2014.

\bibitem{caballero2014dictionary}
J.~Caballero, A.~N. Price, D.~Rueckert, and J.~V. Hajnal, ``Dictionary learning
  and time sparsity for dynamic {MR} data reconstruction,'' \emph{IEEE Trans.\
  Medical Imag.}, vol.~33, no.~4, pp. 979--994, 2014.

\bibitem{Nakarmi2016accelerating}
U.~Nakarmi, Y.~Zhou, J.~Lyu, K.~Slavakis, and L.~Ying, ``Accelerating dynamic
  magnetic resonance imaging by nonlinear sparse coding,'' in \emph{Proc.\
  ISBI}, 2016.

\bibitem{Wang2017parallel}
Y.~Wang, N.~Cao, Z.~Liu, and Y.~Zhang, ``Real-time dynamic {MRI} using parallel
  dictionary learning and dynamic total variation,'' \emph{Neurocomputing},
  vol. 238, pp. 410--419, 2017.

\bibitem{nakarmi2017m}
U.~Nakarmi, K.~Slavakis, J.~Lyu, and L.~Ying, ``{M-MRI}: A manifold-based
  framework to highly accelerated dynamic magnetic resonance imaging,'' in
  \emph{Proc.\ ISBI}, 2017, pp. 19--22.

\bibitem{poddar2018free}
S.~Poddar, Y.~Mohsin, D.~Ansah, B.~Thattaliyath, R.~Ashwath, and M.~Jacob,
  ``Free-breathing cardiac {MRI} using bandlimited manifold modelling,''
  \emph{arXiv preprint arXiv:1802.08909}, 2018.

\bibitem{poddar2018recovery}
S.~Poddar and M.~Jacob, ``Recovery of noisy points on bandlimited surfaces:
  Kernel methods re-explained,'' in \emph{2018 IEEE International Conference on
  Acoustics, Speech and Signal Processing (ICASSP)}.\hskip 1em plus 0.5em minus
  0.4em\relax IEEE, 2018, pp. 4024--4028.

\bibitem{Usman.Manifold.15}
M.~Usman, D.~Atkinson, C.~Kolbitsch, T.~Schaeffter, and C.~Prieto, ``Manifold
  learning based {ECG}-free free-breathing cardiac {CINE MRI},'' \emph{J.\
  Magnetic Resonance Imag.}, vol.~41, no.~6, pp. 1521--1527, 2015.

\bibitem{nakarmi2018mls}
U.~Nakarmi, K.~Slavakis, and L.~Ying, ``{MLS}: {J}oint manifold-learning and
  sparsity-aware framework for highly accelerated dynamic magnetic resonance
  imaging,'' in \emph{Proc.\ ISBI}, 2018, pp. 1213--1216.

\bibitem{ahmed2019free}
A.~H. Ahmed, Y.~Mohsin, R.~Zhou, Y.~Yang, M.~Salerno, P.~Nagpal, and M.~Jacob,
  ``Free-breathing and ungated cardiac cine using navigator-less spiral
  {SToRM},'' \emph{arXiv preprint arXiv:1901.05542}, 2019.

\bibitem{shen2017nonlinear}
Y.~Shen, P.~A. Traganitis, and G.~B. Giannakis, ``Nonlinear dimensionality
  reduction on graphs,'' in \emph{Proc.\ IEEE CAMSAP}, 2017.

\bibitem{silva2006selecting}
J.~Silva, J.~Marques, and J.~Lemos, ``Selecting landmark points for sparse
  manifold learning,'' in \emph{Proc.\ NIPS}, 2006, pp. 1241--1248.

\bibitem{chen2006improved}
Y.~Chen, M.~Crawford, and J.~Ghosh, ``Improved nonlinear manifold learning for
  land cover classification via intelligent landmark selection,'' in
  \emph{Proc.\ IEEE IGARSS}, 2006, pp. 545--548.

\bibitem{Landmark.MinMax}
V.~De~Silva and J.~B. Tenenbaum, ``Sparse multidimensional scaling using
  landmark points,'' Stanford University, Tech. Rep., 2004.

\bibitem{Tu.book.08}
L.~W. Tu, \emph{An Introduction to Manifolds}.\hskip 1em plus 0.5em minus
  0.4em\relax New~York: Springer, 2008.

\bibitem{RSE.13}
K.~Slavakis, G.~B. Giannakis, and G.~Leus, ``Robust sparse embedding and
  reconstruction via dictionary learning,'' in \emph{Proc.\ CISS}, Baltimore:
  USA, Mar. 2013.

\bibitem{slavakis2017bi}
K.~Slavakis, G.~N. Shetty, A.~Bose, U.~Nakarmi, and L.~Ying, ``Bi-linear
  modeling of manifold-data geometry for dynamic-{MRI} recovery,'' in
  \emph{Proc.\ IEEE CAMSAP}, 2017.

\bibitem{Saul.Roweis.03}
L.~K. Saul and S.~T. Roweis, ``Think globally, fit locally: {U}nsupervised
  learning of low dimensional manifolds,'' \emph{J.\ Machine Learning
  Research}, vol.~4, pp. 119--155, 2003.

\bibitem{Rockafellar.convex.analysis}
R.~T. Rockafellar, \emph{Convex Analysis}.\hskip 1em plus 0.5em minus
  0.4em\relax Princeton, NJ: Princeton University Press, 1970.

\bibitem{Elhamifar.Vidal.nips.11}
E.~Elhamifar and R.~Vidal, ``Sparse manifold clustering and embedding,'' in
  \emph{Proc.\ NIPS}, Granada: Spain, Dec. 2011.

\bibitem{slavakis.FMHSDM}
K.~Slavakis and I.~Yamada, ``Fej\'{e}r-monotone hybrid steepest descent method
  for affinely constrained and composite convex minimization tasks,''
  \emph{Optimization}, 2018, {DOI}: \url{10.1080/02331934.2018.1505885}.

\bibitem{friedman2001elements}
J.~Friedman, T.~Hastie, and R.~Tibshirani, \emph{The Elements of Statistical
  Learning}, 2nd~ed.\hskip 1em plus 0.5em minus 0.4em\relax Springer, 2009.

\bibitem{facchinei2015parallel}
F.~Facchinei, G.~Scutari, and S.~Sagratella, ``Parallel selective algorithms
  for nonconvex big data optimization,'' \emph{IEEE Trans.\ Signal Process.},
  vol.~63, no.~7, pp. 1874--1889, 2015.

\bibitem{candes2013unbiased}
E.~J. Candes, C.~A. Sing-Long, and J.~D. Trzasko, ``Unbiased risk estimates for
  singular value thresholding and spectral estimators,'' \emph{IEEE Trans.\
  Signal Process.}, vol.~61, no.~19, pp. 4643--4657, 2013.

\bibitem{wissmann2014mrxcat}
L.~Wissmann, C.~Santelli, W.~P. Segars, and S.~Kozerke, ``{MRXCAT}: Realistic
  numerical phantoms for cardiovascular magnetic resonance,'' \emph{Journal of
  Cardiovascular Magnetic Resonance}, vol.~16, no.~1, p.~63, 2014.

\bibitem{MATLAB.2018b}
``{MATLAB} ({R}2018b),'' {T}he {M}athworks, {I}nc., {N}atick, {M}assachusetts,
  2018.

\bibitem{code.storm}
\BIBentryALTinterwordspacing
S.~Poddar and M.~Jacob, ``{SToRM MATLAB Package},'' accessed in 2018. [Online].
  Available: \url{https://github.com/sunrita-poddar/l2SToRM}
\BIBentrySTDinterwordspacing

\bibitem{code.lassi}
\BIBentryALTinterwordspacing
S.~Ravishankar, B.~E. Moore, R.~R. Nadakuditi, and J.~A. Fessler, ``{LASSI
  MATLAB Package},'' accessed in 2018. [Online]. Available:
  \url{https://web.eecs.umich.edu/~fessler/irt/reproduce/17/ravishankar-17-lra/}
\BIBentrySTDinterwordspacing

\bibitem{code.xdgrasp}
\BIBentryALTinterwordspacing
L.~Feng, L.~Axel, H.~Chandarana, K.~T. Block, D.~K. Sodickson, and R.~Otazo,
  ``{XD-GRASP MATLAB Package},'' accessed in 2019. [Online]. Available:
  \url{https://cai2r.net/resources/software/xd-grasp-matlab-code}
\BIBentrySTDinterwordspacing

\bibitem{ravishankar2011mr}
S.~Ravishankar and Y.~Bresler, ``{MR} image reconstruction from highly
  undersampled k-space data by dictionary learning,'' \emph{IEEE transactions
  on medical imaging}, vol.~30, no.~5, pp. 1028--1041, 2011.

\bibitem{subbarao1993focusing}
M.~Subbarao, T.-S. Choi, and A.~Nikzad, ``Focusing techniques,'' \emph{Optical
  Engineering}, vol.~32, no.~11, pp. 2824--2837, 1993.

\bibitem{wang2004image}
Z.~Wang, A.~C. Bovik, H.~R. Sheikh, E.~P. Simoncelli \emph{et~al.}, ``Image
  quality assessment: From error visibility to structural similarity,''
  \emph{IEEE transactions on image processing}, vol.~13, no.~4, pp. 600--612,
  2004.

\bibitem{lin2000thin}
Y.-C. Tsai, H.-D. Lin, Y.-C. Hu, C.-L. Yu, and K.-P. Lin, ``Thin-plate spline
  technique for medical image deformation,'' \emph{J.\ Medical and Biological
  Eng.}, vol.~20, no.~4, pp. 203--210, 2000.

\bibitem{maleki2013asymptotic}
A.~Maleki, L.~Anitori, Z.~Yang, and R.~G. Baraniuk, ``Asymptotic analysis of
  complex {LASSO} via complex approximate message passing {(CAMP)},''
  \emph{IEEE Trans.\ Information Theory}, vol.~59, no.~7, pp. 4290--4308, 2013.

\bibitem{Lang.complex}
S.~Lang, \emph{Complex Analysis}, 4th~ed.\hskip 1em plus 0.5em minus
  0.4em\relax New~York: Springer, 1999.

\end{thebibliography}

\end{document}